\documentclass[useAMS,usenatbib]{mn2e}

\usepackage{times}
\usepackage{graphicx}
\usepackage{subfigure}
\usepackage{amsmath}
\newcommand{\rd}{{\rm d}}
\newcommand{\gsim}{\mathrel{\hbox{\rlap{\lower.55ex \hbox {$\sim$}}
                   \kern-.3em \raise.4ex \hbox{$>$}}}}
\newcommand{\lsim}{\mathrel{\hbox{\rlap{\lower.55ex \hbox {$\sim$}}
                   \kern-.3em \raise.4ex \hbox{$<$}}}}

\title[Combining radiative transfer and the diffuse ISM]{Combining radiative transfer and diffuse interstellar medium physics to model star formation}
\author[M. R. Bate \& E. R. Keto]{Matthew R. Bate$^{1}$\thanks{E-mail:
mbate@astro.ex.ac.uk} and Eric R. Keto$^{2,3}$\\ 
$^{1}$ School of Physics and Astronomy, University of Exeter, Stocker
Road, Exeter EX4 4QL \\ 
$^{2}$ Harvard-Smithsonian Center for Astrophysics, 160 Garden St., Cambridge, MA 02420, USA \\
$^{3}$ Max-Planck-Institut f\"ur Astronomie, K\"onigstuhl 17, D-69117 Heidelberg, Germany
}

\date{\today}
\begin{document}
\maketitle
\begin{abstract}
We present a method for modelling star-forming clouds that combines two different models of the thermal evolution of the interstellar medium (ISM).  In the combined model, where the densities are low enough that at least some part of the spectrum is optically thin, a model of the thermodynamics of the diffuse ISM is more significant in setting the temperatures.  Where the densities are high enough to be optically thick across the spectrum, a model of flux limited diffusion is more appropriate.  Previous methods either model the low-density interstellar medium and ignore the thermal behaviour at high densities (e.g. inside collapsing molecular cloud cores), or model the thermal behaviour near protostars but assume a fixed background temperature (e.g. $\approx 10$~K) on large-scales.  Our new method treats both regimes.  It also captures the different thermal evolution of the gas, dust, and radiation separately.  We compare our results with those from the literature, and investigate the dependence of the thermal behaviour of the gas on the various model parameters.  This new method should allow us to model the ISM across a wide range of densities and, thus, develop a more complete and consistent understanding of the role of thermodynamics in the star formation process.

\end{abstract}

\section{Introduction}
\label{introduction}

The thermal behaviour of interstellar gas is of fundamental importance for star formation.  To initiate star formation, the thermal pressure must be insufficient to support the gas against gravitational collapse \citep{Jeans1929}.  Further evolution also depends on the thermodynamics and density structure, with a variety of different outcomes being possible such as the formation of a single polytrope or fragmentation into many clouds \citep{Hoyle1953, Hunter1962, Layzer1963, Tohline1980, Rozyczka1983}.  \cite{Larson1985, Larson2005} emphasised the importance of the relationship between temperature and density in the interstellar medium, and \cite*{WhiBofFra1998} emphasised the importance of the transition from molecular cooling to dust cooling.  The typical decrease in gas temperature from $T_{\rm g} \sim 1000$~K at number densities of hydrogen nuclei of $n_{\rm H}\sim 1$~cm$^{-3}$ to $T_{\rm g} \sim 10$~K at densities of $n_{\rm H}\sim 10^4$~cm$^{-3}$ promotes fragmentation, while the transition to an isothermal regime or  temperature that increases with density above $n_{\rm H}\sim 10^4$~cm$^{-3}$ may help to produce a characteristic stellar mass \citep*{Larson1985, Larson2005, Jappsenetal2005, BonClaBat2006, ElmKleWil2008}.  Chemical reactions that control the abundances of gas phase coolants, and therefore radiative equilibrium, may affect this transition and the formation of molecular cloud cores.  At even higher densities, deep within collapsing molecular cloud cores, different phases of collapse are believed to occur as the gas becomes optically thick and later as molecular hydrogen dissociates \citep{Larson1969}.  The treatment of radiative transfer can be crucial for determining whether the core fragments into multiple protostars or not \citep{Bossetal2000, Krumholz2006, WhiBat2006}.

Previous radiation hydrodynamical simulations of star cluster formation demonstrate the importance of radiative feedback from protostars to correctly capture fragmentation and produce realistic numbers of brown dwarfs \citep{Bate2009b, Offneretal2009}.  Whereas barotropic calculations of star cluster formation produce characteristic stellar masses that depend on the initial Jeans mass in the cloud \citep{BatBon2005}, \cite{Bate2009b} showed that radiative feedback reduces this dependence on the initial conditions.  This may help to explain the apparent universality of the stellar initial mass function (IMF), at least in recent epochs \citep[see also][]{Krumholz2011}.  Indeed, some recent radiation hydrodynamical simulations of star cluster formation have been quite successful in reproducing the observed IMF and other properties of stellar systems \citep*{Bate2012, Bate2014, KruKleMcK2012}.

However, to date, radiation hydrodynamical simulations of star formation have been restricted to modelling dense molecular clouds and assuming that the only significant sources of radiation are the protostars themselves.  In this case, the gas temperature at large distances from the protostars is assumed to be $\approx 10$~K.  However, this approximation is generally invalid at densities $n_{\rm H}\lsim 10^4$~cm$^{-3}$ with solar metallicities, and is invalid at even higher densities at lower metallicities, if at all.  

Informed by the vast literature on the physics, thermodynamics, and chemistry of the diffuse interstellar medium (ISM) \citep[see the thorough review provided by][]{Tielens2005} and collapsing clouds with different metallicities \citep[e.g.][]{Omukai2000, Omukaietal2005}, interest has grown in studying the formation and evolution of molecular clouds using three-dimensional hydrodynamical calculations coupled with chemistry.  Initial calculations treated only the formation of molecular hydrogen (\citealt*{DobBonPri2006, DobBon2007, GloMac2007a, GloMac2007b}; \citealt{Micicetal2012}).  More recent work has included more complex chemistry, in particular that involving carbon and oxygen \citep{Gloveretal2010, GloMac2011, GloCla2012a, GloCla2012b, GloCla2012c, Clarketal2012}.  Along with the formation mechanisms of molecular clouds and their chemical and temperature distributions, some of these calculations have begun to investigate star formation, using sink particles to replace collapsing regions of gas (in particular the models of \citeauthor{GloCla2012a}).  At the same time, there have been several studies of the chemical and thermal evolution of low metallicity clouds and their star formation \citep[e.g.][]{Dopckeetal2011, Dopckeetal2013, GloCla2012c}.

Therefore, at the present time, the star formation community has two different classes of hydrodynamical models for star formation.  One class treats the low-density thermochemical evolution in some detail, but does not treat radiative feedback from protostars.  The other the ignores the complicated physics of the diffuse ISM, which is particularly important at low densities and low metallicities, but includes the radiative effects of protostars.  The goal of this paper is to make an attempt to bridge this gap, by combining a model for the physics of the low-density interstellar medium with a method for modelling radiative transfer.  Our main purpose is to develop a model that does a reasonable job of modelling the thermodynamics at both low and high densities in star-forming clouds, with metallicities as low as 1/100 solar.  The aim is not to produce a detailed chemical model.  Rather, we wish to implement the simplest possible chemical model that will provide the abundances of the major coolants of the gas that are necessary to calculate realistic temperatures.

It turns out that a method very similar to that which we present here has recently been developed by \cite{PavZhi2013} and \cite{Pavlyuchenkovetal2015}.  Although we developed our method independently, each of the methods is based on extending a method of radiative transfer that is valid at high densities to include effects that are important at low densities.  Each of the methods treat gas and dust temperatures separately.  \cite{PavZhi2013} base their method on the diffusion approximation for radiative transfer and add heating and cooling terms relevant at lower densities to model the collapse of dense molecular cloud cores.  \cite{Pavlyuchenkovetal2015} replace the diffusion approximation with flux-limited diffusion like we use in this paper.  They do not include cooling processes relevant at very low-densities that we include (e.g. electron recombination and fine structure emission from atomic oxygen), and they only perform one-dimensional calculations, but the underlying methods are very similar.

This paper is primarily a method paper where we describe our implementation and demonstrate its performance in a wide variety of tests, comparing our results to those of others who have performed similar calculations (sometimes using much more complete and/or complex models).  However, we also explore the effects of varying many of the parameters that go into the model in order to better understand which physical processes may be most important for affecting star formation.  Large-scale star formation calculations are beyond the scope of this paper, but we hope to use this new method to perform such calculations in the future.

In Section 2, we provide the fundamental equations and assumptions that go into our model.  The implementation of these equations into a smoothed particle hydrodynamics (SPH) code is described in Section 3.  We present the results from our test calculations, and compare our results with those in the literature in Section 4.  Finally, we draw our conclusions in Section 5.

\section{Method}
\label{sec:method}

\subsection{The flux-limited diffusion approximation}

In a frame co-moving with the fluid, and assuming local thermal equilibrium
(LTE), the equations governing the time-evolution of radiation hydrodynamics
(RHD) can be written
\begin{equation}
\label{rhd1}
\frac{{\rm D}\rho}{{\rm D}t} + \rho \nabla\cdot \mbox{\boldmath $v$} = 0~,
\end{equation}
\begin{equation}
\label{rhd2}
\rho \frac{{\rm D}\mbox{\boldmath $v$}}{{\rm D}t} = -\nabla p + \frac{\mbox{${\chi_{}}_{\rm \scriptscriptstyle F}\rho$}}{c} \mbox{\boldmath $F$}~,
\end{equation}
\begin{equation}
\label{rhd3}
\rho \frac{{\rm D}}{{\rm D}t}\left( \frac{E}{\rho}\right) = -\nabla\cdot \mbox{\boldmath $F$} - \nabla\mbox{\boldmath $ v${\bf :P}} + 4\pi \kappa_{\rm \scriptscriptstyle P} \rho B - c \kappa_{\rm \scriptscriptstyle E} \rho E~,
\end{equation}
\begin{equation}
\label{rhd4}
\rho \frac{{\rm D}}{{\rm D}t}\left( \frac{e}{\rho}\right) = -p \nabla\cdot \mbox{\boldmath $v$} - 4\pi \kappa_{\rm \scriptscriptstyle P} \rho B + c \kappa_{\rm \scriptscriptstyle E} \rho E~,
\end{equation}
\citep*{MihMih1984,TurSto2001,WhiBatMon2005}.  In these equations, ${\rm D}/{\rm D}t \equiv \partial/\partial t + \mbox{\boldmath $v$}\cdot \nabla$ is the convective derivative.  The symbols $\rho, e,$ \mbox{\boldmath $v$} and $p$ represent the material mass density, energy density, velocity, and scalar isotropic pressure, respectively, and $c$ is the speed of light.  The total frequency-integrated radiation density, momentum density (flux) and pressure tensor are represented by $E$, {\boldmath $F$}, and {\bf P}, respectively.  The assumption of LTE allows the rate of emission of radiation from the matter in equations \ref{rhd3} and \ref{rhd4} to be written as the frequency-integrated Planck function, $B$.  Equations \ref{rhd2}--\ref{rhd4} have been integrated over frequency, leading to the flux mean total opacity $\chi_{\rm F}$, and the Planck mean and energy mean absorption opacities, $\kappa_{\rm P}$ and $\kappa_{\rm E}$.  The total opacity, $\chi$, is the sum of components due to absorption $\kappa$ and scattering $\sigma$.

Taking an ideal equation of state for the gas pressure $p=(\gamma -1)u \rho$, where $u=e/\rho$ is the specific energy of the gas, the temperature of the gas is $T_{\rm g} = (\gamma -1) \mu u/{\cal{R}}$ where $\mu$ is the mean molecular weight of the gas and $\cal{R}$ is the gas constant.  The frequency-integrated Planck function is given by $B=(\sigma_{\rm B}/\pi)T_{\rm g}^4$, where $\sigma_{\rm B}$ is the Stefan-Boltzmann constant.  The radiation energy density also has an associated temperature $T_{\rm r}$ from the equation $E=4 \sigma_{\rm B}T_{\rm r}^4/c$.

A common approximation to make in radiation hydrodynamics is the so-called flux-limited diffusion approximation.  For an isotropic radiation field $\mbox{\bf P}=E\mbox{\bf I}/3$.  The Eddington approximation assumes this relation holds everywhere and implies that, in a steady state
\begin{equation}
\mbox{\boldmath $F$} = - \frac{c}{3 \chi \rho}\nabla E.
\label{flux}
\end{equation}
This expression gives the correct flux in optically thick regions, where
$\chi\rho$ is large. However in optically thin regions where 
$\chi\rho \rightarrow 0$ the flux tends to infinity whereas in reality 
$|\mbox{\boldmath $F$}| \le cE$.  Flux-limited diffusion solves this problem
by limiting the flux in optically thin environments to always obey this
inequality.  \citet{LevPom1981} wrote the radiation flux
in the form of Fick's law of diffusion as
\begin{equation}
\label{fld1}
\mbox{\boldmath $F$} = -D \nabla E,
\end{equation}
with a diffusion constant given by
\begin{equation}
\label{fld2}
D = \frac{c\lambda}{\chi\rho}.
\end{equation}
The dimensionless function $\lambda(E)$ is called the flux limiter.  The 
radiation pressure tensor may then be written in terms of the radiation
energy density as
\begin{equation}
\label{fld3}
\mbox{\rm \bf P} = \mbox{ \rm \bf f} E,
\end{equation}
where the components of the Eddington tensor, {\bf f}, are given by
\begin{equation}
\label{fld4}
\mbox{\rm \bf f} = \frac{1}{2}(1-f)\mbox{\bf I} + \frac{1}{2}(3f-1)\mbox{\boldmath $\hat{n}\hat{n}$},
\end{equation}
where $\mbox{\boldmath $\hat{n}$}=\nabla E/|\nabla E|$ is the unit vector
in the direction of the radiation energy density gradient and the dimensionless
scalar function $f(E)$ is called the Eddington factor.  The flux limiter
and the Eddington factor are related by
\begin{equation}
\label{fld5}
f = \lambda + \lambda^2 R^2,
\end{equation}
where $R$ is the dimensionless quantity $R = |\nabla E|/(\chi\rho E)$.

Equations \ref{fld1} to \ref{fld5} close the equations of RHD.  
However, we must still choose
an expression for the flux limiter, $\lambda$.  In this paper, we choose
the flux limiter of \citet{LevPom1981} 
\begin{equation}
\lambda(R) = \frac{2+R}{6 + 3R + R^2}.
\end{equation}

There are a number of problems with using the flux-limited diffusion approximation to 
model radiative processes in star formation.  Many of these are a direct result of the 
approximations made in deriving the method.  
The diffusion approximation is good at high optical depths, but the directional behaviour
of anisotropic radiation at low optical depths is modelled poorly.  This means, for
example, that shadowing is not reproduced.  The flux limiter gives the correct
limiting behaviour for the propagation rate in the optically thin and 
thick regimes, but at intermediate optical depths is dependent on the arbitrary choice 
of the flux limiter.  Finally, most implementations take the grey
approach, ignoring the frequency dependence of the radiative transfer in favour of
mean opacities (as discussed above).
This is likely to be a good approximation when most of the energy is in long-wavelength
radiation, but is a poor approximation near hot massive stars  \citep*{WolCas1987,PreSonYor1995,YorSon2002,EdgCla2003}.

Despite these drawbacks, flux-limited diffusion is expected to do a reasonable job of
modelling radiative transfer in dense star-forming regions that produce 
low-mass stars \citep[e.g.][]{Bate2009b,Offneretal2009,Bate2012} and it is computationally
efficient.

\subsection{The diffuse interstellar medium}

However, other problems arise when modelling low-density environments.  For a start, the
above equations treat radiation and matter, but do not distinguish between different
types of matter which may have different temperatures.  In particular, at solar metallicities,
dust and gas temperatures are only well coupled (by collisions) at gas densities 
above $10^5$~cm$^{-3}$ \citep{BurHol1983, Goldsmith2001,GloCla2012a}.  
At lower densities, the gas and dust temperatures
can be very different from one another with the gas temperature typically exceeding that
of the dust.  There are also sources of heating that affect the matter other than work done 
on the gas and the radiative interaction between the matter and the radiation field that are
included in equation \ref{rhd4}.  
Cosmic rays heat the gas by direct collision \citep*{GolHabFie1969,ODoWat1974,BlaDal1977,GolLan1978}.  
Ultraviolet photons heat the gas indirectly through the photoelectric release of hot electrons
from dust grains \citep{Draine1978, BakTie1994}.
Absorption and emission of radiation that is highly frequency dependent is also 
problematic in a grey treatment.  In particular, gas cooling occurs by
atomic and molecular line cooling which may be optically thin due to Doppler shifts
even when static clouds would be optically thick.  The external interstellar 
radiation (ISR) field \citep{Habing1968,WitJoh1973,Draine1978,Black1994},
attenuated by dust extinction, also heats the dust.

\subsection{Combining a diffuse interstellar medium model with flux-limited diffusion} 
\label{sec:combine}

In this section, we present a method to combine a model of the radiative equilibrium of
the diffuse ISM with flux-limited diffusion to determine the gas, dust, and radiation 
temperatures in both low-density and 
high-density regions of molecular clouds.  We include treatments for all of the effects mentioned
in the previous section.  We also allow for heating due to H$_2$ formation.  
We do not treat photoionisation or heating from X-rays.

We use equations \ref{rhd2}--\ref{rhd4} to model the continuum radiation
field and the gas, and add extra terms to handle cosmic ray heating, the photoelectric 
effect, and radiation that is strongly frequency dependent.  We replace equations \ref{rhd3}
and \ref{rhd4} with
\begin{equation}
\label{rhdnew3}
\begin{split}
\rho \frac{{\rm D}}{{\rm D}t}\left( \frac{E}{\rho}\right) = - & \nabla\cdot \mbox{\boldmath $F$} - \nabla\mbox{\boldmath $ v${\bf :P}} - a c \kappa_{\rm g} \rho \left(T_{\rm r}^4 - T_{\rm g}^4 \right) - \\ & a c \kappa_{\rm d} \rho  \left(T_{\rm r}^4 - T_{\rm d}^4 \right),
\end{split}
\end{equation}
\begin{equation}
\label{rhdnew4}
\begin{split}
\rho \frac{{\rm D}u}{{\rm D}t} = - & p \nabla\cdot \mbox{\boldmath $v$} + a c \kappa_{\rm g} \rho  \left(T_{\rm r}^4 - T_{\rm g}^4 \right) - \\ &   \Lambda_{\rm gd} - \Lambda_{\rm line} + \Gamma_{\rm cr} + \Gamma_{\rm pe} + \Gamma_{\rm H2,g}~,
\end{split}
\end{equation}
where equation \ref{rhdnew4} now describes the evolution of the specific internal energy of the gas, $u$, separately from the dust.  The extra terms at the end of equation \ref{rhdnew4} are  $\Lambda_{\rm gd}$ which takes account of the thermal interaction due to collisions between the gas and the dust, $\Lambda_{\rm line}$ which is the cooling rate per unit volume due to atomic and molecular line emission, $\Gamma_{\rm cr}$ which is the heating rate per unit volume due to cosmic rays,  $\Gamma_{\rm pe}$ which is the heating rate per unit volume due to the photoelectric effect, and $\Gamma_{\rm H2,g}$ which is the heating rate per unit volume due to the formation of molecular hydrogen on dust grains. We also define the radiation constant $a=4 \sigma_{\rm B}/c$ and the dust temperature $T_{\rm d}$.   We define the Rosseland mean gas opacity $\kappa_{\rm g}$ (which is only important above the dust sublimation temperature) and the mean dust opacity $\kappa_{\rm d}$, which may be a Planck mean or Rosseland mean (see Section \ref{sec:opacity}).   

We still need to determine the dust temperature.  As done in most studies of the thermal structure of molecular clouds \citep[e.g.][]{Goldsmith2001}, we assume that the dust is in LTE with the total radiation field (i.e. that its temperature responds quickly to any change).  This replaces a dust thermal energy equation (which would also require the thermodynamic properties of the dust).  Thus, we take
\begin{equation}
\label{dustLTE}
\rho \Lambda_{\rm ISR} + ac\kappa_{\rm d}\rho \left(T_{\rm r}^4 - T_{\rm d}^4\right) + \Lambda_{\rm gd} = 0,
\end{equation}
where $\Lambda_{\rm ISR}$ is the heating of the dust due to the interstellar radiation field (which is taken to be separate from the grey radiation field with temperature $T_{\rm r}$).  By rearranging equation \ref{dustLTE} for $\Lambda_{\rm gd}$ and eliminating this term from equation \ref{rhdnew4}, it is easy to show that equations \ref{rhdnew3} and \ref{rhdnew4} reduce to equations \ref{rhd3} and \ref{rhd4} when $T_{\rm d}=T_{\rm g}$ and $\Lambda_{\rm line}=\Lambda_{\rm gd}=\Gamma_{\rm cr}=\Gamma_{\rm pe}=\Gamma_{\rm H2,g}=0$, and where $\kappa_{\rm P}\rho=(\kappa_{\rm d} + \kappa_{\rm g})\rho$ and $\kappa_{\rm P}=\kappa_{\rm E}$.

We note that the dust cooling rate
\begin{equation}
\label{eq:dustopacity}
\begin{split}
\Lambda_{\rm dust}  = ac\kappa_{\rm d} T_{\rm d}^4 & = 4 \sigma_{\rm B} \left( \int Q_\nu B_\nu {\rm d}\nu \right) \frac{\pi}{\sigma_{\rm B} T_{\rm d}^4} T_{\rm d}^4, \\
& = 4 \pi \int Q_\nu B_\nu {\rm d}\nu
\end{split}
\end{equation}
where $\nu$ is the frequency of the radiation, and $Q_\nu$ is the frequency-dependent dust absorption efficiency.  Therefore when $\Lambda_{\rm gd}=0$, and $T_{\rm r}=0$, and in the absence of extinction, the thermal balance with the interstellar radiation is obtained when
\begin{equation}
\label{eq:normalise}
\Lambda_{\rm dust} = \Lambda_{\rm ISR} = 4 \pi \int Q_\nu J^{\rm ISR}_\nu {\rm d}\nu,
\end{equation}
where $J^{\rm ISR}_\nu$ is the frequency-dependent ISR field.

We note that equations 12 and 13 in \cite{Goldsmith2001} appear to be incorrect.  It is stated that the dust cooling rate is given by
\begin{equation}
\Lambda_{\rm dust} = c \int U_\nu(T_{\rm d})\kappa(\nu) {\rm d}\nu,
\end{equation}
where $U_\nu(T)$ is the Planck energy density and where they took
\begin{equation}
\kappa(\nu) = 3.3 \times 10^{-26} n_{\rm H2} (\nu/\nu_0)^2 ~{\rm cm}^{-1},
\end{equation}
with $\nu_0=3.8\times 10^{11}~{\rm Hz}$, and $n_{\rm H2}$ is the number density of molecular hydrogen.
However, we can write
\begin{equation}
\begin{split}
\Lambda_{\rm dust} & =  c \int U_\nu(T_{\rm d})\kappa(\nu) {\rm d}\nu = 4 \pi \int B_\nu(T_{\rm d}) \kappa(\nu) {\rm d}\nu, \\
& =  \frac{3.3\times 10^{-26} n_{\rm H2} 4 \pi}{{\nu_0}^2} \int \frac{2h\nu^5}{c^2} \frac{1}{e^{h\nu/(kT_{\rm d})} - 1}  {\rm d}\nu, \\
& =  \frac{3.3\times 10^{-26} n_{\rm H2} 8 \pi h}{{\nu_0}^2 c^2} \int \nu^5 \frac{1}{e^{h\nu/(kT_{\rm d})} - 1}  {\rm d}\nu, \\
& =  \frac{3.3\times 10^{-26} n_{\rm H2} 8 \pi}{{\nu_0}^2 c^2} \frac{k^6T_{\rm d}^6}{h^5}\int s^5 \frac{1}{e^s - 1}  {\rm d}s,
\end{split}
\end{equation}
where $h$ is Planck's constant, $k$ is Boltzmann's constant, and we have written $s=h\nu/(kT_{\rm d})$.  The integral can be performed over all frequencies, and can be evaluated as
\begin{equation}
\int^\infty_0 \frac{s^5}{e^s - 1}  {\rm d}s = \Gamma(6) \zeta(6) = 5!~\zeta(6) = \frac{8\pi^6}{63} \approx 122.08
\end{equation}
where $\Gamma(n)=(n-1)!$ is the Gamma function and the Riemann zeta function, $\zeta(2n)$ can be evaluated is
\begin{equation}
\zeta(2n) = \frac{(-1)^{n+1} B_{2n} (2\pi)^{2n}}{2(2n)!}
\end{equation}
so that
\begin{equation}
\label{eq:zeta}
\zeta(6) = \frac{(-1)^{4} B_{6} (2\pi)^{6}}{2(6)!}  = \frac{\pi^6}{945}
\end{equation}
where $B_{2n}$ is a Bernoulli number and $B_6=1/42$.  Thus,
\begin{equation}
\begin{split}
\label{lambda_dust}
\Lambda_{\rm dust} & =  \frac{3.3\times 10^{-26} n_{\rm H2} 8 \pi}{{\nu_0}^2 c^2} \frac{k^6T_{\rm d}^6}{h^5} 122.08, \\
& = 4.22 \times 10^{-31} n_{\rm H2} T_{\rm d}^6 ~~{\rm erg~cm}^{-3}~{\rm s}^{-1}.
\end{split}
\end{equation}
This is a factor of $\approx 62$ times larger (probably $2^6=64$) than equation 13 in \cite{Goldsmith2001} which states $\Lambda_{\rm dust}  =  6.8 \times 10^{-33} n_{\rm H2} T_{\rm d}^6$.  The effect of the greater dust cooling rate is to lower the dust temperatures by a factor of two for a given ISR field.  We note that \cite{GloCla2012b} use $\Lambda_{\rm dust}  =  4.68 \times 10^{-31} n_{\rm H2} T_{\rm d}^6$, which only differs from equation \ref{lambda_dust} by 10\% (presumably due to the choice of different dust opacities).

\subsection{Equations for heating and cooling terms in the ISM}
\label{sec:heatingcooling}

To calculate the extra heating and cooling terms appearing in equations \ref{rhdnew3} to \ref{dustLTE}, we make the same approximations as those made in the past by many others who have studied the thermal structure of molecular clouds.  

\subsubsection{Gas heating equations}
\label{sec:heating}

Following \cite{Goldsmith2001}, and \cite{KetFie2005}, we set the rate of energy transfer from cosmic rays into the gas as
\begin{equation}
\label{eq:cosmicray}
\Gamma_{\rm cr}= 5 \times 10^{-28} n_{\rm H} ~{\rm erg~cm}^{-3}~{\rm s}^{-1}.
\end{equation}
Note that \cite{Goldsmith2001} and \cite{KetFie2005} express their rates as functions of $n_{\rm H2}=n_{\rm H}/2$ for molecular gas.  We will usually refer to $n_{\rm H}$ throughout this paper, since when we allow for both atomic and molecular hydrogen the meaning of $n_{\rm H2}$ may not be clear.
%
%Note that this is much larger than Falgarone \& Puget (1985) who use 1.5x10^-28 n_H  (n_H=2 n_H2 so this is about 3.3 times less).
%

The ISR is used to determine both the heating rate of the dust grains, and the photoelectric heating rate of the gas.  In both cases, the ISR is attenuated due to dust extinction inside a molecular cloud.
To describe the ISR, we use a slight modification to that described in detail by \cite*{ZucWalGal2001}.  They made approximate fits to the ISR given by \cite{Black1994} using the sum of a power-law distribution and five modified blackbody distributions of the form
\begin{equation}
\label{eq:blackbodies}
J^{\rm ISR}_\nu = \frac{2h\nu^3}{c^2} \left( \frac{\lambda_{\rm p}}{\lambda}\right)^p \sum_i \frac{W_i}{e^{h\nu/kT_{\rm i}} - 1},
\end{equation}
where $\lambda$ is the wavelength of the radiation, and the parameters $\lambda_{\rm p}$, $W_i$, and $T_i$ are given in  their Table B.1.  Note that the mid-infrared term in their ISR is a power-law with a cut-off longward of 100~$\mu$m (which is mentioned in the main text of the paper, but not in their appendix).  Although \cite{ZucWalGal2001}'s parameterisation does a good job of fitting \citeauthor{Black1994}'s ISR at most wavelengths, it does not provide a significant ultraviolet (UV) flux which is necessary for photoelectric heating.  To allow us to use one ISR parameterisation for both dust heating and the photoelectric effect, we add the `standard' UV background from equation 11 of \cite{Draine1978} but only in the range $h\nu = 5-13.6$~eV.  Note that to transform \citeauthor{Draine1978}'s parameterisation into the same units as equation \ref{eq:blackbodies} it must be multiplied by $h^2\nu/{\rm eV}$.

%
%c      IF (v.GT.5.0*1.602E-12/h .AND. v.LT.13.6*1.602E-12/h) THEN
%c         E_eV = v*h/1.602E-12
%c         val = val + (1.658E+06*E_eV - 2.152E+05*E_eV**2 + 
%c     &        6.919E+03*E_eV**3)*h*E_eV
%

For the photoelectric heating, we follow the prescription of \cite{BakTie1994}, which has also been used by \cite{Youngetal2004} and \cite{KetCas2008}, that
\begin{equation}
\label{eq:photoelectric}
\Gamma_{\rm pe} = 1.33 \times 10^{-24} \epsilon ~ G({\bf r}) n_{\rm H} ~{\rm erg~cm}^{-3}~{\rm s}^{-1},
\end{equation}
where, as described by \cite{Youngetal2004}, we have assumed that the number density of nucleons $n_{\rm n}  \approx n_{\rm H}+4 ~n({\rm He})$ and that the gas is 25 percent helium by mass so that $n_{\rm n}= 1.33~n_{\rm H}$.  The quantity $G({\bf r})$ is the ratio of the attenuated intensity of high energy ($h\nu>6$~eV) photons to their unattenuated intensity
\begin{equation}
\label{eq:g}
G({\bf r}) = \frac{\displaystyle \int^{4\pi}_{0} \int^{\infty}_{6{\rm eV}} J^{\rm ISR}_{\rm \nu} \exp[-\tau_\nu(\mbox{\boldmath{$r$}},\omega)] {\rm d}\nu {\rm d}\Omega}{\displaystyle  \int^{4\pi}_{0} \int^{\infty}_{6{\rm eV}} J^{\rm ISR}_{\rm \nu}  {\rm d}\nu {\rm d}\Omega },
\end{equation}
where $\tau_\nu(\mbox{\boldmath{$r$}},\omega)$ is the frequency-dependent optical depth from \mbox{\boldmath{$r$}} to the cloud surface along direction, $\omega$, and $\Omega$ is solid angle. The efficiency factor $\epsilon$ is a complicated function of the type of dust grains, radiation intensity, $G(\mbox{\boldmath{$r$}})$, the temperature, and the electron number density.  We use equation 43 of \cite{BakTie1994} 
\begin{equation}
\begin{array}{ll}
\epsilon = & \displaystyle \frac{0.049}{1+4\times 10^{-3}[G(\mbox{\boldmath{$r$}}) ~T_{\rm g}^{1/2}/(n_{\rm e}\phi_{\rm PAH})]^{0.73}}  \\ \\
& \displaystyle +  \frac{0.037 ~(T_{\rm g}/10^4)^{0.7}}{1+2\times 10^{-4}[G(\mbox{\boldmath{$r$}})~ T_{\rm g}^{1/2}/(n_{\rm e}\phi_{\rm PAH})]},
\end{array}
\end{equation}
but with the adjustable parameter $\phi_{\rm PAH}$ that was introduced by \cite{Wolfireetal2003}.  We follow \cite{Wolfireetal2003} and use $\phi_{\rm PAH}=0.55$ whereas the original equation of \cite{BakTie1994} had 
$\phi_{\rm PAH}=1$.  
For the conditions in cold molecular clouds the efficiency factor is nearly constant and can be approximated as $\epsilon=0.05$ \cite[note that in][there is a typographical error giving $\epsilon= 0.5$, and they also neglect the factor of 1.33 in $\Gamma_{\rm pe}$]{KetCas2008}.  However, it is important to use the more complicated form at the lower densities of the warm and cold neutral mediums.  
Ordinarily the electron number density, $n_{\rm e}$, would come from a chemical model, but because we do not have such a model we need to parameterise it.  Using the results of \cite{Wolfireetal2003}, in particular those displayed in their Fig.~10, we use the simple parameterisation
\begin{equation}
n_{\rm e} = n_{\rm H} \max(1\times 10^{-4}, \min(1, 0.008/n_{\rm H})).
\end{equation}

\subsubsection{Gas cooling equations}

We develop a model for cooling in the diffuse ISM based on the results of the detailed model developed by \cite{Wolfireetal2003} \citep[see also][]{GloMac2007a}.  In the warm neutral medium (WNM), with a characteristic temperature of $T\sim 8000$~K, cooling is dominated by Ly$\alpha$ emission from atomic hydrogen, electron recombination with small grains and polycyclic aromatic hydrocarbons (PAHs), and fine structure emission from atomic oxygen.  At temperatures between those of the WNM and the cold neutral medium (CNM; $T\lsim 300$~K), oxygen emission continues to contribute significant cooling but fine-structure emission from ionised carbon also becomes important and dominates in the colder parts of the CNM.  Unlike \cite{Wolfireetal2003} and \cite{GloMac2007a}, our aim is to achieve realistic temperatures without having to develop a detailed chemical model.  Since we are not interested in regions of the ISM with temperatures greater than those of the WNM, we can produce a reasonable fit to the thermal behaviour by treating only the electron recombination, oxygen, and carbon emission.  Following \cite{Wolfireetal2003} and \cite{GloMac2007a}, we use the modified formula of \cite{BakTie1994} 
\begin{equation}
\label{eq:recombination}
\Lambda_{\rm rec} = 4.65 \times 10^{-30} \phi_{\rm PAH} T_{\rm g}^{0.94} \varphi^\beta n_{\rm e} n_{\rm H} ~~{\rm erg~cm}^{-3} ~{\rm s}^{-1},
\end{equation}
where $\beta=0.74/T_{\rm g}^{0.068}$, and 
\begin{equation}
\varphi = \frac{G(\mbox{\boldmath{$r$}}) T_{\rm g}^{1/2}}{n_{\rm e}\phi_{\rm PAH}}.
\end{equation}

For the atomic oxygen fine-structure cooling, we use equation C3 of \cite{Wolfireetal2003}
\begin{equation}
\label{eq:oxygen}
\Lambda_{\rm OI} = 2.5 \times 10^{-27} n_{\rm H}^2 \left(\frac{T_{\rm g}}{\rm 100~K}\right)^{0.4}   \exp\left(\frac{-228}{T_{\rm g}}\right) ~{\rm erg~cm}^{-3}~{\rm s}^{-1}.
\end{equation}
\cite*{KetRawCas2014} calculate the cooling taking into account the abundance of atomic oxygen.

For the atomic carbon fine-structure cooling, we use equation C1 of \cite{Wolfireetal2003} 
\begin{equation}
\label{eq:carbon}
\Lambda_{\rm C^+} = 3.15 \times 10^{-27} n_{\rm H}^2 \exp\left(\frac{-92}{T_{\rm g}}\right)  ~~{\rm erg~cm}^{-3}~{\rm s}^{-1}.
%
% Valid for n_H < 3x10^3 according to Tielens2005
%
\end{equation}
which assumes a carbon abundance relative to hydrogen of $\cal{A}_{\rm C}= \rm 1.4 \times 10^{-4}$. This is almost identical to equation of 2.67 of \cite{Tielens2005}, which was also used by \cite{KetCas2008}, except that in equation \ref{eq:carbon} the dependence on the degree of ionisation, $X_{\rm i}$, has been neglected.  \cite{KetCas2008} assume that the degree of ionisation is equal to the abundance of C$^+$ with respect to H$_2$.

\begin{figure}
\centering \vspace{-0.3cm} \hspace{-0cm}
    \includegraphics[width=8.0cm]{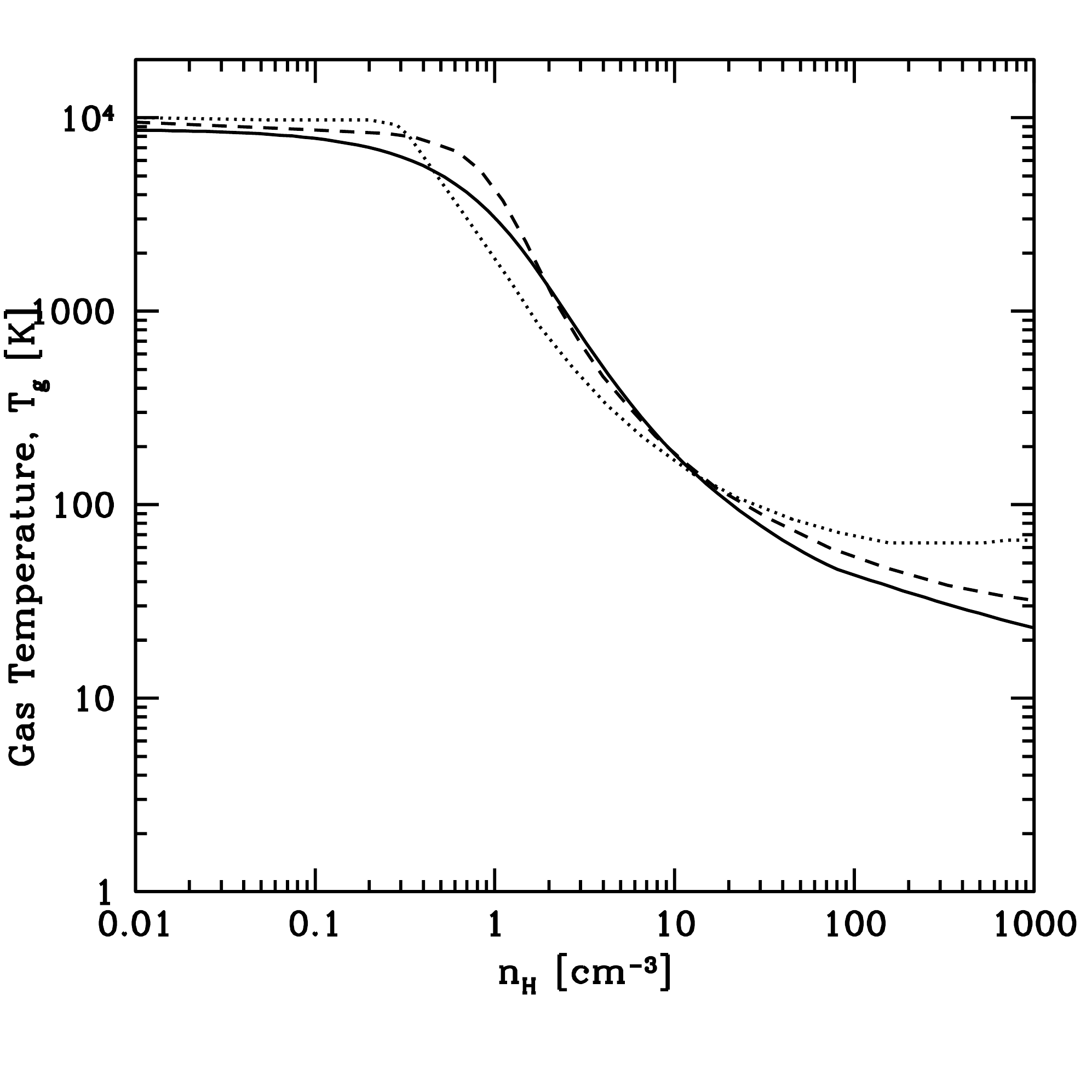} \vspace{-1cm}
\caption{The equilibrium temperature as a function of the number density of hydrogen nuclei, $n_{\rm H}$, in the warm and cold neutral mediums achieved by considering cosmic ray and photoelectric heating (without extinction) balanced by cooling due to electron recombination, and fine structure emission from carbon and oxygen (solid line; equations \ref{eq:cosmicray},  \ref{eq:photoelectric}, \ref{eq:recombination}, \ref{eq:oxygen}, and \ref{eq:carbon}).  The dashed line gives the result obtained by Wolfire et al.~(2003) for solar metallicity gas at a Galactic radius of 8.5 kpc. The dotted line gives the result from the model of Glover \& MacLow (2007a).}
\label{T_n}
\end{figure}

At the low densities of the WNM and CNM, the equations for cosmic ray and photoelectric heating and cooling due to electron recombination and oxygen and carbon cooling can be used to calculate the equilibrium temperature as a function of density.  This equilibrium curve resulting from our chosen parameterisations is shown in Fig.~\ref{T_n} as the solid line and compared to the results obtained by \cite{Wolfireetal2003} for solar metallicities at a Galactic radius of 8.5 kpc using a dashed line.  Given that our model is much simpler than that of \cite{Wolfireetal2003} and that our main interest is in the star formation occurring in higher-density ($n_{\rm H}\gsim 100$) molecular gas where other processes dominate, the level of agreement is satisfactory.  Also plotted in the figure with a dotted line is the result obtained by \cite{GloMac2007a}.  We note that our model gives slightly lower temperatures than \cite{Wolfireetal2003} at densities $n_{\rm H} \gsim 20$~cm$^{-3}$, while the model of \cite{GloMac2007a} gives somewhat higher temperatures.

For the molecular line cooling we follow \cite{KetFie2005} by using the parameterised cooling functions provided by \cite{Goldsmith2001}  for standard abundances given by
\begin{equation}
\label{eq:line}
\Lambda_{\rm line} = \alpha (T_{\rm g} /10~{\rm K})^{\beta}  ~{\rm erg~cm}^{-3}~{\rm s}^{-1},
\end{equation}
where the parameters $\alpha$ and $\beta$ are given in tables in \cite{Goldsmith2001} as functions of $n_{\rm H2}$ (we take $n_{\rm H2}=n_{\rm H}/2$).   We allow the possibility of performing calculations both with standard abundances and with depleted abundances.  In the former case, Table 2 of \cite{Goldsmith2001} gives cooling rates without allowing for depletion onto grains and covers the range $n_{\rm H2}=10^2-10^7$~cm$^{-3}$.  We tried using linear interpolation of the logarithm of the cooling rates at the given points to compute the cooling rate at a particular value of $n_{\rm H2}$, but found that this tended to give discontinuities in the temperature gradients in molecular cloud cores.  Instead we use three-point polynomial interpolation of the logarithm of the cooling rate (taking one point at the nearest tabulated value of $n_{\rm H2}$ below the required density and two points above the required density, except for the highest densities where the reverse is done).  This works well over the entire range of $n_{\rm H2}$ except that it gives a `bulge' in the cooling rate in the range $n_{\rm H2}=10^3-10^4$~cm$^{-3}$.  To fix this, in this range, we calculate two three-point polynomial interpolations (one that has one point below the required value of $n_{\rm H2}$ and two points above, and the other that has two tabulated values below and one above) and then use linear interpolation of the logarithm of these two values to obtain the cooling rate.  At densities $n_{\rm H2}<10^2$,~cm$^{-3}$ we extrapolate the line cooling rate using the three-point polynomial derived from the first three points in the table.  At densities $n_{\rm H2}>10^8$~cm$^{-3}$, we set the line cooling to zero (since it will be negligible compared to the other heating and cooling terms anyway).

To allow for depletion, we use Table 4 of \cite{Goldsmith2001} which gives cooling rates as functions of density over the range $n_{\rm H2}=10^3-10^6$~cm$^{-3}$ and as functions of a depletion factor ranging from 1 to 100.  When depletion is allowed, for $n_{\rm H2}<10^2$~cm$^{-3}$ we use the standard cooling rates divided by the depletion factor, and for $n_{\rm H2}>10^7$~cm$^{-3}$ we use the standard cooling rates (since the cooling rates become less dependent on the depletion factor at high densities, according to \citealt{Goldsmith2001}).  In the intermediate density range we use bilinear interpolation in log-space of the values in Table 4 that are given as functions of density and depletion.  In the ranges $n_{\rm H2}=10^2-10^3$~cm$^{-3}$ and $n_{\rm H2}=10^6-10^7$~cm$^{-3}$ we further use linear interpolation in log-space to achieve smooth transitions between using the depleted cooling rates (from Table 4) and the depleted standard cooling rates ($n_{\rm H2}=10^2-10^3$~cm$^{-3}$) or the standard cooling rates ($n_{\rm H2}=10^6-10^7$~cm$^{-3}$).

\subsubsection{Dust heating and cooling equations}
\label{sec:dust_equations}

For the heating rate of dust grains, we follow \cite{ZucWalGal2001} and \cite{KetFie2005} and calculate the grain heating due to an incident radiation field whose intensity is attenuated by the optical depth averaged over $4\pi$ steradians upon grains of the absorption efficiency, $Q_\nu$.  Thus,
\begin{equation}
\label{eq:grainheating}
\Lambda_{\rm ISR} =  \int_0^{4\pi} \int_0^{\infty} Q_\nu J^{\rm ISR}_\nu \exp[-\tau_\nu(\mbox{\boldmath{$r$}},\omega)] {\rm d}\nu {\rm d}\Omega
\end{equation}
where $\tau_\nu(\mbox{\boldmath{$r$}},\omega)$ is the frequency-dependent optical depth from \mbox{\boldmath{$r$}} to the cloud surface along direction, $\omega$, and we have deliberately omitted to normalise the integral over solid angle because of equation \ref{eq:normalise}.  \cite{ZucWalGal2001} defines $Q_\nu$ in units of cm$^{2}$~H$_2^{-1}$, but in the above equation we redefine $Q_\nu$ to be in units of cm$^2$~g$^{-1}$ by dividing by $\mu m_{\rm p}$, where $m_{\rm p}$ is the proton mass.  For simplicity we use the parameterisation of $Q_\nu$ given by \cite{ZucWalGal2001}.  See section \ref{sec:opacity} for more information regarding the opacities we use.

The thermal interaction between the gas and the dust depends sensitively on the distribution of grain sizes \citep{BurHol1983}.  We use two different rates in Section \ref{sec:calculations}, depending on which tests we are performing.  For most of the calculations, we use the rate of \cite{KetFie2005}
\begin{equation}
\label{eq:gasdust}
\Lambda_{\rm gd} = 2.5 \times 10^{-34} n_{\rm H}^2 T_{\rm g}^{1/2} (T_{\rm g} - T_{\rm d}) ~{\rm erg~cm}^{-3}~{\rm s}^{-1},
\end{equation}
which is similar to that used by \cite{Goldsmith2001} (our rate is a factor of $\sqrt{10}/2 \approx 1.6$ larger). However, this rate is more than an order of magnitude smaller than the rate given in equation 2.15 of \cite{HolMcK1989}
\begin{equation}
\label{eq:gasdust2}
\begin{split}
\Lambda_{\rm gd} = & ~3.8 \times 10^{-33} n_{\rm H}^2 T_{\rm g}^{1/2} (T_{\rm g} - T_{\rm d}) \\
&  \times \left[ 1-0.8 \exp\left(\frac{-75}{T_{\rm g}}\right) \right] ~{\rm erg~cm}^{-3}~{\rm s}^{-1},
\end{split}
\end{equation}
assuming a minimum grain size of 0.01~$\mu$m.  The last term in this equation takes account of the effects of the contributions of gas species other than protons and the effects of particle and grain charges.  This is the rate used by \cite{GloCla2012b}, so we use this rate for the tests in Section \ref{sec:testGC}.

\subsubsection{Metallicity}

In all of the above equations, we have assumed standard abundances and gas-to-dust mass ratios.  To allow the modelling of molecular gas with different metallicities, we assume that $Q_{\nu}$ (and hence $\kappa_{\rm d}$), $\Lambda_{\rm gd}$, $\Lambda_{\rm line}$, and $\Gamma_{\rm pe}$ all scale linearly with the metallicity.  Thus, each of these quantities is multiplied by the factor $Z/{\rm Z}_{\rm \odot}$.  In doing so, we are also assuming that the grain properties are independent of the metallicity.  Note that $\Gamma_{\rm cr}$ does not depend on the metallicity.  Also, the gas opacities, $\kappa_{\rm g}$, already explicitly include the metallicity dependence (see Section \ref{sec:opacity}).

\subsection{Chemistry}
\label{sec:chemistry}

As mentioned in the introduction, our aim with this method is to develop a relatively simple model that captures the most important thermodynamic behaviour of the low-density interstellar medium, not to develop a full chemical model of molecular clouds.  The equations in the previous section almost enable us to do this without modelling any chemistry at all.  However, as we will see in Section \ref{sec:testGC}, we do need to have some model to treat the transformation of C$^+$ into CO.  At low-densities ($n_{\rm H}\approx 10-10^3$~cm$^{-3}$) C$^+$ is the major coolant, while at higher densities CO is the primary coolant until dust takes over \citep[e.g.][]{GloCla2012b}.  Furthermore, particularly at low metallicities, extra heating of the gas can be provided by the formation of molecular hydrogen so we need to treat the evolution of atomic and molecular hydrogen.

\subsubsection{Carbon chemistry}
\label{sec:carbon}

\cite{KetCas2008} provide a very simple model of the abundances of C$^+$, neutral carbon, and CO, including the depletion of CO onto dust grains.  \cite{GloCla2012b} show that the model significantly under-estimates the abundance of neutral carbon, but this is only important in a narrow density range around $n_{\rm H}\approx 10^3$~cm$^{-3}$ and is unimportant for cooling \citep{GloCla2012a}.  Therefore, we implement the model of \citeauthor{KetCas2008} as stated in their paper, except as noted by \cite{GloCla2012b} there is a typographical error in equation 5 of \cite{KetCas2008} which should read
\begin{equation}
\frac{{\rm CO}}{{\rm C}^+} = \frac{6\times 10^{-16} n_{\rm H2}}{1.4 \times 10^{-11} G_0 \exp(-3.2 A_{\rm V})},
\end{equation}
where $G_0$ is the ISR field in units of the Habing flux \citep{Habing1968} and we take $G_0=1$.

The only significant extra quantity that we need to calculate in order to implement the chemical model of \citeauthor{KetCas2008} is the mean visual extinction
\begin{equation}
\label{eq:extinction}
\langle \exp(-A_{\rm V}) \rangle = \frac{1}{4 \pi} \int \exp(-A_{\rm V}) {\rm d}\Omega,
\end{equation}
where we take $A_{\rm V} = \Sigma Q_\nu({\rm V})$ where $\Sigma$ is the column density in g~cm$^{-2}$ (so if the hydrogen is fully molecular the column density of H$_2$ is $N_{\rm H2} = \Sigma/[ \mu m_{\rm H}]$) and we take the frequency of visual light to be that of light with a wavelength of 550~nm.
This is calculated at the same time as the integrals in equations \ref{eq:grainheating} and \ref{eq:g} (see Section \ref{sec:attenuation}).
We use the resulting C$^+$ abundance to scale the fine-structure carbon cooling in equation \ref{eq:carbon}, and we use the CO depletion factor that the model provides as the depletion value to use for the line cooling of \cite{Goldsmith2001}.

\subsubsection{Hydrogen chemistry}
\label{sec:hydrogen}

For the hydrogen chemistry, we only consider H$_2$ formation on grains and dissociation due to cosmic rays and photodissociation.  
\cite{Omukai2000} finds that grain formation dominates over gas phase formation (i.e.\ via H$^-$ or H$_2^+$ ions) for metallicities $Z/{\rm Z}_\odot \gsim 10^{-4}$.  \cite{Glover2003} compares gas-phase and grain-catalysed formation in detail and draws similar conclusions -- for temperatures less than a few hundred Kelvin, H$_2$ formation on grains is expected to dominate for dust-to-gas ratios $\gsim 10^{-3}-10^{-4}$ of that found in the local ISM, over wide ranges of densities and levels of ionization.
We neglect collisional dissociation of H$_2$ as at the temperatures we are dealing with in the low-density interstellar medium it is expected to be negligible compared to photodissociation and dissociation due to cosmic rays \citep[c.f. the rates in][]{LepShu1983}.

In the above sections, when we have used $n_{\rm H2}$ it has been in the context of fully molecular gas for which $n_{\rm H2}=n_{\rm H}/2$.  However, with the introduction of hydrogen chemistry, we need to distinguish between atomic and molecular hydrogen.  We therefore introduce the fraction of molecular hydrogen, $x_{\rm H2}$, which is equal to $n_{\rm H2}/n_{\rm H}=1/2$ when the gas is fully molecular, and is equal to zero when the gas is fully atomic.

For the formation rates of H$_2$ on grains, and the dissociation due to cosmic rays and photodissociation, we use the same model as \cite{Gloveretal2010}.  The formation rate is given by equation 165 in Table B1 of \cite{Gloveretal2010}, which is
\begin{equation}
\label{eq:h2form1}
R_{\rm H2} = ~3.0\times 10^{-18} ~ T_{\rm g}^{0.5} f_{\rm A} f_{\rm B} ~(Z/{\rm Z}_{\rm \odot})  ~~{\rm cm}^3~{\rm s}^{-1},
\end{equation}
where we have assumed that the rate scales linearly with the metallicity due to the variation in the grain abundance, and where
\begin{equation}
f_{\rm A} = \left[ 1.0 + 10^4 \exp(-600/T_{\rm d})\right]^{-1},
\end{equation}
\begin{equation}
f_{\rm B}=  [1.0 + 0.04(T_{\rm g}+T_{\rm d})^{0.5} + 0.002 T_{\rm g} + 8\times 10^{-6}T_{\rm g}^2]^{-1}.
\end{equation}
The number density of H$_2$ then evolves as
\begin{equation}
\left. \frac{{\rm d} n_{\rm H2}}{{\rm d}t}\right|_{\rm form} = R_{\rm H2} ~\left[ n_{\rm H} \left( {1-2 x_{\rm H2}}\right) \right]^2.
\end{equation}

For the magnitude of the dissociation of molecular hydrogen due to cosmic rays, we use the rate given in Table 2 of \cite{Berginetal2004}
\begin{equation}
\label{eq:h2form2}
\left. \frac{{\rm d} n_{\rm H2}}{{\rm d}t}\right|_{\rm cr}  = 1.2\times 10^{-17} x_{\rm H2} ~n_{\rm H}  ~~{\rm cm}^{-3}~{\rm s}^{-1}.
\end{equation}

For photodissociation of molecular hydrogen we use the same model as \cite{GloMac2007a} and \cite{Gloveretal2010}, which is based on the work of \cite{DraBer1996}.  We take the photodissociation rate of H$_2$ in optically thin gas to be
\begin{equation}
R_{\rm pd,0} = 5.6 \times 10^{-11} ~{\rm s}^{-1},
\end{equation}
where we have assumed a \cite{Draine1978} UV interstellar radiation field.  We take into account attenuation of the radiation due to dust absorption, and also self-shielding of the H$_2$ by line absorption due to other H$_2$ molecules.  For the former we use
\begin{equation}
\label{eq:fdust}
f_{\rm dust} = \langle \exp(-A_{\rm V}) \rangle^{3.74},
\end{equation}
which is simply a power of the mean visual extinction that is already required for the carbon chemistry (equation \ref{eq:extinction}).  For the self-shielding we use 
\begin{equation}
\label{eq:self}
f_{\rm shield} = \frac{0.965}{(1+x/b_5)^2} + \frac{0.035}{(1+x)^{1/2}}\exp \left[-8.5 \times 10^{-4}(1+x)^{1/2} \right],
\end{equation}
where $x=N_{\rm H2}/(5 \times 10^{14}~{\rm cm}^{-2})$, and $b_5 = b/(10^5~{\rm cm}~{\rm s}^{-1})$, where $b$ is the Doppler broadening parameter (which we take to be unity in the tests below).  The magnitude of the fully shielded H$_2$ photodissociation rate is then
\begin{equation}
\left. \frac{{\rm d} n_{\rm H2}}{{\rm d}t}\right|_{\rm pd}  = R_{\rm pd,0} f_{\rm dust}f_{\rm shield} x_{\rm H2} ~n_{\rm H}.
\end{equation}

Once the total rate of change of molecular hydrogen has been obtained, the rate of change of the H$_2$ fraction is evolved as
\begin{equation}
\frac{{\rm d} x_{\rm H2}}{{\rm d}t} = \frac{1}{n_{\rm H}} \frac{{\rm d} n_{\rm H2}}{{\rm d}t}.
\end{equation}

The formation of molecular hydrogen on dust grains releases approximately 4.5~eV of energy per molecule.  A fraction of this will be radiatively lost while the remainder will heat the gas, with the relative fractions depending on the collisional de-excitation rate which depends on density.  Following \cite{GloMac2007a}, we assume that the heating rate of the gas is
\begin{equation}
\Gamma_{\rm H2,g} = \frac{7.2 \times 10^{-12}}{(1+n_{\rm cr}/n_{\rm H})} \left. \frac{{\rm d} n_{\rm H2}}{{\rm d}t}\right|_{\rm form} ~{\rm erg~s}^{-1}~{\rm cm}^{-3},
\label{eq:h2dust}
\end{equation}
where
\begin{equation}
\frac{1}{n_{\rm cr}} = \frac{1-2x_{\rm H2}}{n_{\rm cr,H}} + \frac{2x_{\rm H2}}{n_{\rm cr,H2}},
\end{equation}
and 
\begin{equation}
\begin{split}
& \log n_{\rm cr,H}= 3.0 - 0.461 ~T_4 - 0.327 ~T_4^2, \\
& \log n_{\rm cr,H2}=4.845 - 1.3 ~T_4 + 1.62 ~T_4^2,
\end{split}
\end{equation}
with $T_4 = T_{\rm g}/10^4$~K and where the first equation, accounting for H$_2$-H interactions, is an order of magnitude less than the value given by \cite{LepShu1983}, as recommended by \cite{Martinetal1996}, and for H$_2$-H$_2$ interactions the equation is taken from \cite{ShaKan1987}. 

In addition to heating due to molecular hydrogen formation, we investigated (using calculations similar to those presented in Section \ref{sec:testGC}) the effect of including gas heating during H$_2$ photodissociation and due to UV pumping of H$_2$ \citep{GloMac2007a}.  However, we found that it is usually insignificant.  The extra heating only has a significant effect when the gas is initially highly molecular {\it and} has a low metallicity ($Z \lsim 0.1~{\rm Z}_\odot$).  Such circumstances are highly unrealistic because the low extinction favours prior destruction of the H$_2$.  Even then, the heating only affects the outer, low-density ($n_{\rm H} \lsim 10^3$~cm$^{-3}$) parts of the clouds (because of H$_2$ self-shielding) which are even less likely to be molecular than the inner parts of the clouds.  Since we find this heating to be insignificant in realistic situations, and for the sake of simplicity, we do not include this effect.

%
%When we do include it, we use the same parameterisation as \cite{GloMac2007a}.  We assume that each photodissociation deposits 0.4~eV of heat into the gas.  We further assume that, as well as dissociating some of the H$_2$, the FUV radiation also produces vibrationally excited H$_2$ via radiative pumping that, at high densities, heats the gas.  Adopting a radiative pumping rate that is 8.5  times the photodissociation rate and assuming an average energy transfer of $2/(1+n_{\rm cr}/n_{\rm H})$~eV to the gas, we obtain an overall heating rate due to UV radiation of
%\begin{equation}
%\begin{split}
%\Gamma_{\rm H2,pd} =  & ~\left(6.4 \times 10^{-13} + \frac{2.7 \times 10^{-11}}{1+n_{\rm cr}/n_{\rm H}} \right) \\
%& \times  \left. \frac{{\rm d} n_{\rm H2}}{{\rm d}t}\right|_{\rm pd} ~{\rm erg~s}^{-1}~{\rm cm}^{-3}.
%\end{split}
%\end{equation}

\section{Implementation}
\label{sec:implementation}

The calculations presented in this paper were performed 
using a three-dimensional smoothed particle
hydrodynamics (SPH) code based on the original 
version of \citeauthor{Benz1990} 
(\citeyear{Benz1990}; \citealt{Benzetal1990}), but substantially
modified as described in \citet*{BatBonPri1995},
\citet*{WhiBatMon2005}, \citet{WhiBat2006},
\cite{PriBat2007}, and 
parallelised using both OpenMP and the message passing interface (MPI).

Gravitational forces between particles and a particle's 
nearest neighbours are calculated using a binary tree.  
The cubic spline kernel is used and the smoothing lengths of particles are variable in 
time and space, set iteratively such that the smoothing
length of each particle 
$h = 1.2 (m/\rho)^{1/3}$ where $m$ and $\rho$ are the 
SPH particle's mass and density, respectively
\cite[see][for further details]{PriMon2007}.  The SPH equations are 
integrated using a second-order Runge-Kutta-Fehlberg 
integrator \citep{Fehlberg1969} with individual time steps for each particle
\citep{BatBonPri1995}.
To reduce numerical shear viscosity, we use the
\cite{MorMon1997} artificial viscosity
with $\alpha_{\rm_v}$ varying between 0.1 and 1 while $\beta_{\rm v}=2 \alpha_{\rm v}$
\citep[see also][]{PriMon2005}.

\subsection{Implementation of the new method}

In this section, we describe the detailed implementation of the method described in Section \ref{sec:method}.  The two main new elements are the implicit integration of equations \ref{rhdnew3} to \ref{dustLTE}, and the calculation of the dust attenuation which is required to compute the local ISR that appears in equations \ref{eq:g}, \ref{eq:grainheating}, and the extinction in equation \ref{eq:extinction}.

\subsubsection{Solving the energy equations}

In solving the energy equations \ref{rhdnew3} to \ref{dustLTE}, we closely follow the implementation of the flux-limited diffusion method of \cite{WhiBatMon2005} and \cite{WhiBat2006}.  The energy equations are very similar to those for the pure flux-limited diffusion method, except that the dust and gas now have different temperatures and there are some additional heating and cooling terms.  \cite{WhiBatMon2005} developed an implicit method for solving equations \ref{rhd3} and \ref{rhd4} based on writing the equations in terms of the specific radiation energy,  $\xi=E/\rho$, and the specific internal energy of the gas, $u$.  For each SPH particle $i$, implicit expressions were derived for these two quantities at time step $n+1$.  These two equations were combined so as to solve for $\xi^{n+1}_i$ and $u^{n+1}_i$.  The values from the previous time step, $\xi^{n}_i$ and $u^{n}_i$, were used as the initial guesses for $\xi^{n+1}_i$ and $u^{n+1}_i$, and Gauss-Seidel iteration was performed over all particles that were being evolved until the quantities converged to a given tolerance.  The same basic method is used here.

During each Gauss-Seidel iteration, the dust temperature must be solved for first.  Equation \ref{dustLTE} is solved directly for $T_{\rm d}$ of each particle based on its quantities from the previous time step or iteration.  This is a root finding problem, for which Newton-Raphson iteration converges very quickly.  It involves computing the lefthand side of equation \ref{dustLTE} and its derivative with respect to $T_{\rm d}$, which is straightforward except for the fact that $\kappa_{\rm d}$ is a function of $T_{\rm d}$. However, since $\kappa_{\rm d}$ is stored in a table as a function of $T_{\rm d}$ (see Section \ref{sec:opacity}), we simply use a numerical derivative to calculate ${\rm d}(\kappa_{\rm d})/{\rm d}(T_{\rm d})$.

Once the dust temperatures have been found, we solve for $\xi^{n+1}_i$ and $u^{n+1}_i$ in a similar manner to \cite{WhiBatMon2005}, but including the additional terms.  However, we found it necessary to solve the equations in two different ways in the low-density and high-density regimes, reflecting the fact that equations \ref{rhdnew3}--\ref{dustLTE} can be combined in two ways.  In the low-density regime, when the gas and dust are poorly coupled, we solve
\begin{equation}
\label{rhdnew5}
\begin{split}
\rho \frac{{\rm D}}{{\rm D}t}\left( \frac{E}{\rho}\right) = - & \nabla\cdot \mbox{\boldmath $F$} - \nabla\mbox{\boldmath $ v${\bf :P}} - a c \kappa_{\rm g} \rho \left(T_{\rm r}^4 - T_{\rm g}^4 \right)  \\ & + \rho \Lambda_{\rm ISR} +  \Lambda_{\rm gd} ,
\end{split}
\end{equation}
\vspace{-12pt}
\begin{equation}
\label{rhdnew6}
\begin{split}
\rho \frac{{\rm D}u}{{\rm D}t} = - & p \nabla\cdot \mbox{\boldmath $v$} + a c \kappa_{\rm g} \rho  \left(T_{\rm r}^4 - T_{\rm g}^4 \right) \\ & -  \Lambda_{\rm gd} - \Lambda_{\rm line} + \Gamma_{\rm cr} + \Gamma_{\rm pe}  + \Gamma_{\rm H2,g}~,
\end{split}
\end{equation}
where we have obtained \ref{rhdnew5} by rearranging equation \ref{dustLTE} to solve for the term $ac\kappa_{\rm d}\rho \left(T_{\rm r}^4 - T_{\rm d}^4\right)$ and replaced this term in equation \ref{rhdnew3}.  This works well in the poorly coupled regime because the $\Lambda_{\rm gd}$ term is very small.  Furthermore, in the low-density regime the gas temperature is usually $T_{\rm g} \ll 1000$~K and hence $\kappa_{\rm g}\approx 0$.  Thus, the two equations are almost uncoupled.  However, in the well coupled regime, we find it better to solve 
\begin{equation}
\label{rhdnew7}
\begin{split}
\rho \frac{{\rm D}}{{\rm D}t}\left( \frac{E}{\rho}\right) = - & \nabla\cdot \mbox{\boldmath $F$} - \nabla\mbox{\boldmath $ v${\bf :P}} - a c \kappa_{\rm g} \rho \left(T_{\rm r}^4 - T_{\rm g}^4 \right)  \\ & - a c \kappa_{\rm d} \rho  \left(T_{\rm r}^4 - T_{\rm d}^4 \right),
\end{split}
\end{equation}
\vspace{-12pt}
\begin{equation}
\label{rhdnew8}
\begin{split}
\rho \frac{{\rm D}u}{{\rm D}t} = - & p \nabla\cdot \mbox{\boldmath $v$} + a c \kappa_{\rm g} \rho  \left(T_{\rm r}^4 - T_{\rm g}^4 \right) + ac\kappa_{\rm d}\rho \left(T_{\rm r}^4 - T_{\rm d}^4\right)   \\ & + \rho \Lambda_{\rm ISR}   - \Lambda_{\rm line} + \Gamma_{\rm cr} + \Gamma_{\rm pe}  + \Gamma_{\rm H2,g}~,
\end{split}
\end{equation}
where we have obtained equation \ref{rhdnew8} by rearranging equation \ref{dustLTE} to solve for $\Lambda_{\rm gd}$ and replaced this term in equation \ref{rhdnew4}.  These equations are almost identical to those solved by \cite{WhiBatMon2005}, except for the last four terms in equation \ref{rhdnew8}, and these all tend to be very small when the gas and dust are well coupled.  This form is better in the well-coupled regime because the $\Lambda_{\rm gd}$ term in equations \ref{rhdnew5} and \ref{rhdnew6} can become very large when the gas and dust are well coupled even if the difference in the gas and dust temperatures is very small ($|T_{\rm g}-T_{\rm d}|\ll 0.1$~K).  In this case, the Gauss-Seidel iterations can fail to converge and/or the radiation energy density can become negative.  We use equations \ref{rhdnew5} and \ref{rhdnew6} when $n_{\rm H2}(Z/Z_\odot) < 10^{11}$~cm$^{-3}$ and $| T_{\rm r} - T_{\rm d} | >1$~K, otherwise we use equations \ref{rhdnew7} and \ref{rhdnew8}.

We solve equations \ref{rhdnew5} and \ref{rhdnew6} or equations \ref{rhdnew7} and \ref{rhdnew8} implicitly in a similar manner to \cite{WhiBatMon2005}.  For equations  \ref{rhdnew7} and \ref{rhdnew8},  we write
\begin{equation}
\label{eqn:GSxi}
\begin{split}
\xi^{n+1}_i =  & ~\xi_i^n + \rd t \sum_j \frac{m_j}{\rho_i \rho_j} b c \left( \rho_i
\xi_i^{n+1} - \rho_j \xi_j^{n+1} \right) \frac{\nabla W_{ij}}{r_{ij}}  \\
& - \rd t \left( \nabla v_i \right) f_i \xi_i^{n+1} \\
& -  \rd t a c \kappa_{{\rm g},i} \left[ \frac{ \rho_i
\xi_i^{n+1}}{a} - \left( \frac{ u_i^{n+1}}{c_{{\rm v},i}} \right)^4 \right] \\
& - \rd t a c \kappa_{{\rm d},i} \left[ \frac{ \rho_i
\xi_i^{n+1}}{a} - T_{{\rm d}i}^4 \right],
\end{split}
\end{equation}
where
\begin{equation}
b= \displaystyle {\lambda_{i} \over \kappa_{i} \rho_{i}} +
{\lambda_{j} \over \kappa_{j} \rho_{j}} 
\end{equation}
for brevity, and 
\begin{equation}
\label{eqn:GSu}
\begin{split}
u^{n+1}_i = & ~u_i^n + \rd t~(pdV^n_i) +  \rd t a c \kappa_{{\rm g},i} \left[ \frac{ \rho_i
\xi_i^{n+1}}{a} - \left( \frac{ u_i^{n+1}}{c_{{\rm v},i}} \right)^4 \right] \\
& - ~10^{-33} \rd t  \frac{n_{{\rm H2},i}^2 }{\rho_i} T_{{\rm g},i}^{1/2} \left( \frac{ u_i^{n+1}}{c_{{\rm v},i}} - T_{{\rm d},i}\right)  \frac{Z}{{\rm Z}_\odot} \\ 
& -\rd t  \frac{\alpha_i}{\rho_i} \left(\frac{T_{{\rm g},i}}{10~{\rm K}}\right)^{\beta_i} \frac{Z}{{\rm Z}_\odot}\\
& + \rd t \frac{\Gamma_{{\rm cr},i}}{\rho_i} ~~~ + \rd t \frac{\Gamma_{{\rm pe},i}}{\rho_i} \frac{Z}{{\rm Z}_\odot}~~~ + \rd t \frac{\Gamma_{{\rm H2,g},i}}{\rho_i} ~,
\end{split}
\end{equation}
in which we have taken equation \ref{eq:gasdust} rather than \ref{eq:gasdust2}.
The equations obtained when solving equations \ref{rhdnew5} and \ref{rhdnew6} are very similar (and simpler) to the above equations so we omit them for the sake of brevity.  Note that for an ideal gas, the temperature of the gas is given by $T_{\rm g}=u/c_{\rm v}$, where $c_{\rm v}$ is the specific heat capacity.  In fact, our equation of state is considerably more complex (see Section \ref{sec:eos}) and $c_{\rm v}$ is in fact simply the ratio of $u/T_{\rm g}$ which is obtained from a pre-calculated table of values that gives this ratio as a function of $u$ and $\rho$.  Equation \ref{eqn:GSu} includes the quantity $pdV^n_i$ which is an explicit term (i.e. evaluated at time step $n$)  for the work done on the gas and the thermal contribution from the artificial viscosity.  In \cite{WhiBatMon2005}, these terms were included implicitly, but from \cite{Bate2010} onwards we have used the explicit term here rather than the implicit term because we found empirically that using the explicit term results in better energy conservation in star formation calculations. The explicit term for particle $i$ is calculated as
\begin{equation}
\begin{split}
& pdV_i   =  \frac{P_i}{\Omega_i \rho_i^2} \sum_{j} m_j \mbox{\boldmath $v$}_{ij} \cdot \nabla_i W_{ij}(h_i) \\
& -  \frac{1}{2} \sum_j \frac{m_j}{\bar{\rho}_{ij}}  \bar{\alpha}_{ij} v_{\rm sig} (\mbox{\boldmath $v$}_{ij} \cdot \hat{\mbox{\boldmath $r$}}_{ij}) \mbox{\boldmath $v$}_{ij} \cdot \left[ \frac{\nabla_i W_{ij}(h_i)}{2\Omega_i} + \frac{\nabla_i W_{ij}(h_j)}{2\Omega_j}  \right]
\end{split}
\end{equation}
where the the second term is only applied between approaching particles ($\mbox{\boldmath $v$}_{ij} \cdot \hat{\mbox{\boldmath $r$}}_{ij}<0$), the viscous signal velocity is $v_{\rm sig}= \bar{c}_{\rm s} - 2 \mbox{\boldmath $v$}_{ij} \cdot \hat{\mbox{\boldmath $r$}}_{ij}$, and the average sound speed and densities of particles $i$ and $j$ are given by $\bar{c}_{\rm s}= \frac{1}{2} (c_{{\rm s},i} + c_{{\rm s},j})$, and $\bar{\rho}_{ij}= \frac{1}{2} (\rho_i + \rho_j)$, respectively.  The viscosity parameters evolve for each particle based on the method of \cite{MorMon1997}, and their average is $\bar{\alpha}_{ij}= \frac{1}{2} (\alpha_i + \alpha_j)$.  

Within each Gauss-Seidel iteration, for each particle $i$, equations \ref{eqn:GSxi} and \ref{eqn:GSu} are solved simultaneously for $u_i^{n+1}$ and then $\xi_i^{n+1}$.  This could be done exactly for the equations used by \cite{WhiBatMon2005}.  Note, however, that equation \ref{eqn:GSu} includes the quantities $T_{{\rm g},i}^{1/2}$ and $(T_{{\rm g},i})^{\beta_i}$ in the collisional dust-gas term and the line cooling term, and the heating rate due to molecular hydrogen formation, $\Gamma_{\rm H2,g}$, also involves the gas temperature.  For a purely implicit solution of $u_i^{n+1}$, the terms involving $T_{{\rm g},i}$ should be written as $T_{{\rm g},i}= u_i^{n+1}/c_{{\rm v},i}$, but this would introduce fractional powers of $u_i^{n+1}$, making direct solution impractical.  Instead, in equation \ref{eqn:GSu} we take $T_{{\rm g},i}= u_i^n/c_{{\rm v},i}$ and rely on the fact that $u_i^n$ is updated in every Gauss-Seidel iteration so that if $u_i^{n+1}$ converges, so too does $u_i^n$.  Empirically, this seems to work well as long as large decreases in the gas temperature (which would result in a large decrease in the emission line cooling) are avoided from one iteration to the next --- we limit $u^{n+1}_i\ge 0.8 u^i_i$.

Apart from these changes, the simultaneous solution for $u_i^{n+1}$ and then $\xi_i^{n+1}$ follows the method of \cite{WhiBatMon2005} and will not be repeated here.  Briefly, in general, equations \ref{eqn:GSxi} and \ref{eqn:GSu} are combined to eliminate $\xi_i^{n+1}$ and produce a quartic equation for $u_i^{n+1}$ that can be solved analytically.  When solving the implicit version of equation \ref{rhdnew6}, we use Newton-Raphson iteration to obtain $u_i^{n+1}$. Once $u_i^{n+1}$ is obtained, $\xi_i^{n+1}$ is found simply by rearranging \ref{eqn:GSxi}.  This is done for all active particles for each iteration, and the iterations are repeated until a desired tolerance is reached.

\subsubsection{Calculating the ISR attenuation}
\label{sec:attenuation}

It is necessary to calculate the local intensity of the ISR (attenuated due to dust extinction) and the mean visual extinction in order to evaluate the $\Lambda_{\rm ISR}$ and $\Gamma_{\rm pe}$ terms and to model the chemistry (Section \ref{sec:chemistry}).  This is potentially very difficult computationally because it involves integrating over all angles at each point in the simulation. To make this computationally tractable, we use a similar method to the TREECOL method of \cite*{ClaGloKle2012}.  They propose calculating the mass surrounding a particle in angular cones (and thereby calculating the column density and optical depths) using the same tree which is frequently used in SPH codes to calculate gravity and neighbouring particles.  The algorithm essentially sums over the column densities of nearby particles and more distant tree nodes that intersect the angular cone being considered.  To cover the full $4\pi$ steradians uniformly, they used angular cones defined using the HEALPix\footnote{http://healpix.jpl.nasa.gov/} library functions \citep{Gorskietal2005}.  

We implement a very similar method to that of \cite{ClaGloKle2012}, with the following differences.  The SPH code used by \citeauthor{ClaGloKle2012} used an oct-tree.  An oct-tree is constructed by a purely spatial decomposition of the particle distribution and the size and shape of each node in the tree is well-defined (in three dimensions it is a cube with a size that is purely a function of its level in the tree hierarchy).  This allowed \citeauthor{ClaGloKle2012}  to roughly approximate the size and shape of each node projected onto the sphere as a square, thereby allowing them to easily calculate whether or not a node contributed to the mass lying in a particular angular cone.  In turn, this allowed them to construct a method that conserved the total mass (i.e. average column density).  However, the SPH code we use for this paper uses a binary tree based on a nearest neighbour decomposition \citep{Benzetal1990}.  Thus, the nodes do not have well-defined shapes or sizes.  We still implement a method that guarantees the average column density is exactly preserved, but for simplicity we assume that the projected shape of each node is a circle.  Similarly, we assume that the projection of the solid angle covered by each HEALPix direction is a circle with area $4\pi/N$ steradians, where $N$ is the chosen number of HEALPix directions.  The binary tree defines a size for the $n$th node as
\begin{equation}
\label{treenode}
s_n = \max \left(\frac{m_2}{m_1+m_2}r_{12}+s_1, \frac{m_1}{m_1+m_2}r_{12}+s_2 \right),
\end{equation}
where the subscripts 1 and 2 identify the two sub-nodes, $r_{ij}=| \mbox{\boldmath $r$}_i- \mbox{\boldmath $r$}_j|$, and $m_i$  and $\mbox{\boldmath $r$}_i$ are the mass and position of the centre of mass of the sub-nodes, respectively.  If the sub-node is an individual SPH particle, its size is zero in equation \ref{treenode}.  As mentioned above, in general the tree nodes do not have well-defined shapes.  For the purposes of calculating the column density we assume their projection is circular, but we are free to scale their effective radius by an arbitrary factor so that the radius of the circle is $r_n=f s_n$, where $f$ is expected to be of order unity.  We explore the effects of choosing different values of $f$ in Section \ref{sec:ISR} and conclude that $f=0.5$ is a good choice.  If the node is an individual SPH particle, we take the radius to be twice the particle's smoothing length, $r_n = 2h_n$.  From a particular particle, $p$, at location, $\mbox{\boldmath $r$}_p$, the vector between the particle and the node is $\mbox{\boldmath $r$}_{np} = \mbox{\boldmath $r$}_n - \mbox{\boldmath $r$}_p$. We then define the angular radius of the node as viewed from the particle by $a_n = \arctan(r_n/ r_{np})$.  We denote the HEALPix direction as $\hat{\mbox{\boldmath $l$}}$, which we treat as the centre of the HEALPix circle of angular radius $a_{\rm pix} = \sqrt{4/N}$ (the $\pi$'s cancel).  The angular distance between the centres of these two circles is $d=\arccos(\mbox{\boldmath $r$}_{\rm np} \cdot \hat{\mbox{\boldmath $l$}}/r_{np})$.

To calculate the contribution of each tree node or particle to the column density in each HEALPix area, we calculate the overlapping area of the two circles.  For two circles of radii $r$ and $R$ whose centres are separated by distance $d$ this can be computed as 
\begin{equation}
\label{eq:overlap}
\begin{split}
A = & ~~r^2 \arccos \left( \frac{d^2 + r^2 - R^2}{2 d r} \right) \\
& + R^2 \arccos \left( \frac{d^2 + R^2 - r^2}{2 d R} \right) \\
& - \frac{1}{2} \sqrt{(-d+r + R) (d+r - R) (d-r + R) (d+r + R)},
\end{split}
\end{equation}
where we take $r=a_{\rm pix}$ and $R=a_n$.  The contribution of the node to the average column density of the HEALPix area is then $m_n/(\pi r_n^2) A / (4\pi/N)$ where $m_n/(\pi r_n^2)$ is the column density of the node and $A / (4\pi/N)$ is the fraction that is assigned to that HEALPix area.  The computational efficiency can be improved by treating trivial cases separately.  If $a_n+a_{\rm pix} \le d$ there is no overlap.  If $d+a_{\rm pix} \le a_n$ then the HEALPix area is entirely covered by the node and the contribution is simply $m_n/(\pi r_n^2)$.  If $d+a_n \le a_{\rm pix}$ then the node lies entirely within the HEALPix area and the contribution is $m_n/(4\pi/N)$.

Once we have determined the average column density in a particular HEALPix area, we can calculate the visual extinction $A_{\rm V}$ in that direction.  We define $Q_{\rm V}=Q_\nu(V)$ at a visual wavelength of 550~nm.  Rather than performing the integrals appearing in equations \ref{eq:g} and  \ref{eq:grainheating} during an SPH calculation, we pre-calculate tables of values.  For equation \ref{eq:grainheating}, the attenuated dust heating in a single direction as a function of $\log(A_{\rm V})$ 
\begin{equation}
\label{eq:ext}
\lambda^{\rm ISR}(A_{\rm V})=  \int_0^{\infty} Q_\nu J^{\rm ISR}_\nu \exp\left[-\frac{A_{\rm V}}{1.086} \frac{Q_\nu(\nu)}{Q_{\rm V}}\right] ~{\rm d}\nu,
\end{equation}
where the factor of $1.086 = 2.5 \log(e)$ appears because extinction $A$ is related to optical depth $\tau$ by $A = -2.5 \log(e^{-\tau})$.   The integral is actually calculated in ${\rm d}(\ln\nu)$ rather than ${\rm d}\nu$.   This table can be interpolated to find the heating rate for any particular value of the visual extinction $A_{\rm V}$.  The total heating rate at a location is then provided by averaging the contributions from all directions $\hat{\mbox{\boldmath $l$}}_i$ as
\begin{equation}
\label{eq:grainheating2}
\Lambda_{\rm ISR} = \frac{1}{N} \sum_i^N \lambda^{\rm ISR}(A_{{\rm V}i}),
\end{equation}
where the directions are chosen using the HEALPix library functions.
The same is done to compute the quantity $G({\bf r})$ for the photoelectric heating (equation \ref{eq:g}).  We pre-calculate a table of values along a single line of sight containing
\begin{equation}
\label{eq:g2}
g(A_{\rm V}) = \frac{\displaystyle \int^{\infty}_{6{\rm eV}} J^{\rm ISR}_{\rm \nu}\exp\left[-\frac{A_{\rm V}}{1.086} \frac{Q_\nu(\nu)}{Q_{\rm V}}\right]  {\rm d}\nu}{\displaystyle  \int^{\infty}_{6{\rm eV}} J^{\rm ISR}_{\rm \nu}  {\rm d}\nu  }.
\end{equation}
The total value $G({\bf r})$ at a location (equation \ref{eq:g}) is then provided by averaging the contributions from all directions $\hat{\mbox{\boldmath $l$}}_i$ as
\begin{equation}
\label{eq:g3}
G( \mbox{\boldmath $r$}) = \frac{1}{N} \sum_i^N g(A_{{\rm V}i}).
\end{equation}
For equations \ref{eq:extinction} and \ref{eq:fdust}, we simply have
\begin{equation}
\langle \exp(-A_{\rm V}) \rangle =  \frac{1}{N} \sum_i^N \exp(-A_{{\rm V}i}).
\end{equation}
To calculate the average column density of molecular hydrogen $\langle N_{\rm H2} \rangle$ which is required to evaluate the self-shielding of molecular hydrogen (equation \ref{eq:self}) we use the same traversal of the tree that is used to calculate the extinction and $G( \mbox{\boldmath $r$})$, but rather than calculate the full column density the contribution of each particle or tree node is multiplied by its molecular fraction $x_{\rm H2}$.

It should be noted that even using the same tree that is used to calculate gravitational forces to estimate the extinction and other directional averages, obtaining the averages in many directions can still require a substantial amount of computation.  Furthermore, when the code is run in parallel using MPI with domain decomposition, the contributions of each domain to the column densities along each ray must be added together in order to calculate the quantities in each direction.  This would not be necessary if one simply wanted the average column density to a particle (since the column density adds linearly), but because calculating the extinction involves a nonlinear function, $\exp(-\tau)$, the column density in each direction must be computed separately.  Thus, when running with MPI domain decomposition there is also a substantial memory overhead which depends on the chosen number of directions.

\subsection{Equation of state and boundary conditions}
\label{sec:eos}

As in \cite{Bate2009b,Bate2012}, the calculations presented in this paper were performed using radiation hydrodynamics
with an ideal gas equation of state for the gas pressure
$p= \rho T_{\rm g} \cal{R}/\mu$.  
The thermal evolution takes into account the translational,
rotational, and vibrational degrees of freedom of molecular hydrogen 
(assuming a 3:1 mix of ortho- and para-hydrogen; see
\citealt{Boleyetal2007}).  It also includes molecular
hydrogen dissociation, and the ionisations of hydrogen and helium.  
The hydrogen and helium mass fractions are $X=0.70$ and 
$Y=0.28$, respectively.  For this composition, the mean molecular weight of the gas is initially
$\mu = 2.38$.
The contribution of metals to the equation of state is neglected.

In previous calculations using the flux-limited diffusion method of \cite{WhiBatMon2005}, it was 
necessary to set matter and radiation temperature boundary conditions.  
For calculations contained within a particular volume
\citep[e.g. the collapse of a molecular cloud core;][]{WhiBat2006, Bate2010,Bate2011,BatTriPri2014}
the matter and radiation temperatures were fixed on ghost particles outside the boundary at the initial 
(low) temperature (e.g.~10~K).  
For calculations of clouds with free boundaries \citep[e.g.][]{Bate2009b,Bate2012,Bate2014},
all particles with densities less than a particular value (typically $10^{-21}$~g~cm$^{-3}$) had their 
matter and radiation temperatures set to the initial values (again $\approx 10$~K). This matter was two 
orders of magnitude less dense than that in the initial cloud and, thus, these boundary particles 
surrounded the region of interest in which the star cluster formed.  In both cases, this essentially 
meant that the clouds were embedded in an external
radiation field with this boundary temperature.

The new method presented here eliminates the need for these arbitrary temperature boundary conditions.
Now the temperature of the gas, dust and radiation are all set consistently at both high and low densities.

\begin{figure}
\centering \vspace{-0.3cm} \hspace{-0cm}
    \includegraphics[width=8.5cm]{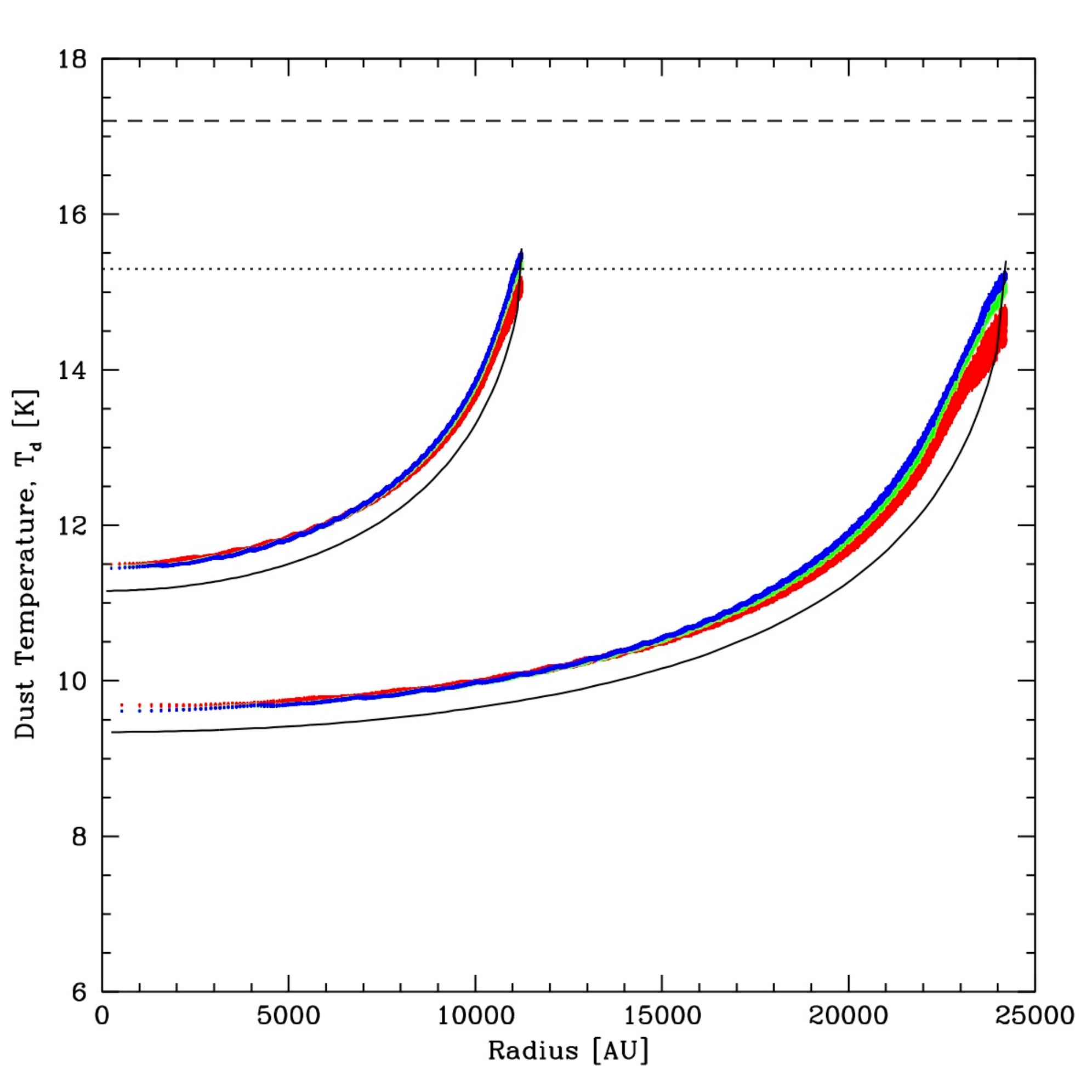} \vspace{-0.5cm}
\caption{The dust temperature as a function of radius inside two uniform-density molecular cloud cores that are subject to an external interstellar radiation (ISR) field. The clouds have the same densities, but different masses: 1~M$_\odot$ (upper distributions) and 10~M$_\odot$ (lower distributions).  The exact solutions are plotted using the solid black lines.  For the SPH calculations, a point is plotted for each SPH particle ($2.6 \times 10^5$ for each cloud).  The different colours give the results obtained when the code uses 12 (red), 48 (green), or 192 (blue) HEALPix directions to calculate the mean extinction.  Dust within the clouds receives less of the external radiation due to extinction and, therefore, is colder.  The number of HEALPix directions used has little impact on the results, except in the very outer part of the clouds.  The equilibrium temperature of dust that is subject to the full ISR is 17.2~K (horizontal dashed line), and the equilibrium temperature of dust that received exactly half of this radiation would be 15.3~K, which is lower by a factor of $2^{1/6}$ (horizontal dotted line).  As expected, dust at the edges of the clouds is close to this temperature.  }
\label{DT_uniform}
\end{figure}

\begin{figure}
\centering \vspace{-0.3cm} \hspace{-0cm}
    \includegraphics[width=8.5cm]{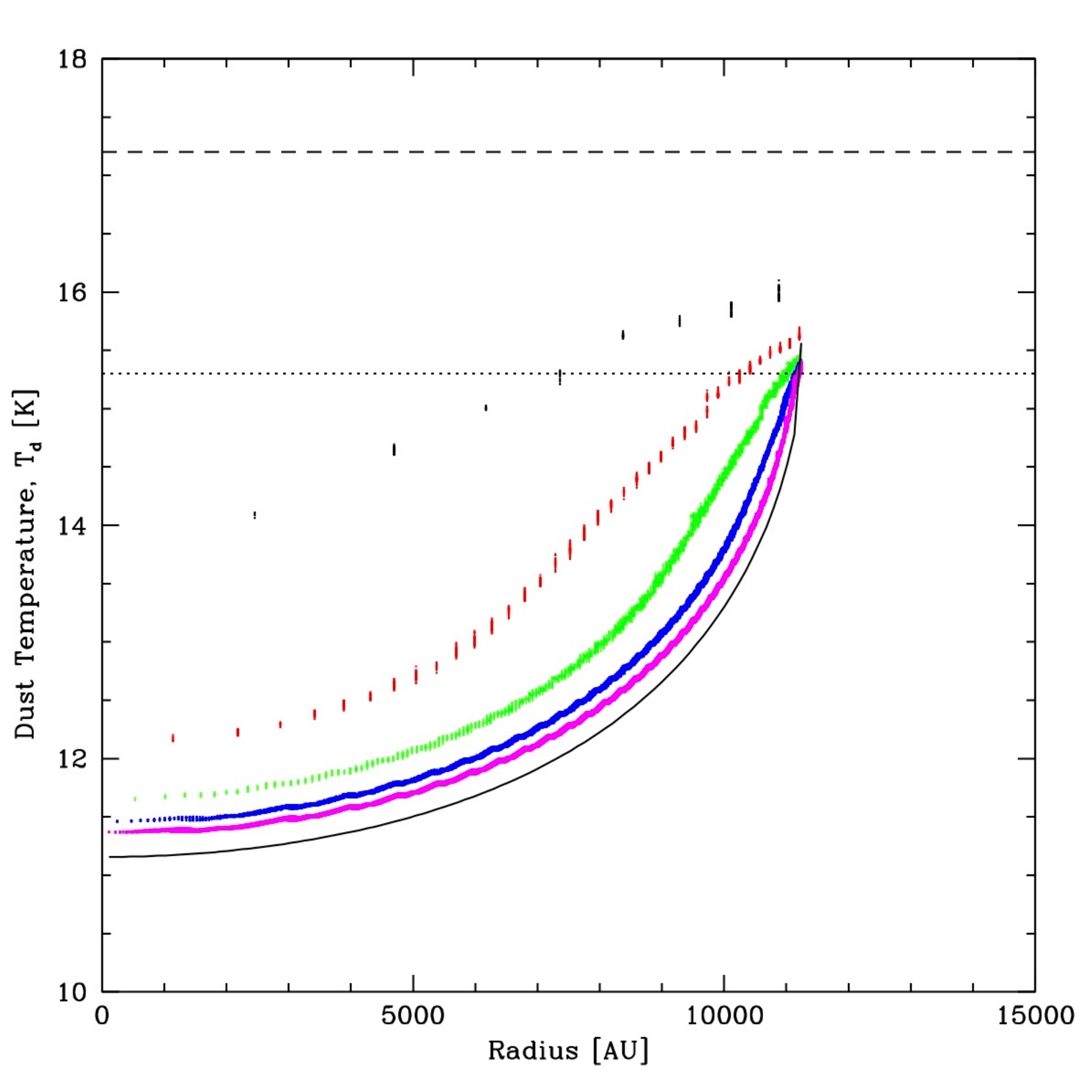} \vspace{-0.5cm}
\caption{The dust temperature as a function of radius inside the 1~M$_\odot$ uniform-density molecular cloud core that is subject to external ISR. The exact solution is plotted using the solid black line.   The results from five SPH calculations are illustrated, and a point is plotted for each SPH particle. The calculations each use 48 HEALPix directions to calculate the mean extinction, but different numbers of SPH particles are used to model the cloud -- from top to bottom: 280 (black), 2608 (red), $2.6 \times 10^4$  (green) $2.6 \times 10^5$ (blue), and 2.6 million (magenta).  It can be seen that using fewer particles tends to result in warmer dust temperatures (i.e. an underestimate of the extinction), but the dust temperature appears to be slowly converging toward the exact solution as the number of particles is increased.  As long as $\gsim 3\times 10^4$ particles are used the maximum error in the mean temperature at any radius is $\lsim 1$~K. The equilibrium temperature of dust that is subject to the full ISR is 17.2~K (horizontal dashed line), and the equilibrium temperature of dust that received exactly half of this radiation would be 15.3~K, which is lower by a factor of $2^{1/6}$ (horizontal dotted line).  As expected, with sufficient resolution, the dust at the edges of the clouds is close to this temperature. }
\label{DT_Npart}
\end{figure}

\begin{figure*}
\centering \vspace{-0.3cm} \hspace{-0cm}
    \includegraphics[width=8.5cm]{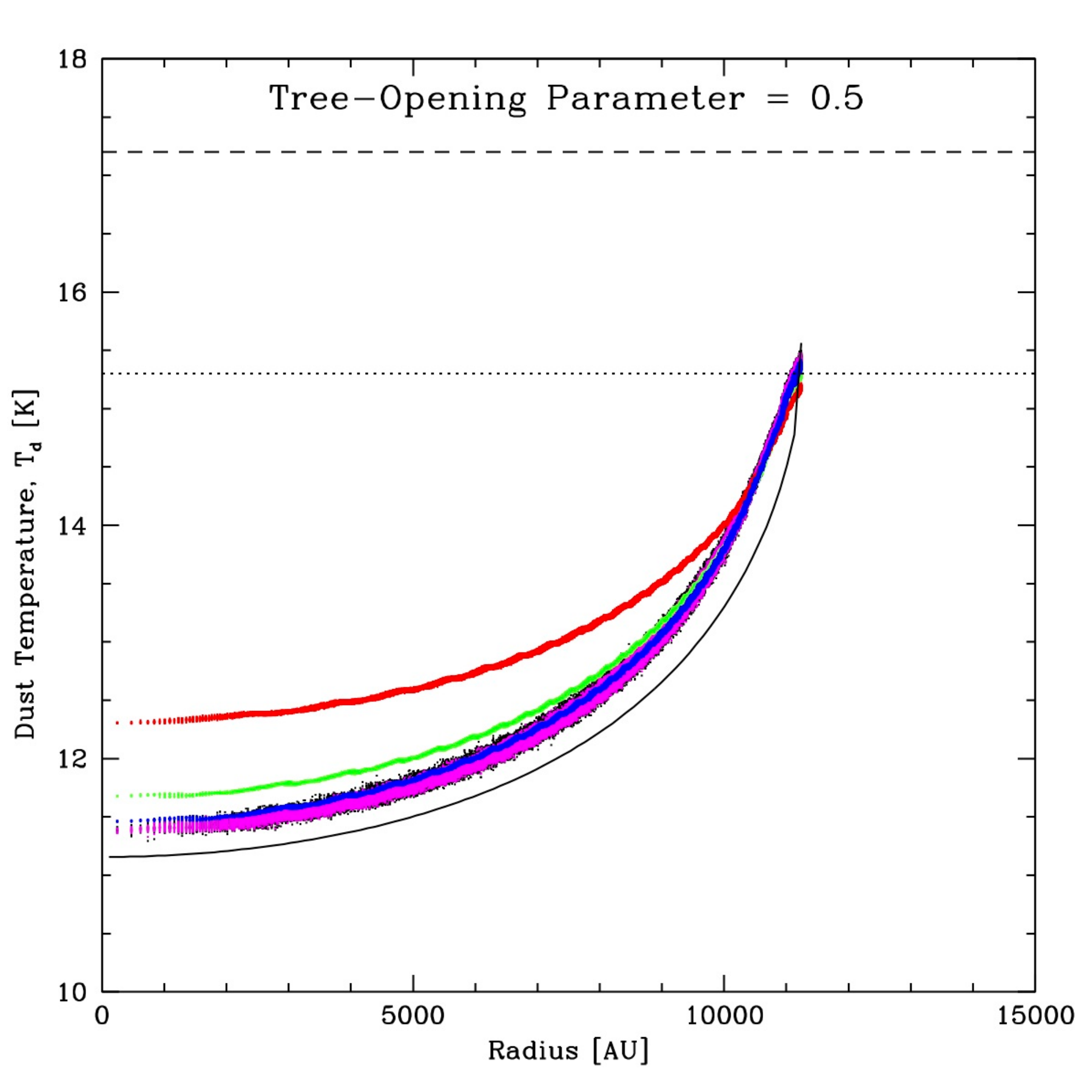}
    \includegraphics[width=8.5cm]{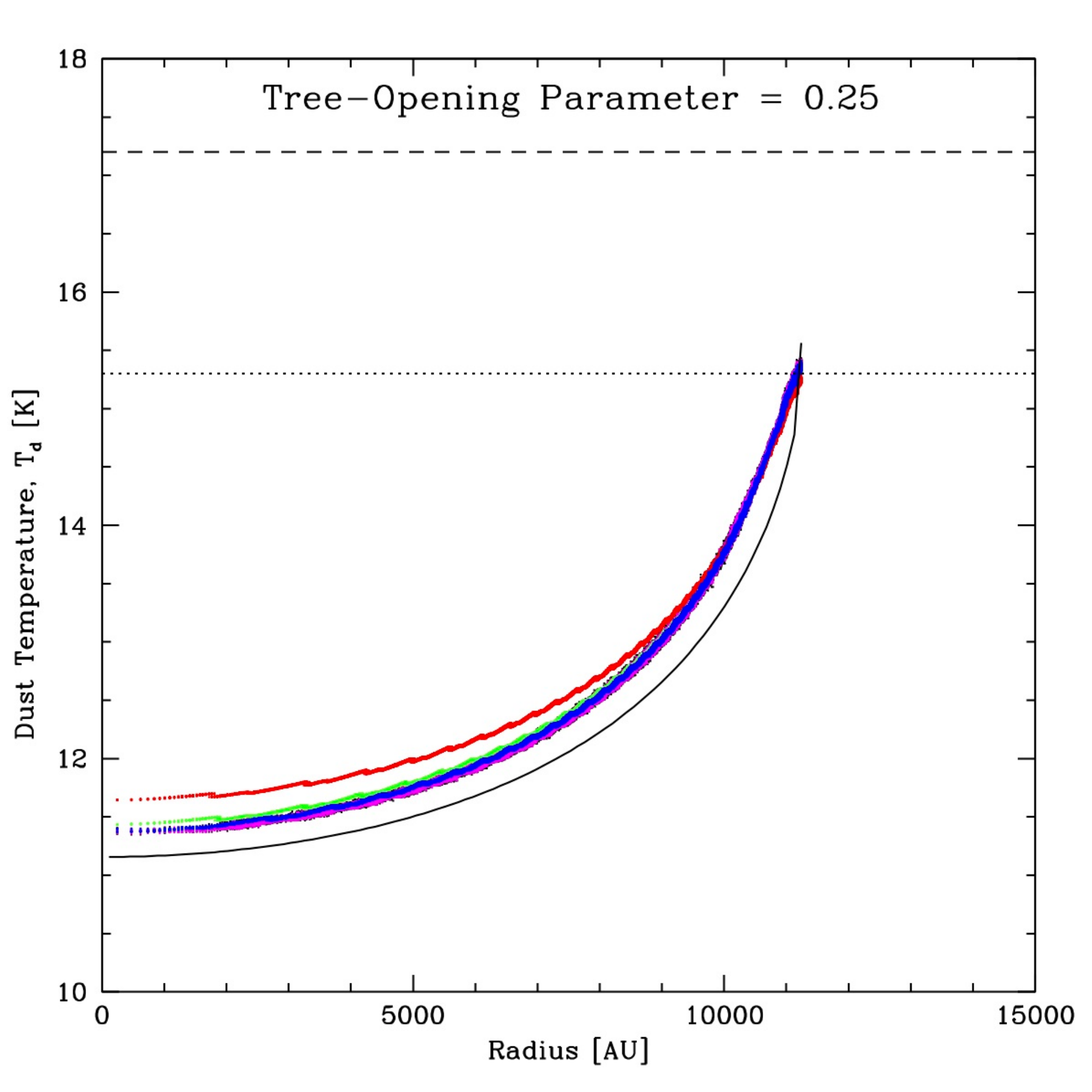} \vspace{-0.5cm}
\caption{The dust temperature as a function of radius inside the 1~M$_\odot$ uniform-density molecular cloud core that is subject to external ISR. The exact solution is plotted using the solid black line. The results from ten SPH calculations are illustrated, and a point is plotted for each SPH particle. The calculations each use 48 HEALPix directions to calculate the mean extinction and $2.6 \times 10^5$ SPH particles, but the tree-opening parameter and the effective sizes of the tree nodes are varied.  In the left panel, the tree-opening parameter is 0.5, while in the right panel it is 0.25.  In each panel the effective size of the tree nodes decreases from top to bottom, using scaling factors of $f=2.0$ (red), 1.0 (green), 0.5 (blue), 0.1 (magenta), and 0.01 (black).   It can be seen that using very large tree-opening angles and/or effective node sizes results in warmer dust temperatures (i.e.\ an underestimate of the extinction).  If the effective node size is too small the scatter increases.  With a tree-opening parameter of 0.5, an effective node size of half the actual node size (i.e.\ $f=0.5$) gives the best trade off between accurately computing the mean radial temperature distribution and minimising the scatter.  The equilibrium temperature of dust that is subject to the full ISR is 17.2~K (horizontal dashed line), and the equilibrium temperature of dust that received exactly half of this radiation would be 15.3~K, which is lower by a factor of $2^{1/6}$ (horizontal dotted line).  As expected the dust right at the edges of the clouds is close to this temperature. }
\label{DT_acc_size}
\end{figure*}

\subsection{Opacities and metallicity}
\label{sec:opacity}

\cite{WhiBat2006}, \cite{Bate2009b, Bate2010, Bate2011, Bate2012}, and 
\cite{BatTriPri2014} assumed solar metallicity gas and used the Rosseland
mean opacity tables of \citet{PolMcKChr1985} for interstellar dust and, at higher 
temperatures when the dust is destroyed, the gas opacity
tables of \citet{Alexander1975} (the IVa King model)  \citep[see][for further details]{WhiBat2006}.
\cite{Bate2014} performed calculations with varying opacities.  He used the same dust opacities,
scaled linearly in proportion to the metallicity, but replaced the
gas opacity tables with the metallicity-dependent tables of \cite{Fergusonetal2005} with $X=0.70$.
It is worth noting that one of the main conclusions of \cite{Bate2014} was that the results of star formation
calculations are very insensitive to the opacities that are used.

In this paper, because the equations treating dust heating and photoelectric heating of the gas require 
integrals over the dust opacity as a function of frequency, we cannot continue simply to use the Rosseland mean
opacities of \citet{PolMcKChr1985}.  At low-densities (i.e. when the optical depth is low and the dust is essentially in thermal equilibrium with the interstellar radiation field), the Planck mean is more appropriate than the Rosseland mean and we desire consistency between the frequency-dependent dust opacity, $Q_\nu(\nu)$, and the grey opacity, $\kappa_{\rm d}$ appearing in equations \ref{rhdnew3} and \ref{dustLTE}.  In other words, equation \ref{eq:dustopacity} should be satisfied.  Therefore, we calculate Planck mean opacities directly from $Q_\nu(\nu)$ before the code begins to evolve the SPH calculation and the values are stored in a table as a function of dust temperature.  Any frequency dependent opacities can be used in principle, but for simplicity, we use the parameterisation provided by \cite{ZucWalGal2001} of the opacities from \cite{OssHen1994}.  We use these values for $\kappa_{\rm d}$ whenever the dust temperature is less than 100~K.  We assume that higher dust temperatures ($T_{\rm d}>100$~K) will only be encountered at high densities (i.e. near protostars) when the gas and dust are optically thick and thermally well-coupled.  In this case, Rosseland mean opacities are appropriate and we use those of \cite{PolMcKChr1985} as we have in our earlier calculations mentioned above.

For the gas, we continue to use the 
Rosseland mean opacities of \cite{Fergusonetal2005} for $\kappa_{\rm g}$, since the gas
opacity only becomes important in the highly optically-thick regions inside, or very near to, a star, for which
the Rosseland mean is appropriate.

\section{Calculations}
\label{sec:calculations}

\subsection{Testing the method for calculating ISR attenuation}
\label{sec:ISR}

One of the tests that \cite{ClaGloKle2012} used for the TREECOL method was to calculate the temperature structure of two uniform-density spherical clouds when subjected to a \cite{Black1994} ISR field.  They took densities of $10^{-19}$~g~cm$^{-3}$ for clouds with masses of 1 M$_\odot$ and 10~M$_\odot$ and used $2.6 \times 10^5$ SPH particles.  We perform the same tests here.  For this test, we set the internal radiation to zero (i.e. $E=T_{\rm r}=0$) and we turn off the coupling between the gas and the dust (i.e. $\Lambda_{\rm gd}=0$) so that equation \ref{dustLTE} is only solving for the equilibrium temperature of the dust subject to the attenuated external ISR.  The ISR is as discussed in Section \ref{sec:heating}, except that we exclude the additional UV flux which is important for photoelectric heating.  We have calculated the exact radial temperature distribution  by analytically calculating the column density along 192 HEALPix directions (using 48 provides an almost identical result) and used these values to calculate the attenuated dust heating (equations \ref{eq:ext} and \ref{eq:grainheating2}) and the equilibrium dust temperature as functions of radius.

In Fig.~\ref{DT_uniform} we give the distribution of the equilibrium dust temperature as a function of radius for each of the two clouds.  One point is plotted for each SPH particle.  Different colours give the results obtained when the code uses 12, 48, or 192 HEALPix directions to calculate the mean extinction.  The number of HEALPix directions used has little impact on the results, except in the very outer part of the clouds when using only 12 directions.  However, using more directions takes longer to compute, so from this point on we use 48 HEALPix directions.  The equilibrium temperature of dust that is subject to the full ISR is 17.2~K for our chosen ISR.  By considering equation \ref{lambda_dust}, we expect a dust particle that is subject to one hemisphere of heating to have an equilibrium temperature that is a factor of $2^{1/6}$ lower (i.e. 15.3~K) and, as expected, this is essentially equal to the temperature at the edge of each of the clouds.  However, within the clouds we find that the dust temperatures are up to $\approx 1$~K warmer than given by the exact solutions (solid black lines).  We note that the central temperatures in our clouds are in good agreement with those found by \cite{ClaGloKle2012}, but they obtained substantially lower temperatures at the outer edges of their clouds ($13-14$~K).  The reason for this is not clear since, as just mentioned, our edge temperatures are what we would expect.  Their ISR may have been different to ours, although both are nominally based on that of \cite{Black1994}.

In Fig.~\ref{DT_Npart}, we plot the dust temperatures from five calculations of the 1-M$_\odot$ cloud that each use 48 HEALPix directions, but that have different numbers of SPH particles: 280, 2608, $2.6\times 10^4$, $2.6 \times 10^5$, and 2.6 million.  It can be seen that the dust temperature tends to be over-estimated when a smaller number of particles is used, but the dust temperature appears to be slowly converging toward the exact solution as the number of particles is increased. The error in the dust temperature is approximately halved each time the number of SPH particles is increased by an order of magnitude (i.e. the error decreases at about the same rate that the linear spatial resolution increases).  Comparing the $2.6\times 10^4$ particle calculation with the exact solution, the maximum difference between the mean temperatures at any radius is only 1~K.  \cite{ClaGloKle2012} did not discuss how their method behaves with different numbers of particles. 

We used a cubic lattice SPH particle distribution to generate the results presented in Figs.~\ref{DT_uniform} and \ref{DT_Npart}.  We have also tried a random particle distribution.  Using $2.6 \times 10^5$ randomly-placed SPH particles to model the 1-M$_\odot$ cloud we obtained an almost identical mean radial temperature distribution to the case that used a cubic lattice, but with a temperature scatter of up to $\pm 0.2$~K.  However, given that this error is similar to the systematic difference between the temperature distributions obtained from the $2.6 \times 10^5$ and 2.6 million particle cubic lattice calculations, we conclude that the accuracy of the solution depends more on the number of particles used than on the details of how they are distributed.

In Fig.~\ref{DT_acc_size}, we investigate the effects on the calculation of the dust temperature of varying the tree-opening criterion and the effective radius of the tree nodes, the latter of which is parameterised by the factor $f$ (see Section \ref{sec:attenuation}).  During the walk through the tree to calculate gravity, nodes are opened if $s_n/r_{np}$ is larger than a critical value, usually taken to be 0.5.  We perform calculations using a tree-opening parameter of 0.5 and 0.25 (the latter of which means more nodes are opened during the tree walk).  For each case, we perform five SPH calculations of the 1-M$_\odot$ cloud that each use 48 HEALPix directions and $2.6 \times 10^5$ SPH particles, but we vary the effective radius of the node that is used when calculating the column density.  We take factors of $f= 2.0$, 1.0, 0.5, 0.1, and 0.01.  It can be seen from Fig.~\ref{DT_acc_size} that using very large tree-opening angles and/or effective node radii results in warmer dust temperatures (i.e.\ an underestimate of the extinction).  We also note that almost identical results are obtained when the product of the opening parameter and the factor $f$ is a constant (e.g.\ using a tree-opening parameter of 0.5 and $f=1$ gives almost identical results to using a tree-opening parameter of 0.25 and $f=2$).  The calculation of the extinction becomes very poor when the product of these two quantities exceeds $\approx 0.5$.  If the effective node size is too small, the calculations provide reasonable mean temperatures at a given radius, but the scatter increases (e.g.\ the black points in the left panel that were produced using $f=0.01$).  With the typical tree-opening parameter of 0.5, setting the factor $f=0.5$ gives the best trade off between accurately computing the mean radial temperature distribution and minimising the scatter. 

The typical resolutions employed in SPH calculations of star cluster formation vary from $\sim 10^3$ particles per solar mass (\citealt*{BonBatVin2003}; \citealt{Bonnelletal2011}) to $\sim 10^5$ particles per solar mass  \citep*{BatBonBro2003,Bate2012}.  The results of this test suggest that the typical errors in the dust temperature due to finite numbers of SPH particles and HEALPix directions and variations of the tree parameters should be $\lsim 1$~K as long as $\gsim 3\times 10^4$ particles per solar mass are used and large effective node radii are avoided.

\begin{figure}
\centering \vspace{-0.3cm} \hspace{-0cm}
    \includegraphics[width=8.5cm]{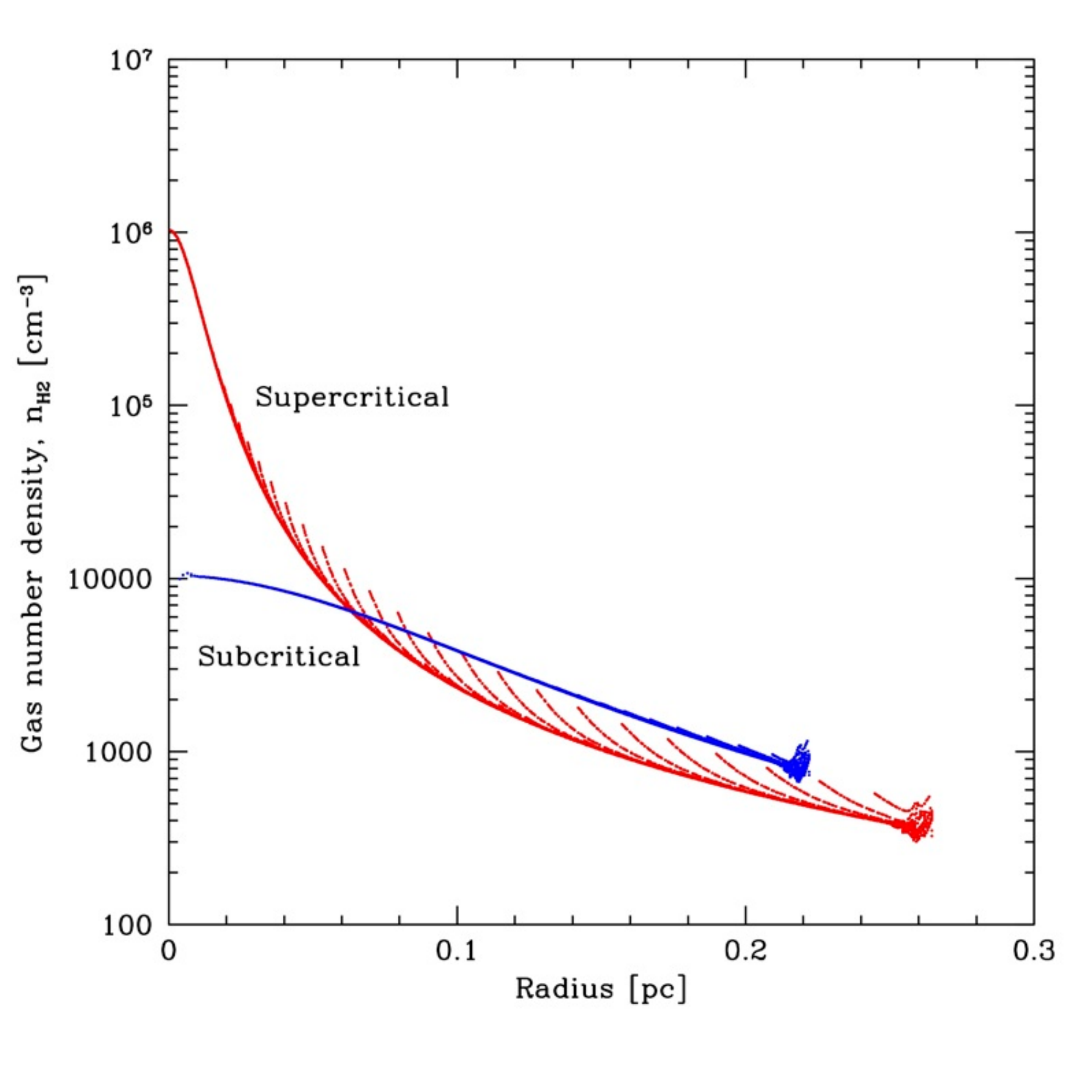} \vspace{-1cm}
\caption{Plots of the SPH molecular hydrogen number density, $n_{\rm H2}$, as functions of radius inside the subcritical (blue) and supercritical (red) 5~${\rm M}_\odot$ Bonner-Ebert spheres.  A point is plotted for each SPH particle.  The `spikes' in the density for the supercritical case are an artefact of the particles being set up on a radially-deformed cubic lattice.}
\label{KF_density}
\end{figure}

\begin{figure*}
\centering \vspace{-0.3cm} \hspace{-0cm}
    \includegraphics[width=8.5cm]{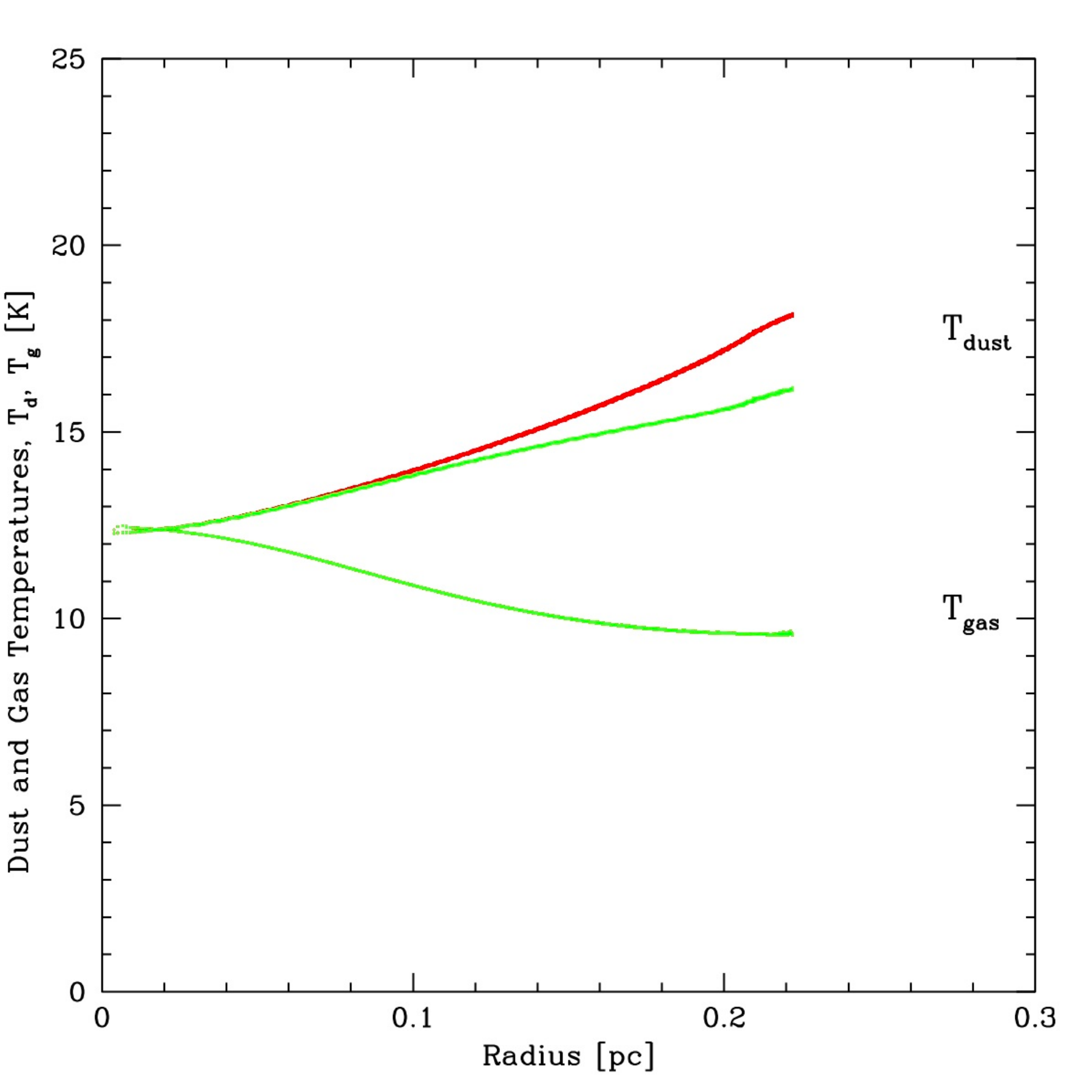} \vspace{-0.0cm}
    \includegraphics[width=8.5cm]{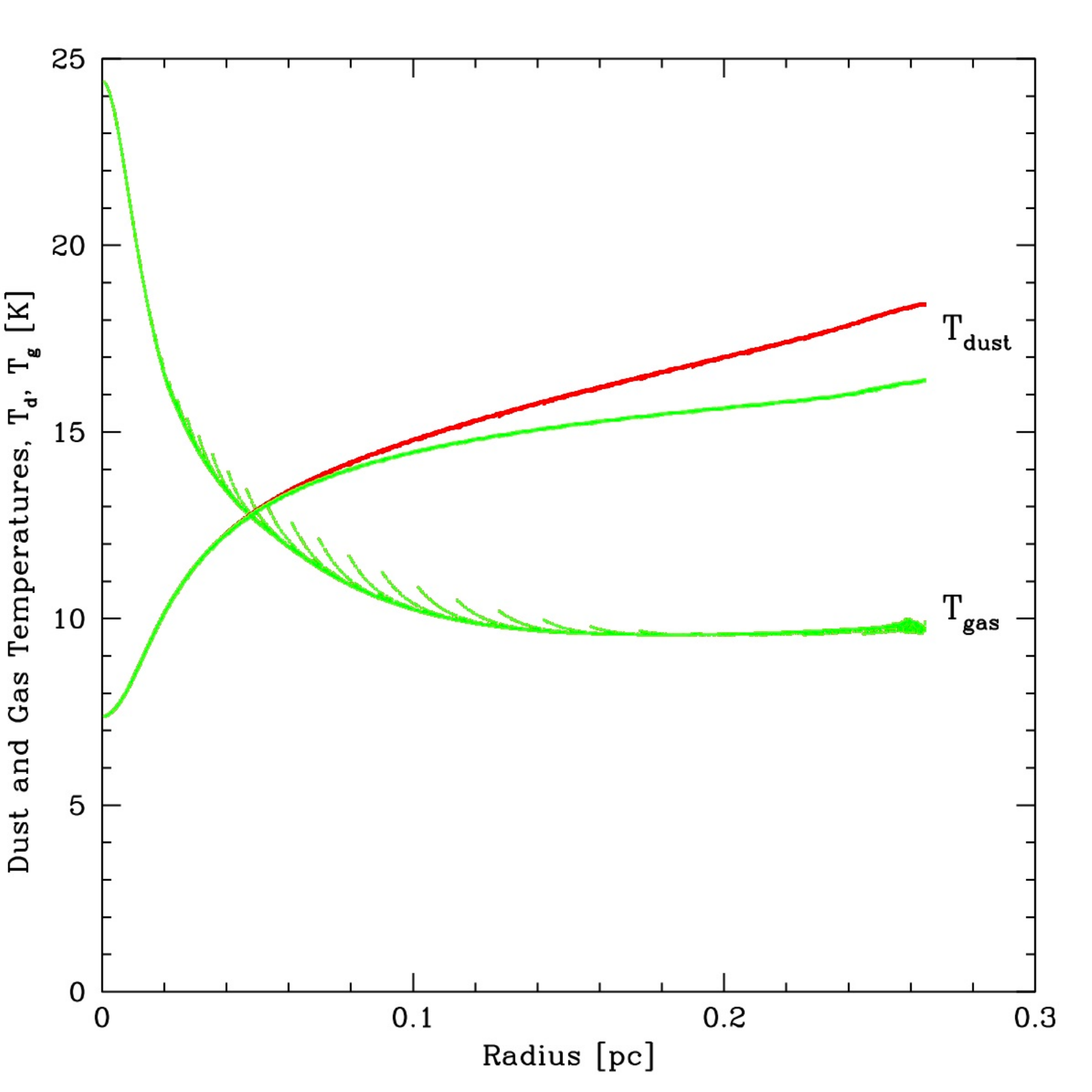} \vspace{-0.0cm}
\caption{The dust (upper) and gas (lower) temperatures as functions of radius inside the subcritical (left) and supercritical (right) 5~${\rm M}_\odot$ Bonner-Ebert spheres.  The ISM physics includes cosmic ray heating and molecular line cooling of the gas, and the dust is in thermal equilibrium with an external ISR that excludes (green) or includes (red) the UV contribution (see Section \ref{sec:heating}). The dust temperature drops as the extinction increases at smaller radii.  The gas temperature rises because the line cooling becomes less effective at higher densities.  The UV ISR only affects the dust temperature in the low-extinction outer parts of the cloud. The small `spikes' in the gas temperature in the right plot at radii $r\approx 0.03-0.16$~pc are due to the artefacts in the calculation of the SPH density due to the particles being set up on a radially-perturbed cubic lattice.  }
\label{KF05_1}
\end{figure*}

\begin{figure*}
\centering \vspace{-0.3cm} \hspace{-0cm}
    \includegraphics[width=8.5cm]{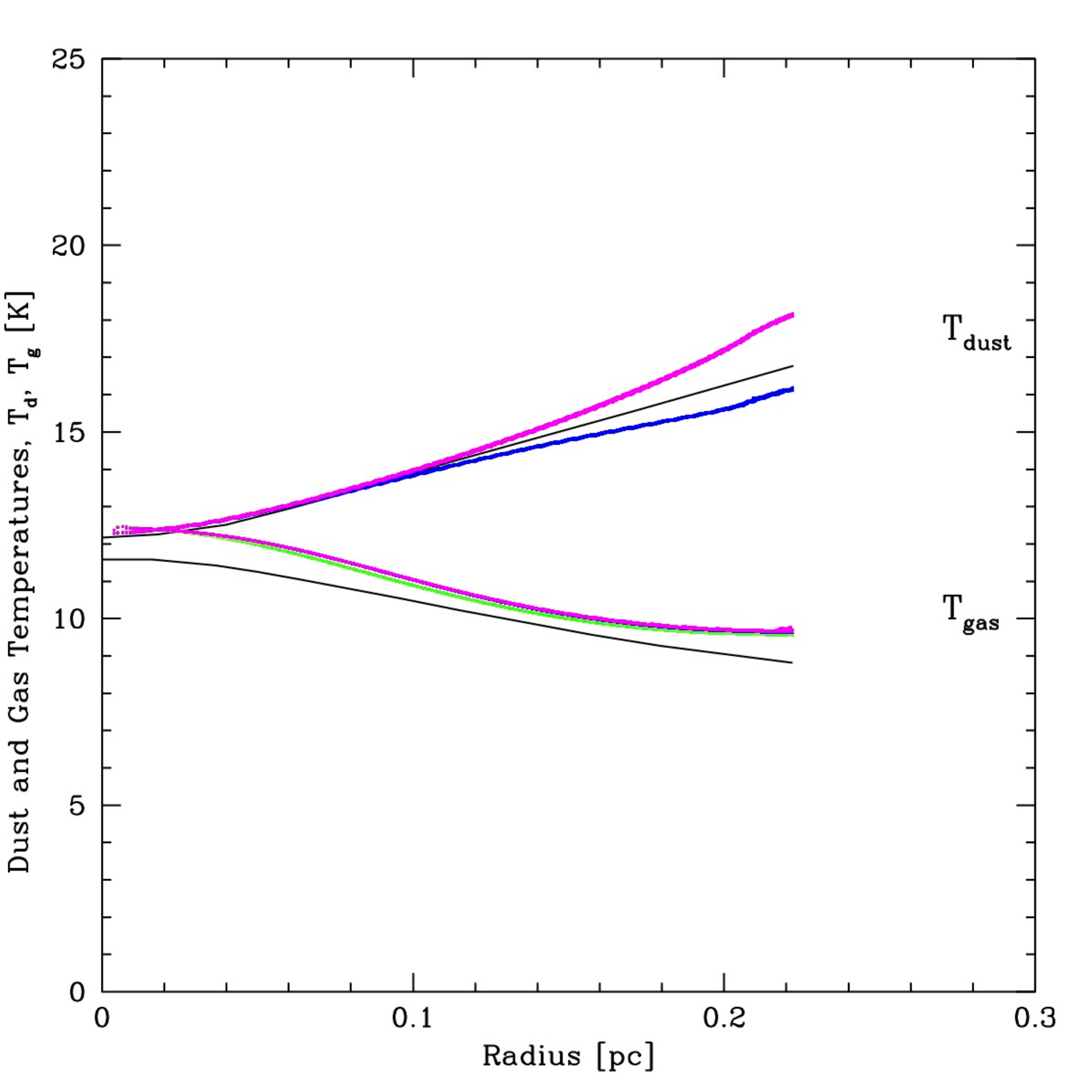} \vspace{-0.0cm}
    \includegraphics[width=8.5cm]{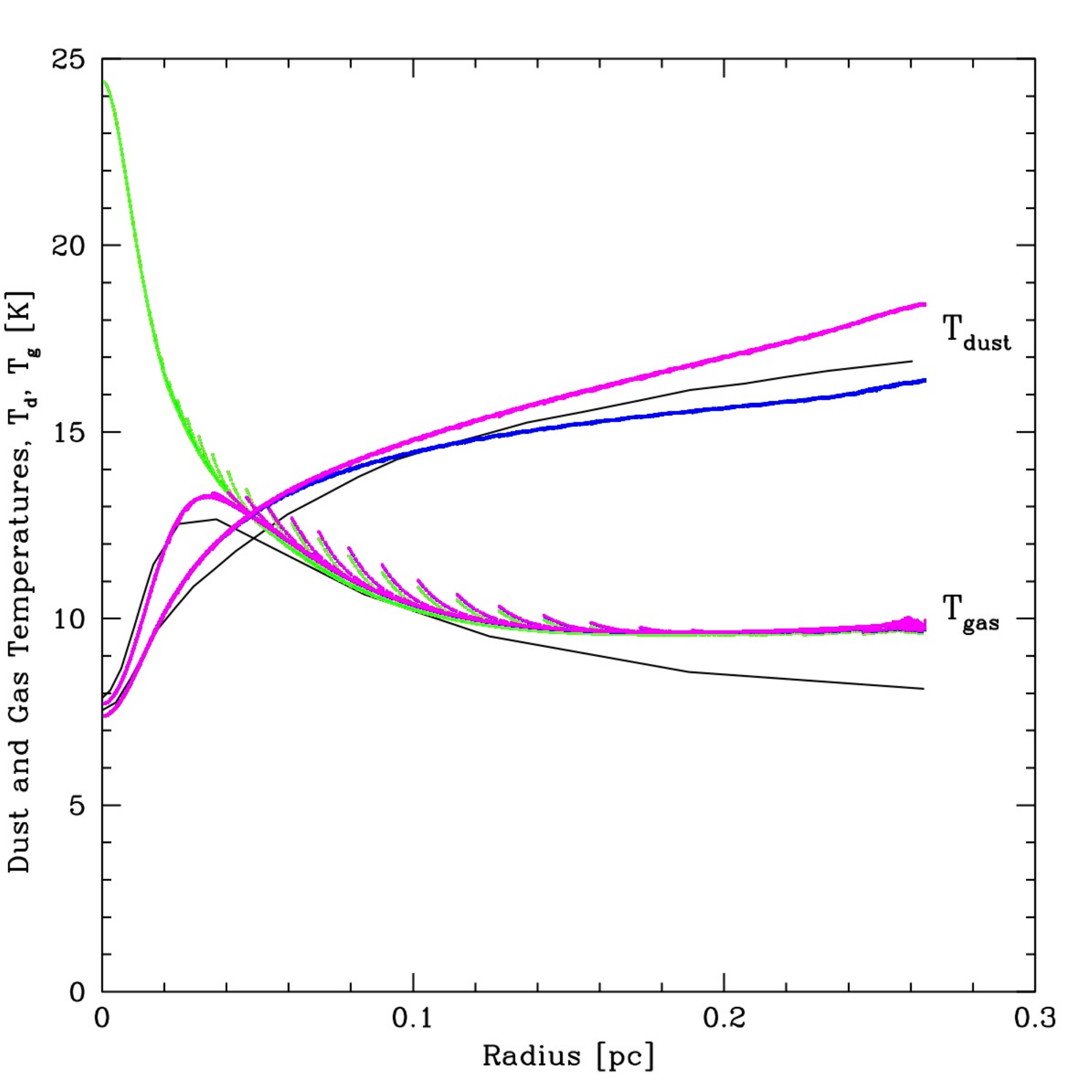} \vspace{-0.0cm}
\caption{The dust (upper) and gas (lower) temperatures as functions of radius inside the subcritical (left) and supercritical (right) 5~${\rm M}_\odot$ Bonner-Ebert spheres.  Red and green points (mostly obscured) are the same as in Fig.~\ref{KF05_1}.  For the blue (without the UV ISR) and magenta (with the UV ISR) points, the ISM physics is exactly the same, except that it includes thermal collisional coupling between the gas and the dust.  The black solid lines give the results obtained by Keto \& Field (2005) without including the UV ISR.  The effect is that at high densities, the gas essentially adopts the dust temperature.  The small `spikes' in the gas temperature in the right plot at radii $r\approx 0.03-0.16$~pc are due to the artefacts in the calculation of the SPH density due to the particles being set up on a radially-perturbed cubic lattice.   }
\label{KF05_2}
\end{figure*}

\begin{figure*}
\centering \vspace{-0.3cm} \hspace{-0cm}
    \includegraphics[width=8.5cm]{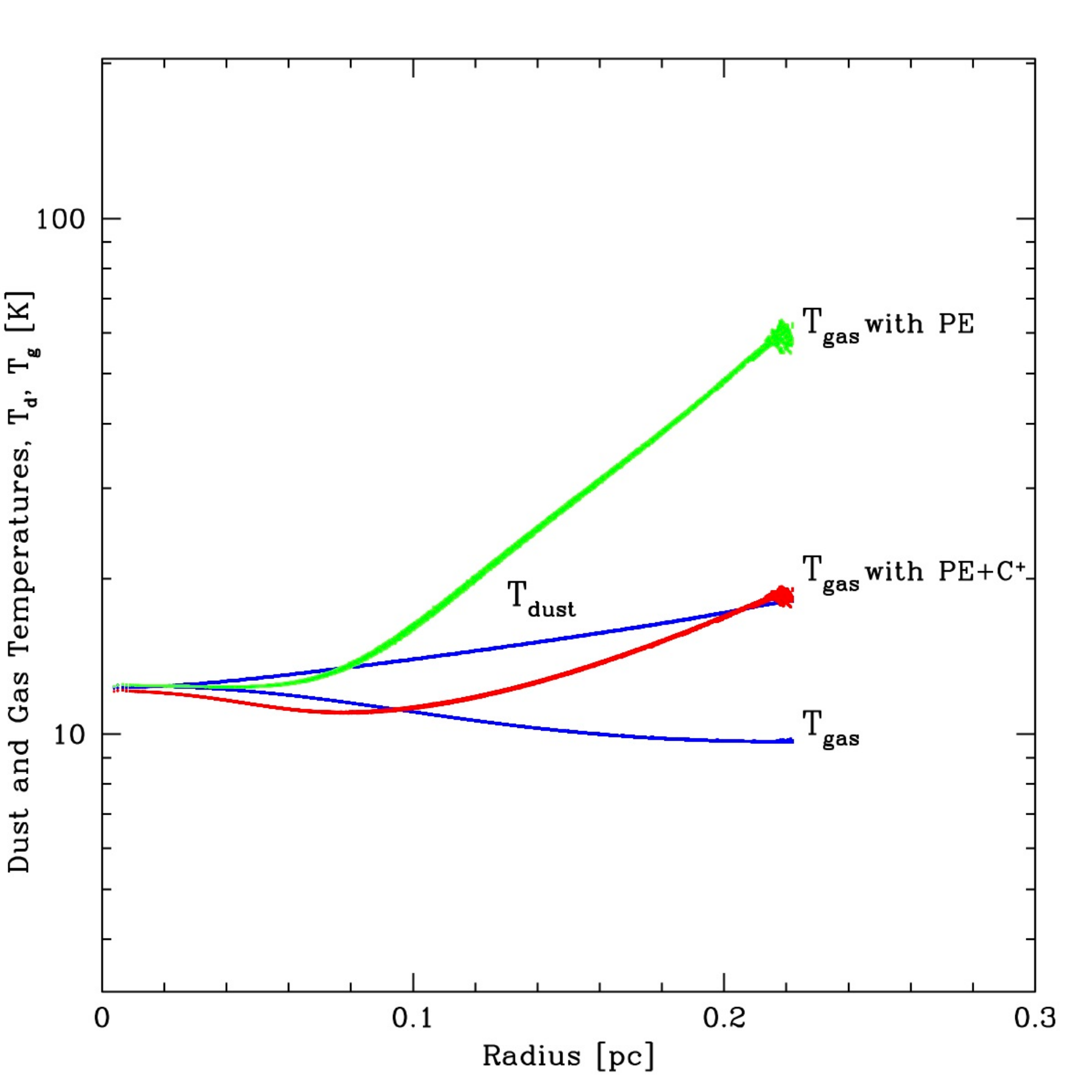} \vspace{-0.0cm}
    \includegraphics[width=8.5cm]{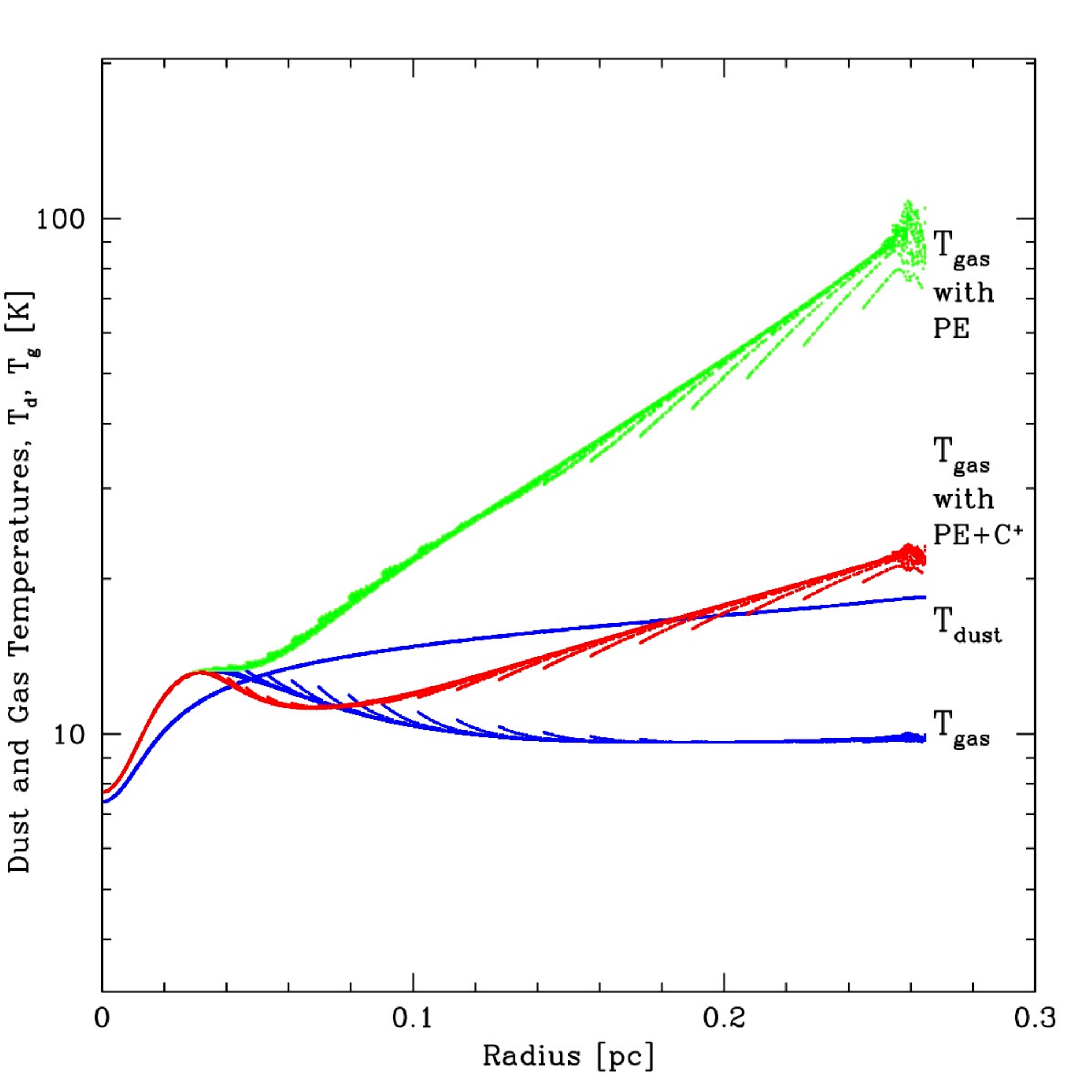} \vspace{-0.0cm}
\caption{The effect of photoelectric heating and C$^+$ cooling of the gas on the dust and gas temperatures as functions of radius inside the subcritical (left) and supercritical (right) 5~${\rm M}_\odot$ Bonner-Ebert spheres.  The dust temperatures are essentially unaffected by the extra gas heating and cooling processes.  The blue points are the same as in Fig.~\ref{KF05_2}, where the gas is subject to cosmic ray heating, molecular line emission, and collisions with the dust, while the dust is subject to heating from the ISR (including the UV), thermal emission, and collisions with the gas.  The green points include the same physical effects as the blue points, but with the addition of photoelectric gas heating (equation \ref{eq:photoelectric}).  The photoelectric heating has an enormous effect on the gas temperature in the outer, low-density parts of the clouds. However, when the C$^+$ cooling is included (red points), the effect of the photoelectric heating on the gas temperature is greatly reduced.}
\label{KF05_5}
\end{figure*}

\begin{figure*}
\centering \vspace{-0.3cm} \hspace{-0cm}
    \includegraphics[width=8.5cm]{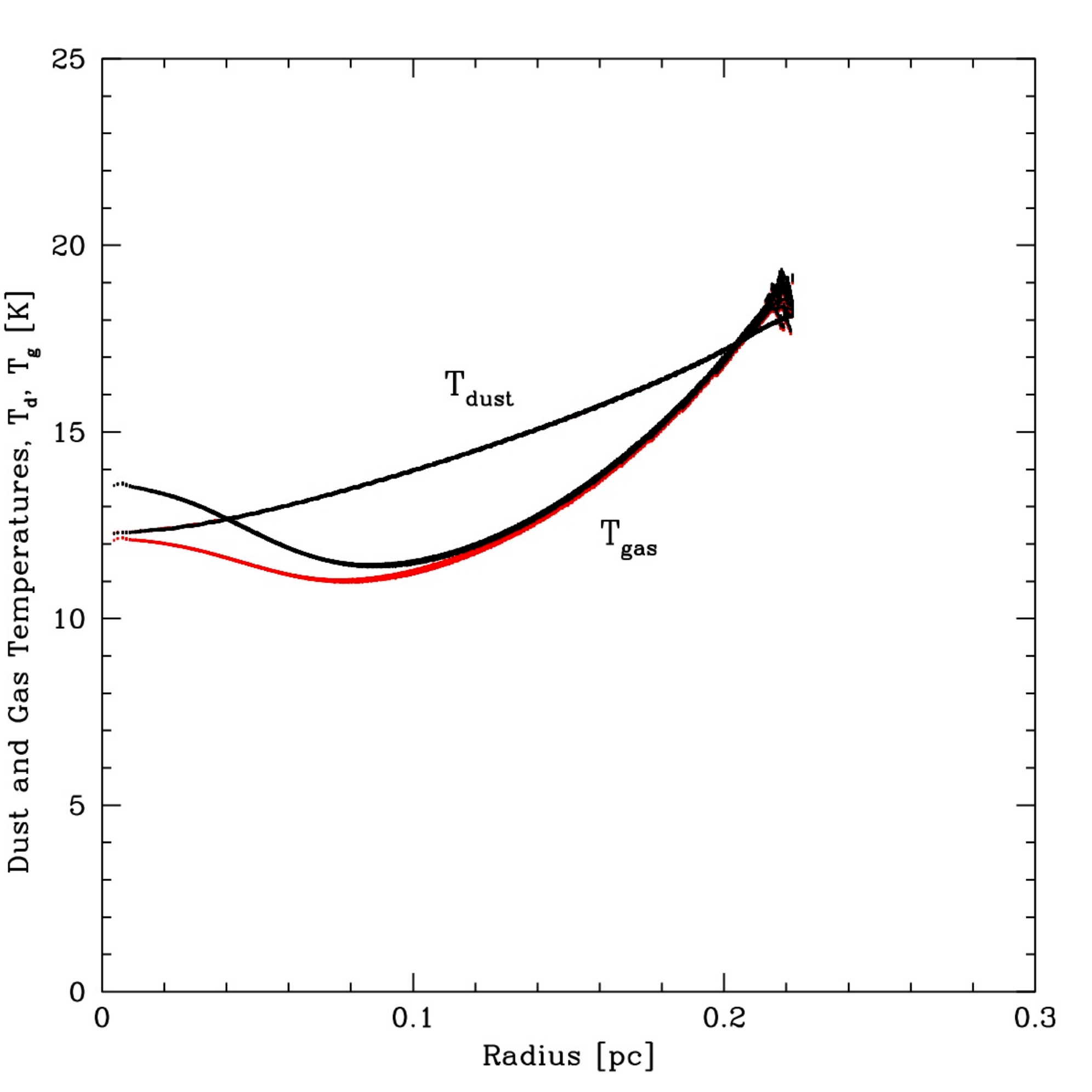} \vspace{-0.0cm}
    \includegraphics[width=8.5cm]{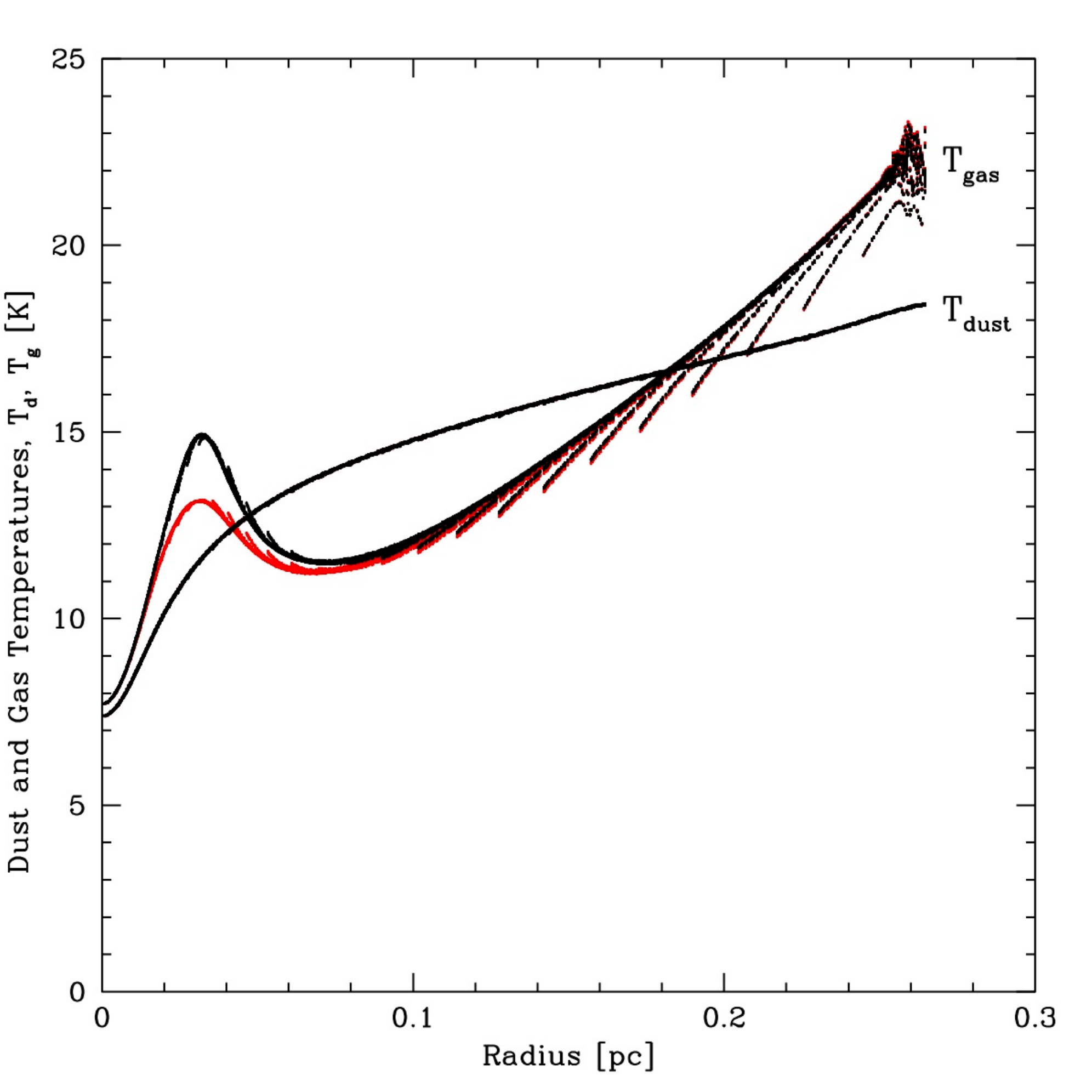} \vspace{-0.0cm}
\caption{The effect of CO depletion on the gas on the dust and gas temperatures as functions of radius inside the subcritical (left) and supercritical (right) 5~${\rm M}_\odot$ Bonner-Ebert spheres.  The red points are the same as in Fig.~\ref{KF05_5}, where the gas is subject to cosmic ray and photoelectric heating, molecular and C$^+$ line emission, and collisions with the dust, while the dust is subject to heating from the ISR (including the UV), thermal emission, and collisions with the gas.  The black points include the same physical processes as the red points, but also allow for CO depletion.  This reduces the molecular line cooling at high densities which raises the gas temperature slightly at densities $n_{\rm H2} = 10^4-10^5$~cm$^{-3}$.  At higher densities the depletion has little effect because the cooling is dominated by dust emission through collisions with the dust. }
\label{KF05_6}
\end{figure*}

\begin{figure*}
\centering \vspace{-0.3cm} \hspace{-0cm}
    \includegraphics[width=8.5cm]{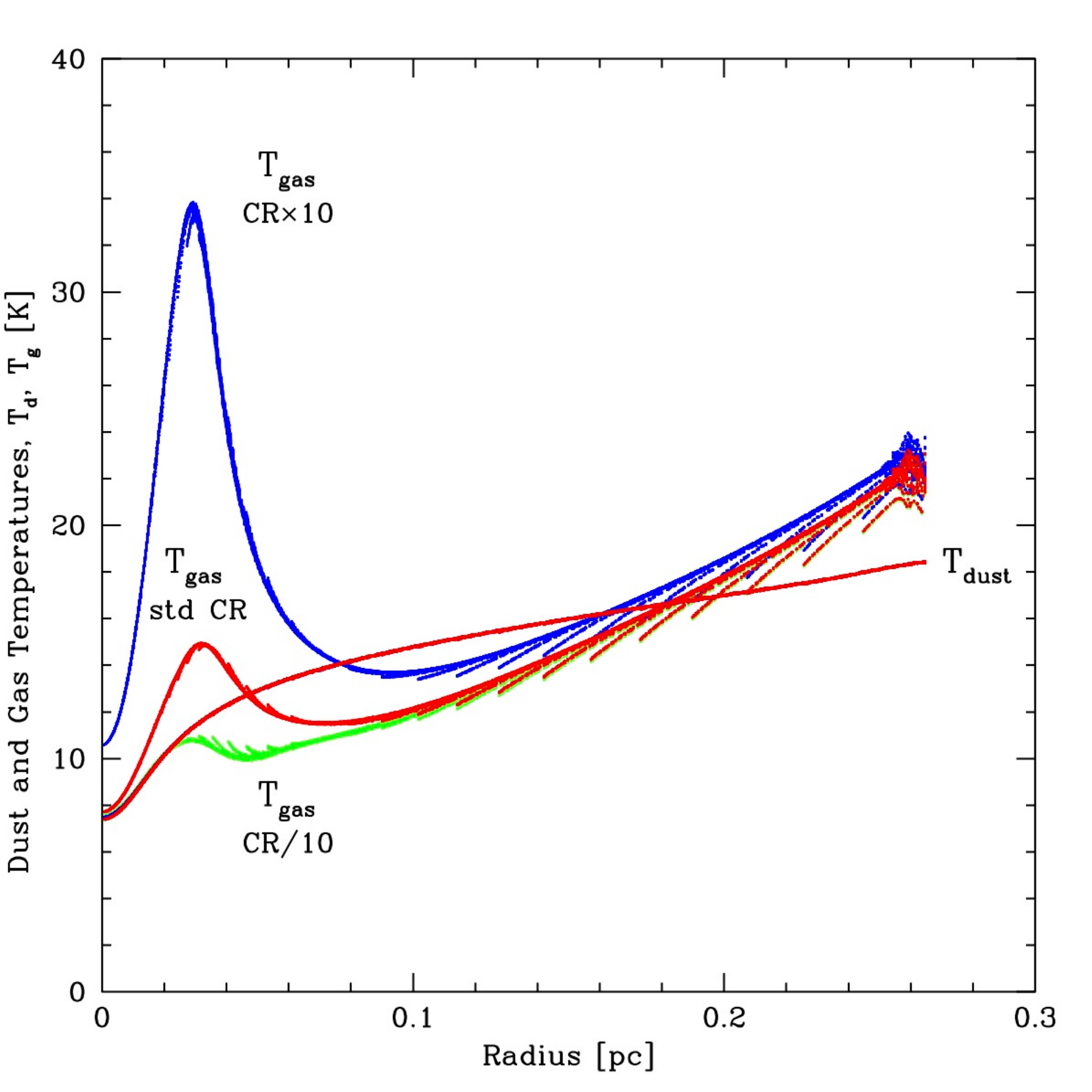} \vspace{-0.0cm}
    \includegraphics[width=8.5cm]{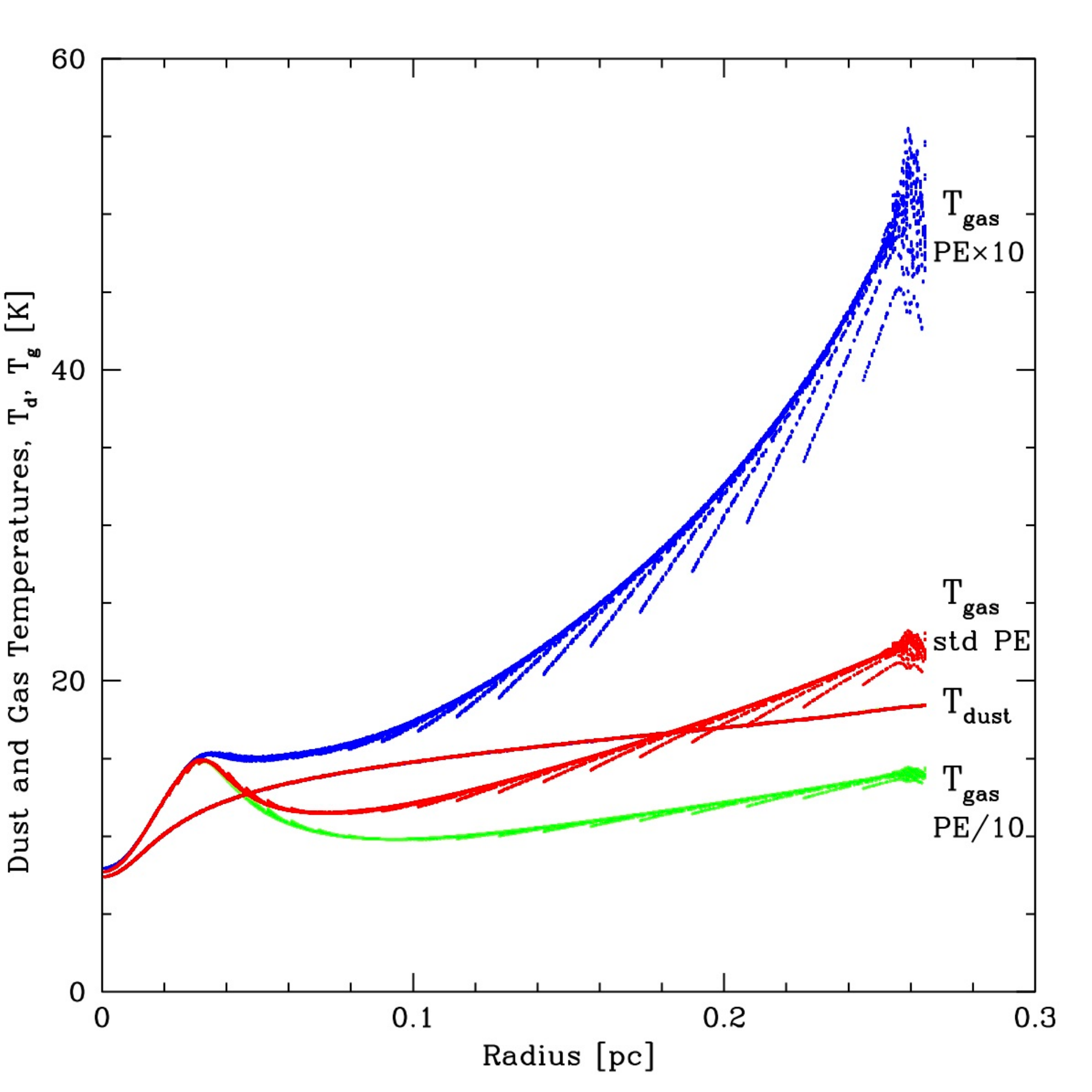} \vspace{-0.0cm}
\caption{The effects of changing the strengths of the cosmic ray (left) and photoelectric (right) heating of the gas on the dust and gas temperatures as functions of radius inside the supercritical 5~${\rm M}_\odot$ Bonner-Ebert sphere.  In both panels the red points are the same as in the black points in the right panel of Fig.~\ref{KF05_6}.  The other points include the same physical effects as the red points, but with one order of magnitude less (green) or more (blue) cosmic ray (left) or photoelectric (right) heating.  Changing the level of cosmic ray heating has a large effect on the gas temperature at intermediate densities, but not at high densities (where the gas and dust are well coupled) or low densities (where photoelectric heating dominates).  Changing the levels of photoelectric heating has a large effect on the gas temperature only in the outer regions of the core.  The dust temperatures are unaffected in either case.  }
\label{KF05_CR_PE}
\end{figure*}

\subsection{Equilibrium dust and gas temperatures}

More realistic than using a uniform-density sphere is to use Bonner-Ebert spheres.  To allow comparison with earlier work, we use the same 5-M$_\odot$ Bonner-Ebert spheres that were used by \cite{KetFie2005} --- a (marginally) subcritical case with a central density of $n_{\rm H2}=10^4$~cm$^{-3}$, and a supercritical case with a central density of $n_{\rm H2}=10^6$~cm$^{-3}$.  The former case has a ratio of the inner to outer density of 14, while the ratio of the latter is 3000.  Their radii are 0.223 and 0.265~pc, respectively.  In these tests, we assume the hydrogen is fully molecular.  We model both clouds with $3\times 10^5$ SPH particles.  

The SPH particles were set up using a cubic lattice which was then deformed radially to obtain the required cumulative radial mass profile. In Fig. \ref{KF_density} we provide the SPH density as a function of radius for of each SPH particle for both cores.  The small `spikes' in the density in the supercritical case are due to the lattice structure affecting the density calculation of some particles in this case with a very strong density gradient.  They do not appear in the subcritical case, which has a much shallower density profile.  In both cases, there is also some `noise' in the density near the outer boundary (which is made of reflected ghost particles).

In this section, we investigate the effects of each of the physical heating and cooling mechanisms on the temperatures of the gas and dust.  We begin by including only cosmic ray heating and molecular line cooling for the gas, and the dust is taken to be in thermal equilibrium with the ISR.  We then add in the effects of gas-dust collisions, photoelectric heating of the gas, and cooling from C$^+$.  For the first two cases, we also investigate the effects of including and excluding the UV flux.

\subsubsection{Cosmic ray heating, molecular line cooling, and dust radiative equilibrium}

In Fig.~\ref{KF05_1}, we plot the dust and gas temperatures as functions of radius inside the two Bonner-Ebert spheres.  These calculations include only cosmic ray heating and undepleted molecular line cooling of the gas, and the dust is in thermal equilibrium with the external ISR.  There is no gas-dust collisional coupling, photoelectric effect, or cooling due to C$^+$, oxygen, or recombination lines.  There are two calculations for each case --- one that excludes the UV contribution to the ISR, and one that includes it.

In the outer parts of the cores the dust temperature exceeds the gas temperature as the local ISR is only weakly attenuated by the dust, and the cosmic ray flux and line cooling are such that the low-density equilibrium gas temperature is $\approx 10$~K.  Including the UV contribution to the ISR boosts the dust temperature by up to 2~K in the outer parts, but the total energy in the UV flux is small compared to the total ISR flux and the UV does not penetrate very far into the cores.  In the inner parts of the cores, the gas temperature rises as the effectiveness of the line cooling decreases \citep{Goldsmith2001}.  On the other hand, the dust temperature decreases, particularly in the supercritical case, because the dust extinction attenuates the ISR.

\subsubsection{The effects of gas-dust collisions}

In Fig.~\ref{KF05_2}, we include the same physical processes as in the previous section, but we now also turn on the transfer of thermal energy between the gas and the dust due to collisions.  The strength of this interaction depends on the square of the density (equation \ref{eq:gasdust}), so we expect it to have little impact at low densities, but a significant impact as the density increases.  Indeed, this can be clearly seen in Fig.~\ref{KF05_2}, in which we plot the same calculations as in Fig.~\ref{KF05_1} for reference, but we now also include the case with gas-dust coupling in blue (without the UV ISR) and magneta (with the UV ISR).  The gas-dust thermal coupling has no significant effect on the temperatures in the subcritical core because the densities are too low.  In the supercritical case, the dust temperature is unaffected, but the gas temperature at $r<0.04$~pc (densities $n_{\rm H2} \gsim 2 \times 10^4$) is `dragged down' by the interaction with the cold dust so that at the centre the gas and dust temperatures are both $\approx 7.5$~K.

Our numerical results are in reasonable agreement with those presented in Figs. 4 and 5 of \cite{KetFie2005}, who use the same models for the ISM physics as we use here, without the UV ISR (black solid lines in Fig.~\ref{KF05_2}).  The main difference is that the gas temperatures at low densities are slightly lower in \cite{KetFie2005}.  The gas temperature at low density is set by the balance between cosmic ray heating and the line cooling rates.  The latter is dependent on the method used to interpolate from the tables of \cite{Goldsmith2001}.  Since the cosmic ray heating and line cooling rates are both simple functions of $n_{\rm H}$ (equations \ref{eq:cosmicray} and \ref{eq:line}), it is easy to solve for the central gas temperature of 12.4~K in the subcritical case (where the gas-dust coupling is negligible). The slightly lower gas temperature ($\approx 11.5$~K) obtained by \cite{KetFie2005} was found to be due to less accurate interpolation used by \citeauthor{KetFie2005} of the line cooling functions of \cite{Goldsmith2001}.  Thus, the different interpolation of the line cooling produces the different gas temperatures.

\begin{figure}
\centering \vspace{-0.3cm} \hspace{-0cm}
    \includegraphics[width=8.5cm]{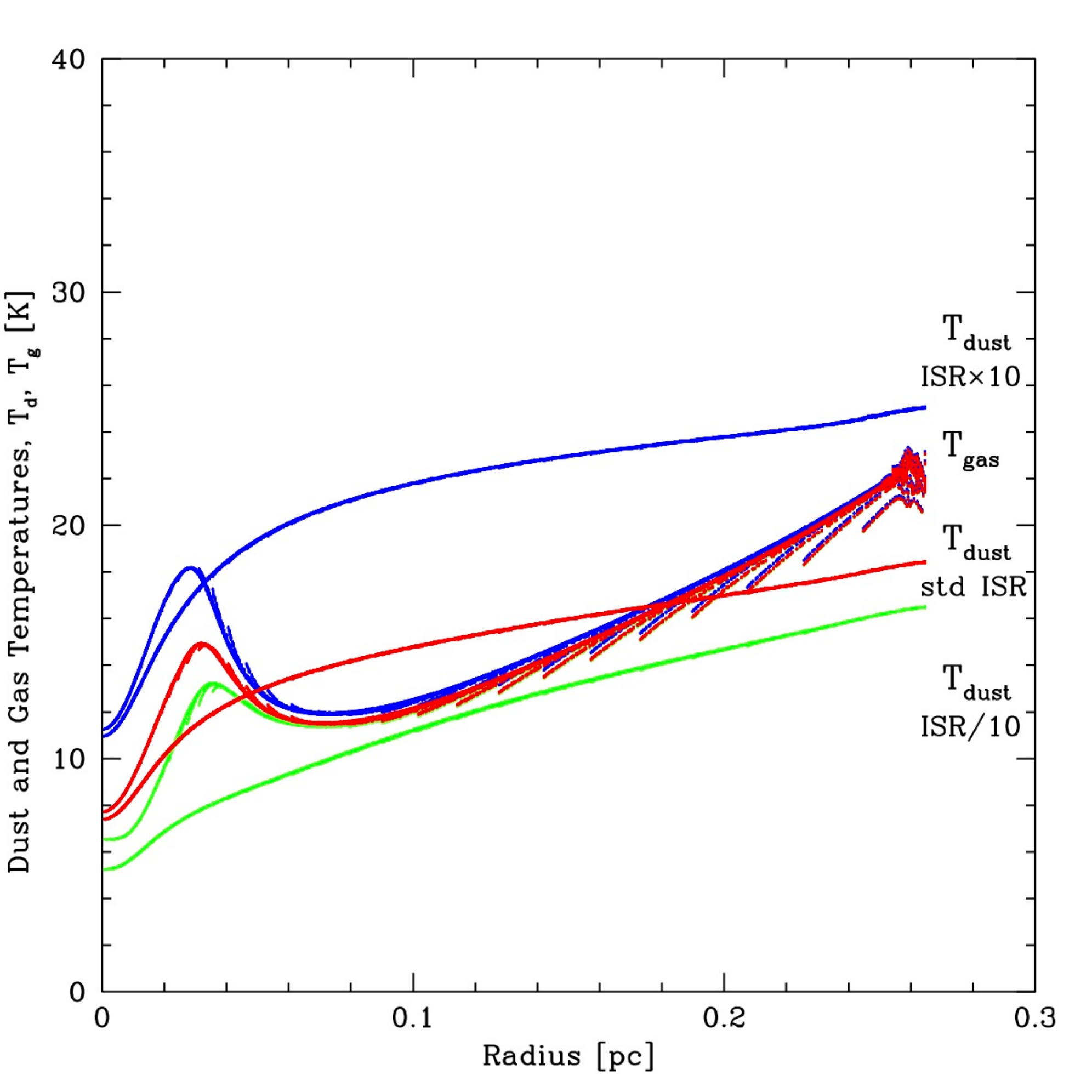} \vspace{-0.0cm}
\caption{The effects of changing the strength of the interstellar radiation (ISR) field on the dust and gas temperatures as functions of radius inside the supercritical 5~${\rm M}_\odot$ Bonner-Ebert sphere.  The red points are the same as the black points in the right panel of Fig.~\ref{KF05_6}.  The other points include the same physical effects as the red points, but with one order of magnitude less (green) or more (blue) ISR heating.  Changing the level of ISR by an order of magnitude has a 50\% effect ($10^{1/6}$) on the dust temperatures.  At high densities the different dust temperature affects the gas temperature in a similar manner, but at low-densities the gas temperature is unaffected as it is set by the photoelectric heating and C$^+$ cooling.}
\label{KF05_ISR}
\end{figure}

\begin{figure}
\centering \vspace{-0.3cm} \hspace{-0cm}
    \includegraphics[width=8.5cm]{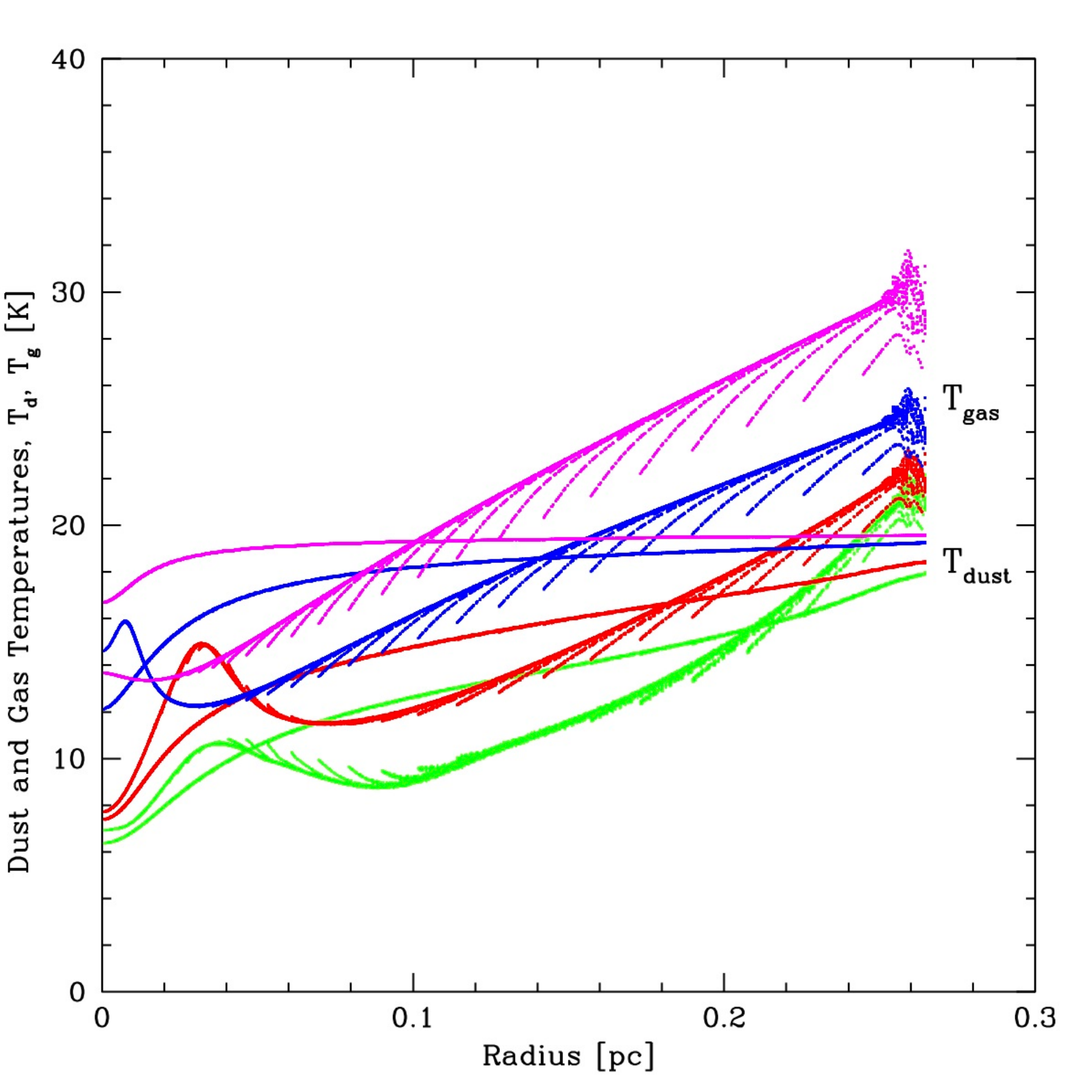} \vspace{-0.0cm}
\caption{The effects of changing the metallicity on the dust and gas temperatures as functions of radius inside the supercritical 5~${\rm M}_\odot$ Bonner-Ebert sphere.  Photoelectric heating has been included. From top to bottom, the calculations have metallicities of 1/100 (magenta points, top curves), 1/10 (blue points), solar metallicity (red points), and 3 times solar metallicity (green points).  The red points are the same as in the black points in the right panel of Fig.~\ref{KF05_6}.    The dust temperatures are greater with lower metallicities because the ISR is less attenuated by extinction.  The gas temperatures are greater with lower metallicities because the UV ISR is less attenuated by extinction (leading to stronger photoelectric heating) and also because the gas emission line cooling is reduced.}
\label{KF05_Metal}
\end{figure}

\subsubsection{The effects of photoelectric heating and carbon chemistry}

In Fig.~\ref{KF05_5}, in addition to the other physical processes, we turn on photoelectric heating of the gas due to UV ISR photons liberating electrons from dust grains (equation \ref{eq:photoelectric}).  The resulting gas and dust temperatures (green) are compared to those without the photoelectric heating (blue).  The dust temperature is essentially unchanged, but the gas temperature in the outer parts of the cores, at densities such that the gas and dust are thermally decoupled ($n_{\rm H2}\lsim 10^4$~cm$^{-3}$), is much hotter than without photoelectric heating.  In fact the photoelectric heating is so strong that the cosmic ray heating is irrelevant in the low-density gas.  

Whereas \cite{KetFie2005} did not consider photoelectric heating, as discussed in Section \ref{sec:chemistry}, \cite{KetCas2008} included photoelectric heating and a simple carbon chemistry model which allowed them to include both cooling from C$^+$ at low densities and treat the depletion of CO at high densities.  Therefore, in Fig.~\ref{KF05_5}, we keep the photoelectric heating on, but also add cooling from C$^+$ (red).  It can be seen that the C$^+$ cooling dramatically lowers the temperatures in the outer parts of the cloud; the temperature is still somewhat higher than it was without either the photoelectric heating or the C$^+$ cooling, but much less than if $C^+$ cooling is omitted.  We note that recombination and oxygen cooling are both insignificant at these densities and that the temperatures are essentially identical whether they are included or not.

Finally, in Fig.~\ref{KF05_6} we turn on the effects of CO depletion based on the equilibrium prescription of \cite{KetCas2008}.  This results in only slightly higher gas temperatures at intermediate densities ($n_{\rm H2}=10^4-10^5$~cm$^{-3}$) where the densities are still too low for dust cooling to dominate, but the densities are high enough that C$^+$ is ineffective.

\subsubsection{The effects of changing the ISM parameters}

The above sections used the diffuse ISM model parameters described in Sections \ref{sec:combine} and \ref{sec:heatingcooling}.  In this section, we investigate the dependence of the gas and dust temperatures on the strengths of the various external heating sources.  We only use the supercritical Bonner-Ebert sphere in these tests because this has the larger range of densities.  In the left panel of Fig.~\ref{KF05_CR_PE} we explore the effects of changing the cosmic ray heating rate (equation \ref{eq:cosmicray}) by an order of magnitude in each direction.  In the right panel of Fig.~\ref{KF05_CR_PE} we explore the effects of change the gas photoelectric heating rate (equation \ref{eq:photoelectric}) by an order of magnitude in each direction.  It can be seen that over particular density ranges, either of these changes the gas temperature by approximately half an order of magnitude in each direction because the gas line cooling scales roughly as the square of the gas temperature for the molecular lines \citep{Goldsmith2001}.  The gas temperatures in the outer parts of the cloud are much more strongly affected by photoelectric heating than cosmic ray heating.  On the other hand, at number densities $n_{\rm H2}\gsim 10^5$~cm$^{-3}$ strong cosmic ray heating can raise the gas temperature significantly above the dust temperature whereas the photoelectric heating has no effect this deep within the core.

In Fig.~\ref{KF05_ISR}, we probe the effects of changing the ISR field (equation \ref{eq:blackbodies}) by an order of magnitude in each direction, excluding the UV contribution which is held fixed.  The calculations include photoelectric heating of the gas and C$^+$ cooling.  The increased ISR primarily affects the dust temperature, but because the dust emission scales as the sixth power of its temperature (equation \ref{lambda_dust}), this only has a 50\% effect on the dust temperatures.  Deep within the core where the gas becomes thermally coupled to the dust, the warmer or cooler dust also results in warmer or cooler gas, respectively.

In Fig.~\ref{KF05_Metal}, we investigate the effects of changing the metallicity of the core, performing additional calculations with metallicities of 1\%, 10\%, and 3 times solar.  It can be seen that the highest metallicity results in the coldest temperatures for both the gas and the dust due to the combination of several effects.  The increased metallicity increases the ISR attenuation, decreasing the dust temperature inside the core.  Similarly, the UV is prevented from penetrating as far into the core, reducing the effects of photoelectric heating of the gas, and the gas line emission is enhanced.  Finally, there is an increase in the effectiveness of collisional thermal coupling between the gas and the dust.  With the lowest metallicity, the dust almost becomes isothermal since there is almost no attenuation of the ISR, and the gas is warmer because the photoelectric heating is strong while emission line cooling is weak.

\begin{figure}
\centering \vspace{-0.3cm} \hspace{-0cm}
    \includegraphics[width=8.5cm]{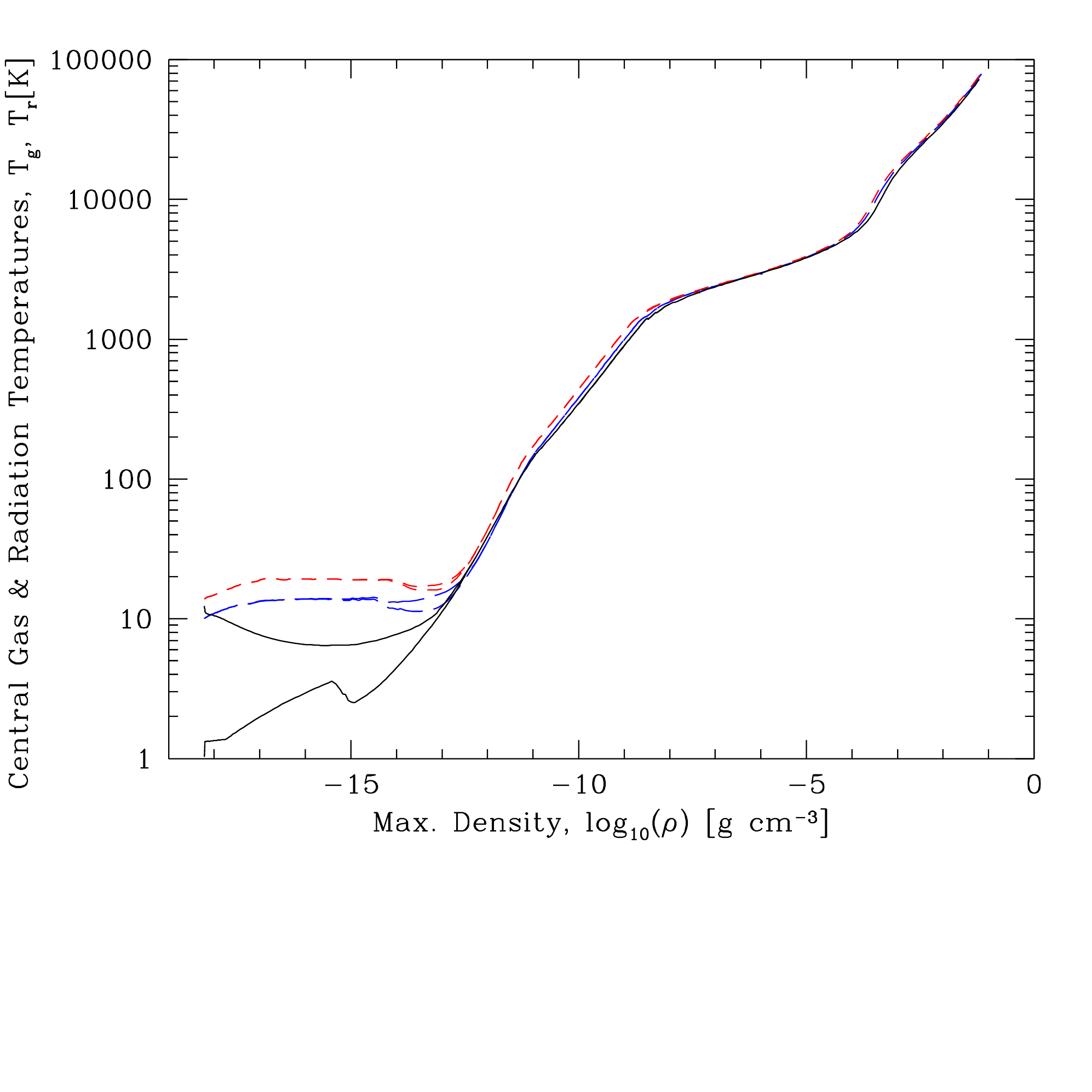} \vspace{-2.0cm}
\caption{The evolution of the central gas and radiation temperatures as functions of the maximum (central) density during a radiation hydrodynamical calculation of the collapse of an unstable spherical 5~${\rm M}_\odot$ Bonner-Ebert sphere.  The radiation temperature is always less than or equal to the gas temperature in this graph.  After the central density exceeds $10^{-17}$~g~cm$^{-3}$, the central dust temperature is indistinguishable from that of the gas. The black solid lines give the results from a calculation using the method described in this paper, while the red dashed lines and blue long-dashed lines give the results using the pure flux-limited diffusion (FLD) method of Whitehouse \& Bate (2006).  The diffuse ISM model used in the former calculation determines the initial gas and dust temperatures, which vary with radius, and the internal radiation temperature is arbitrarily set to a low initial value (1~K).  In the FLD calculations the initial temperatures of the matter (gas and dust) and internal radiation are arbitrary and we set them to 14~K (red dashed) and 10~K (blue long-dashed).  As expected, it is mainly the low-density evolution that is affected by including the diffuse ISM model.  }
\label{BEcollapse}
\end{figure}

\subsection{The evolution of a collapsing Bonner-Ebert sphere}

In this section, as opposed to the static tests from the previous sections, we report the results from a hydrodynamical calculation of the collapse of a Bonner-Ebert sphere and compare them to the results obtained using the pure flux-limited diffusion method of \cite{WhiBatMon2005} and \cite{WhiBat2006}.  We begin with a somewhat unstable Bonner-Ebert sphere, similar to those studied in the previous section.  We choose a 5-M$_\odot$ core with an inner to outer density contrast of 20 and a radius of 0.1~pc.  We model the core with $3\times 10^5$ SPH particles and use spherical reflective boundary conditions (modelled using ghost particles).

In Fig.~\ref{BEcollapse} we plot the central temperatures as functions of the central (maximum) density for the two methods.  All of the initial temperatures in the pure flux-limited diffusion calculations are arbitrary.  As is standard practice in such calculations, we set the matter and radiation temperatures to be in thermal equilibrium initially.  We perform two separate calculations and with two different initial temperatures of 10~K and 14~K.  For the calculation performed using the method described in this paper, the initial gas and dust temperatures are not arbitrary --- they are set by our model of the diffuse ISM.  The initial dust temperature increases from 8~K in the centre to 17~K at the outer edge of the molecular cloud core.  The initial gas temperature is slightly warmer than the dust at the centre (10~K) and cooler at the outside ($\approx 15$~K).  The initial central dust and gas temperatures are quite similar because the central density of the core is quite high ($\rho = 6 \times 10^{-19}$~g~cm$^{-3}$ or $n_{\rm H2}=1.5\times 10^5$~cm$^{-3}$) so that there is significant thermal coupling between the dust and the gas (we use equation \ref{eq:gasdust} in these calculations).  As the collapse proceeds, the central dust and gas temperatures quickly converge, so we do not plot the central dust temperature in Fig.~\ref{BEcollapse}.  The increases in extinction and density during the collapse mean that both the dust and gas temperatures drop to a minimum of 6.5~K when the central density is $10^{-16}$~g~cm$^{-3}$ before the compressional heating starts to increase the central temperature again.  The initial radiation temperature should be set to a low value, because the heating of the initial conditions is via external radiation rather than internal radiation.  As long as a small value is chosen the exact value is unimportant, but we wish to avoid setting it to zero so as to avoid numerical problems.  We chose a value of 1~K.

\begin{figure*}
\centering \vspace{-0.3cm} \hspace{-0cm}
    \includegraphics[width=8.5cm]{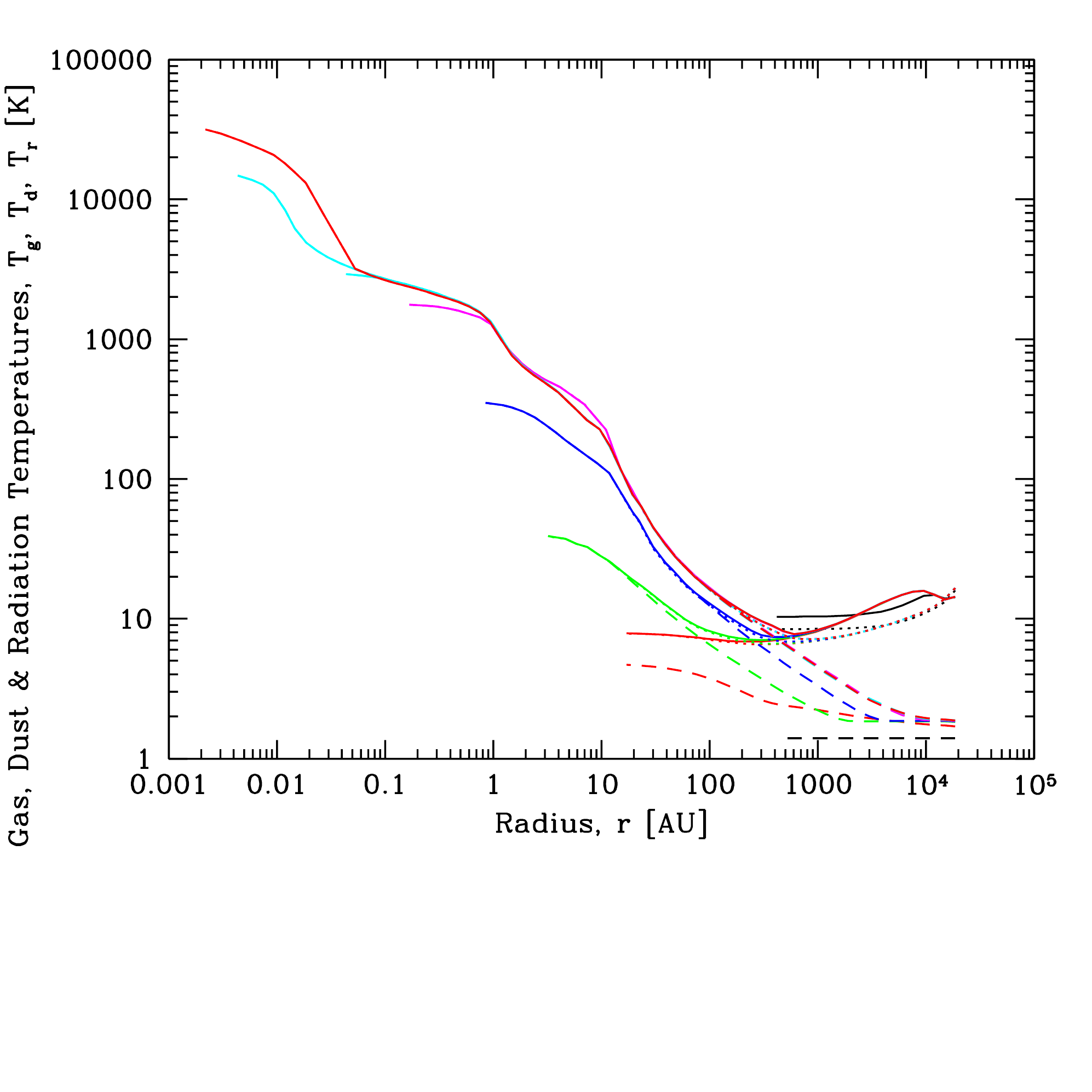} \vspace{0.0cm}
    \includegraphics[width=8.5cm]{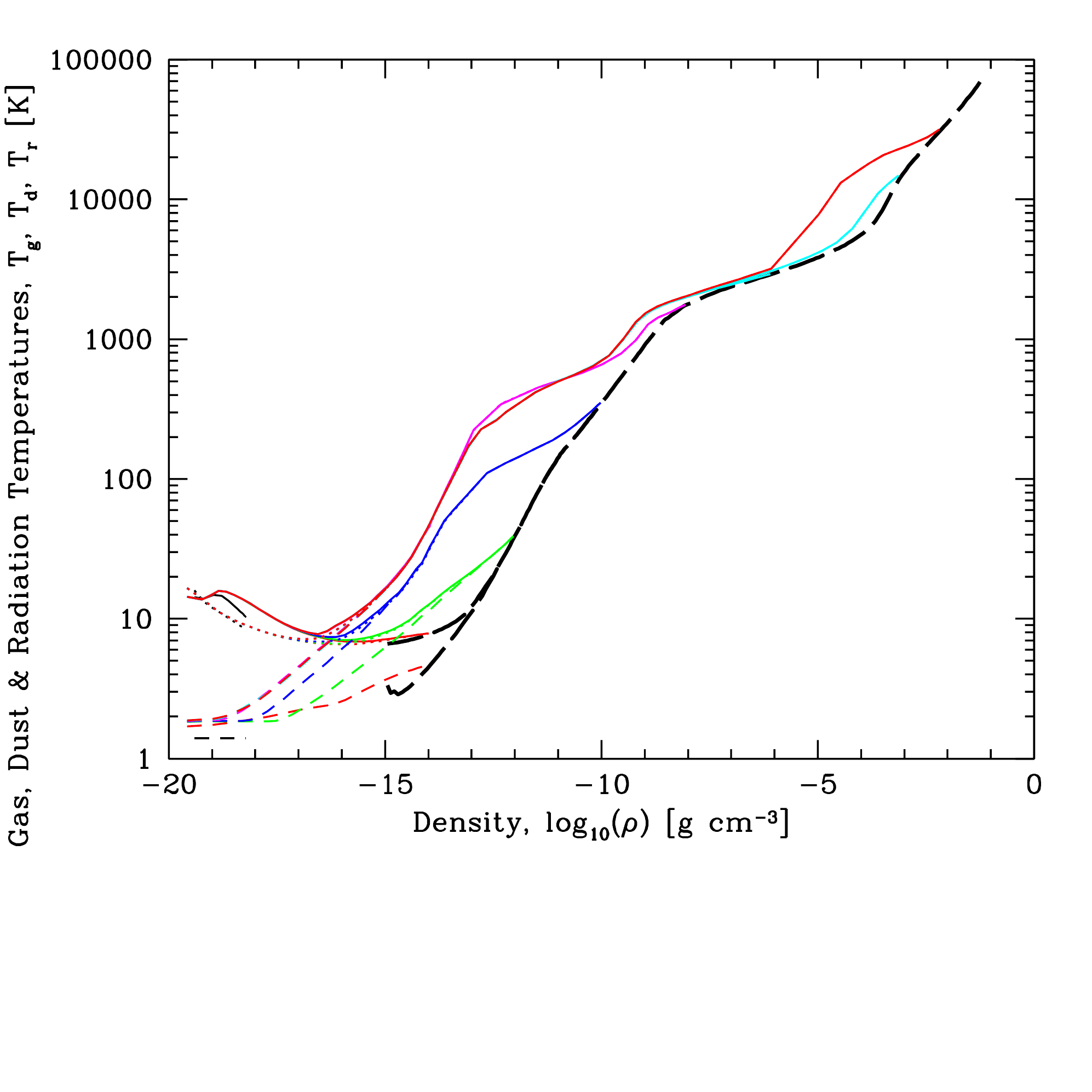} \vspace{-2.0cm}
\caption{The evolution of the gas (solid), dust (dotted), and radiation (dashed) temperatures as functions of radius (left panel) and maximum density (right panel) during a radiation hydrodynamical calculation of the collapse of an unstable spherical 5~${\rm M}_\odot$ Bonner-Ebert sphere.  Each colour gives the state of the calculation when the maximum density reaches a different value, with the initial conditions given in black.  The radiation temperature is always less than or equal to the gas temperature in this graph.  The thick black long-dashed lines in the right panel give the evolution of the central gas and radiation temperatures with central density from Fig.~\ref{BEcollapse} for reference (only for $\rho>10^{-15}$~g~cm$^{-3}$).  It can be seen that for a given density the gas is usually hotter latter in the collapse than earlier due to radiation from the protostar heating the outer parts of the core.  The formation of the stellar core is apparent in the upper-left of the left panel (initial radius $\approx 0.01$~AU).}
\label{BEcollapse_profiles}
\end{figure*}

As may be expected, Fig.~\ref{BEcollapse} shows that it is mainly the low-density phase of evolution of the clouds that differs between the two methods (densities $\lsim 10^{-13}$~g~cm$^{-3}$).  Once the central regions of the core become optically-thick to infrared radiation (the so-called opacity limit for fragmentation; \citealt{LowLyn1976,Rees1976}) all of the temperatures converge.  The evolution obtained with the pure flux-limited diffusion calculation with the lower initial temperature (10~K) is closest to the evolution obtained using the method presented here, presumably because the central temperature in this model is closest to that given by the ISM initial conditions.

In Fig.~\ref{BEcollapse_profiles} we plot temperature profiles as functions of radius (left panel) and density (right panel) at various points during the collapse.  Gas is shown with solid lines, dust with dotted lines, and radiation with dashed lines.  At either a given radius or density the temperatures tend to rise during the collapse as the protostar emits more radiation.  The formation of the stellar core at late times (initial radius $\approx 0.01$~AU and temperature $>10^4$~K) is clearly visible in the left panel.  Except in the outer parts of the core, these profiles are very similar to those obtained using pure flux-limited diffusion.  The main difference in the outer parts of the core (radii $\gsim 10^3$~AU) is that the gas, dust, and radiation all have different temperatures with the new method, whereas with pure flux-limited diffusion all are very close to the initial arbitrary temperature (e.g.~10~K; see \citealt{WhiBat2006}).

\begin{figure}
\centering \vspace{-0.3cm} \hspace{-0cm}
    \includegraphics[width=8.5cm]{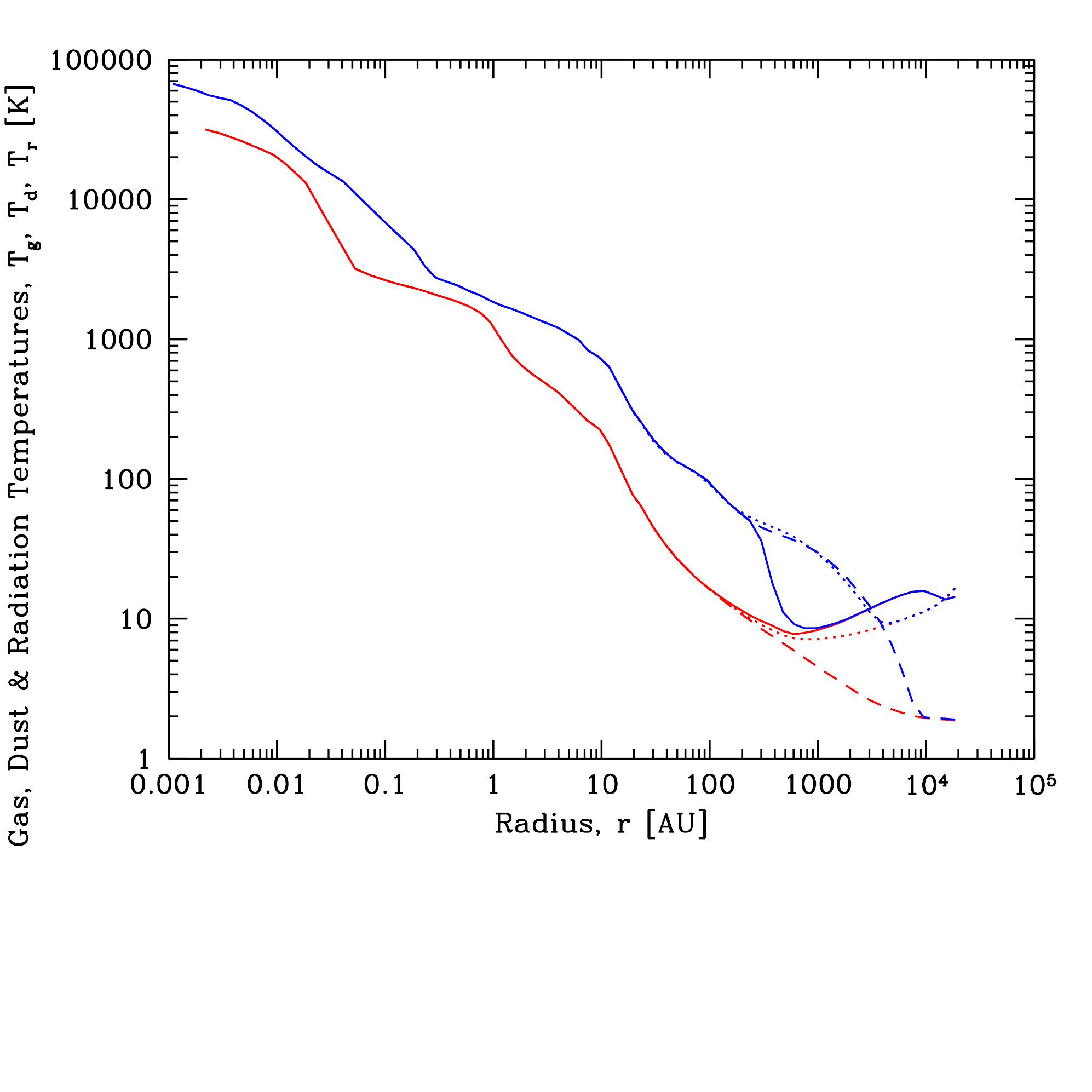} \vspace{-2.0cm}
\caption{The gas (solid), dust (dotted), and radiation (dashed) temperatures as functions of radius when the collapsing spherical 5~${\rm M}_\odot$ Bonner-Ebert sphere has reached a maximum density of $0.01$~g~cm$^{-3}$ (lower, red lines), and 1.76 years later at the end of the calculation when the maximum density is $0.06$~g~cm$^{-3}$ (upper, blue lines).  Near the end of the calculation, after the stellar core forms, the radiation and dust temperatures become hotter than the gas temperature at radii of $400-3000$~AU because of the radiation emitted by the protostar.  }
\label{BEcollapse_end}
\end{figure}

After stellar core formation as radiation works its way outwards from the centre the gas, dust, and radiation temperatures become inverted at intermediate radii in the core (radii $400-3000$~AU; Fig.~\ref{BEcollapse_end}).  Before this point the gas temperature is the hottest and the radiation temperature the coldest at these radii.  However, the radiation emitted by the protostar quickly dominates that from the external radiation field at these radii.  The dust adopts a new thermodynamic equilibrium with the combined radiation field, being heated to temperatures of $15-40$~K, but the gas which is poorly coupled to either the radiation or the dust at these low densities remains at temperatures of $8-12$~K.  Such effects are impossible to capture unless the thermal evolution of the gas and dust are treated separately.

\subsection{The evolution of an isolated turbulent molecular cloud}
\label{sec:testGC}

\cite{GloCla2012a,GloCla2012c} studied the thermodynamics of turbulent molecular clouds with different metallicities.  In particular, they performed calculations of $10^4$~M$_\odot$ clouds, typically modelled using SPH with 2 million particles.  Their initial conditions consisted of a uniform-density sphere of number density $n_{\rm H}=300$~cm$^{-3}$, giving a radius of approximately 6~pc.  The cloud was given initial `turbulent' motions with a power spectrum $P(k) \propto k^{-4}$ where $k$ is the wavenumber.  The energy in the turbulence was initially set equal to the magnitude of the gravitational potential energy of the cloud, giving an initial root-mean-square velocity of around 3 km~s$^{-1}$.  The turbulence was allowed to decay freely.  The calculations were performed using a confining pressure of $p_{\rm ext}= 1.2 \times 10^4$ K~cm$^{-3}$ which was implemented in the SPH equations by subtracting the external pressure from the pressure used in the momentum equation.  We use the same set up as \citeauthor{GloCla2012a}, although of course we are unable to use exactly the same initial velocity field.  We use our full diffuse ISM model to obtain the results below, but we begin by excluding the hydrogen chemistry and heating due to the formation of molecular hydrogen (Section \ref{sec:hydrogen}).  In the first set of results, we have also excluded molecular depletion (because \citeauthor{GloCla2012a} did not include depletion), but since we find that switching depletion on or off has little effect on the results it is included in most of the subsequent calculations.  For the dust-gas thermal coupling we switch to equation \ref{eq:gasdust2} because this is the equation \citeauthor{GloCla2012b} used.

\subsubsection{Results excluding hydrogen chemistry}

\begin{figure}
\centering \vspace{-0.3cm} \hspace{-0cm}
    \includegraphics[width=9cm]{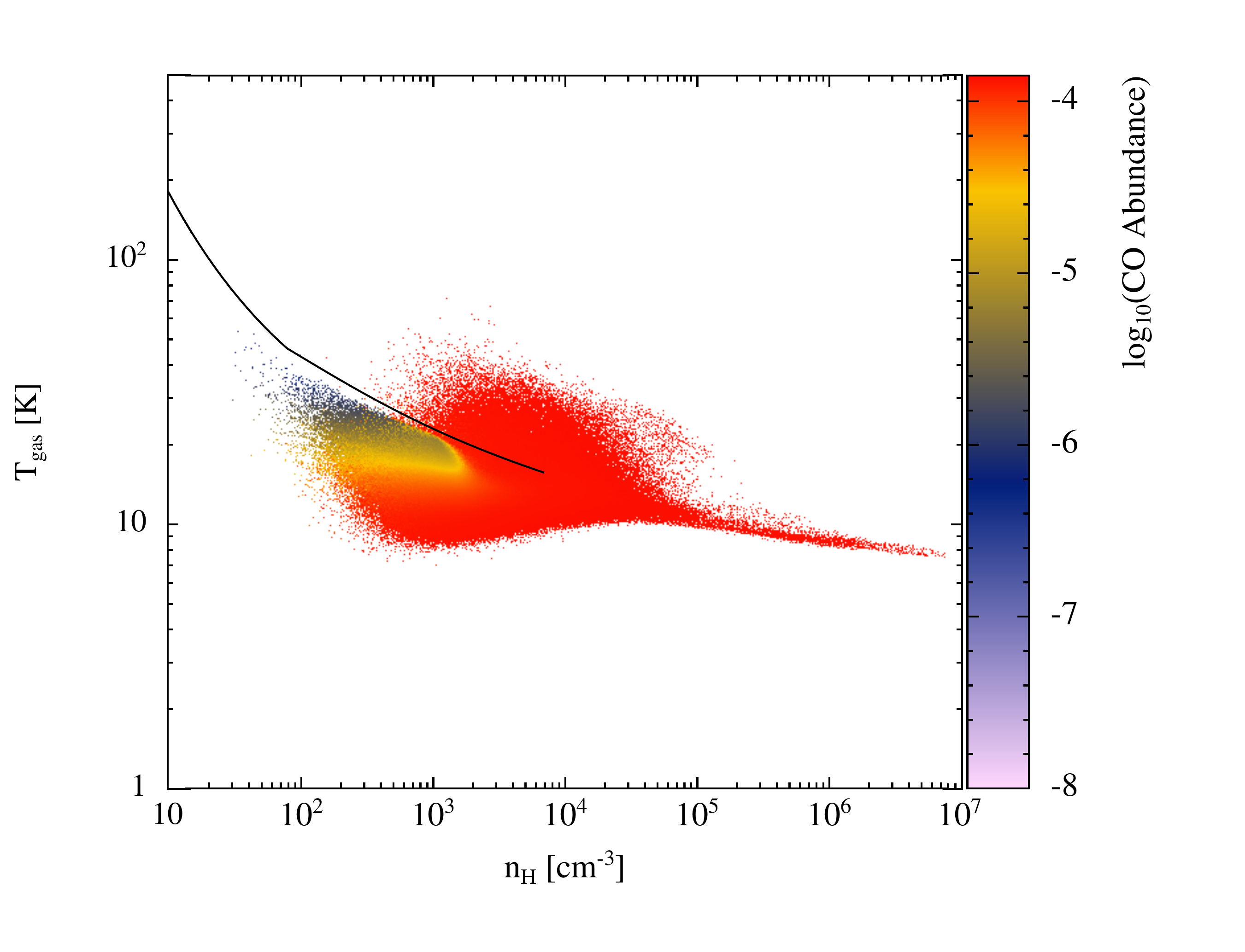} \vspace{-0.0cm}
    \includegraphics[width=9cm]{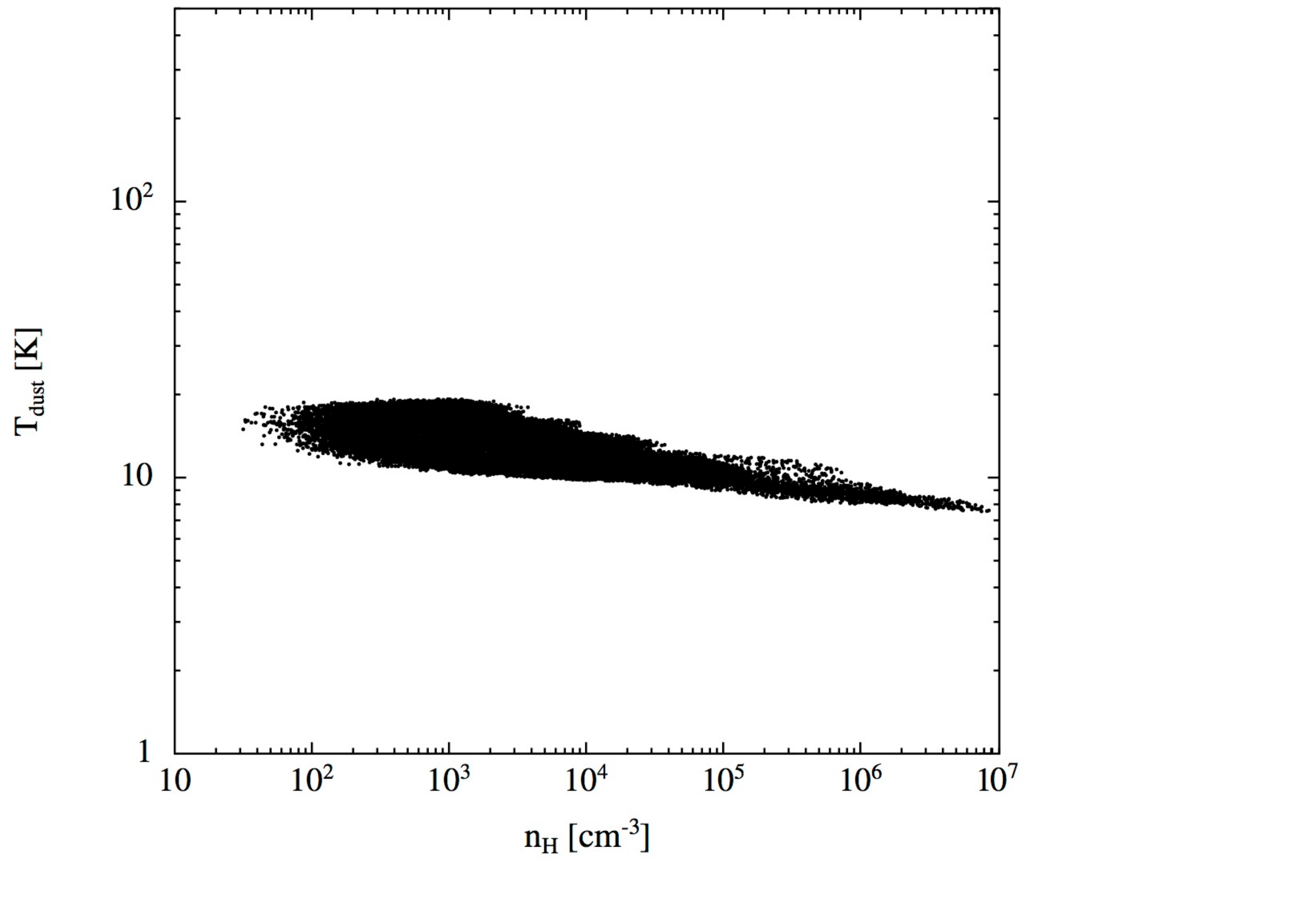} \vspace{-0.6cm}
\caption{Phase diagrams showing the gas (top panel) and dust (lower panel) temperatures as functions of the gas density.  The colours in the gas temperature plot indicate the average CO abundance of each point in temperature-density space.  The black solid line in the gas temperature plot is the equilibrium curve from Fig.~\ref{T_n}. }
\label{GC2012a}
\end{figure}

In Fig.~\ref{GC2012a} we plot the gas (upper panel) and dust (lower panel) temperatures as functions of the number density of hydrogen nuclei, $n_{\rm H}$, for a calculation performed at solar metallicity.  The snapshot is taken just before the first sink particle is formed.  The colours in the gas temperature plot indicate the average CO abundance of each point in temperature-density space. This is the equivalent of Fig.~2 in \cite{GloCla2012a} or the upper-right panel of Fig.~4 in \cite{GloCla2012c}.  Since \citeauthor{GloCla2012a} did not include molecular depletion on to grains, we have turned this off in our calculation.  The distribution of gas temperatures with density shows the same general features seen in the papers of \citeauthor{GloCla2012a}.  At low densities ($n_{\rm H}\lsim 10^3$~cm$^{-3}$) the gas temperature rises towards $\sim 100$~K as the number density decreases towards $n_{\rm H}\sim 10$~cm$^{-3}$.  The temperatures at densities $n_{\rm H}\lsim 100$~cm$^{-3}$ are primarily determined by the models of photoelectric heating, and cooling due to electron recombination and emission from ionised oxygen and carbon (equations \ref{eq:photoelectric} to \ref{eq:carbon}).  The gas temperatures lie beneath the equilibrium curve (solid black line from Fig.~\ref{T_n}) because the equilibrium curve assumes no extinction whereas in fact the photoelectric heating is generally attenuated by the dust extinction (depending on the location of the gas).  Our maximum temperatures in this density regime are somewhat cooler than those obtained by \citeauthor{GloCla2012a} due to our slightly different models in this regime as already demonstrated in Fig.~\ref{T_n}.  At densities $n_{\rm H}\approx 100-10^4$~cm$^{-3}$ there is a large dispersion in the gas temperatures with a range of $T_{\rm g} \approx 8-50$~K obtained at densities $n_{\rm H}\approx 10^3-10^4$~cm$^{-3}$.  In this region, although there is still significant photoelectric heating, the low-density coolants just mentioned become ineffective.  Instead, the main cooling is from molecular lines, in particular CO \citep[e.g.][]{Goldsmith2001,GloCla2012a}, which depends strongly both on density and temperature.  At the same time, work from the gas motions becomes important (either heating or cooling the gas, depending on whether the gas is being compressed or expanding, respectively), as does heating in shocks.  Together these effects lead to the wide range in temperatures.  Finally, at densities $n_{\rm H}\gsim 10^5$~cm$^{-3}$ thermal coupling of the gas to the dust becomes important and the gas temperature converges towards that of the dust as the density increases.

\begin{figure}
\centering \vspace{-0.3cm} \hspace{-0cm}
    \includegraphics[width=9cm]{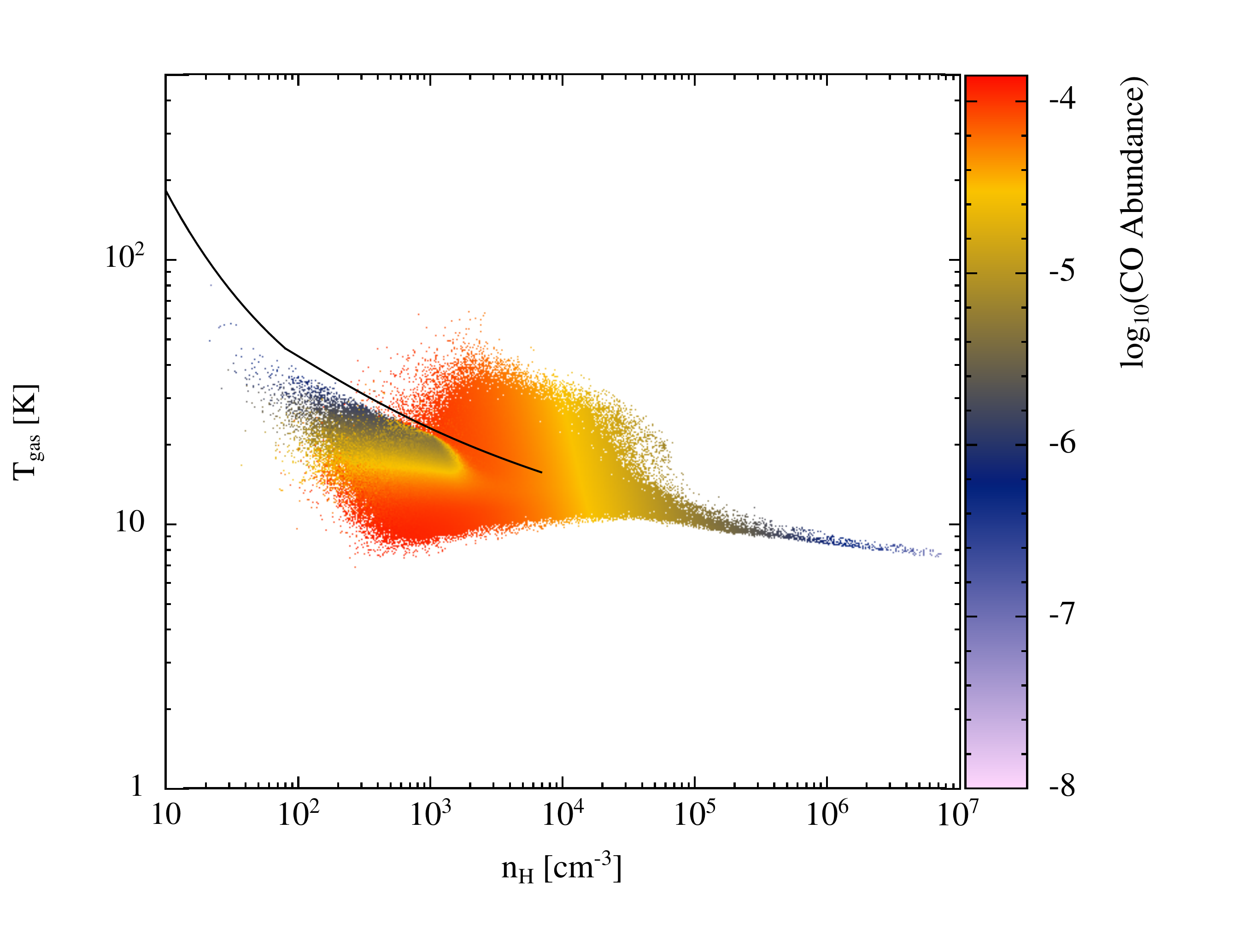} \vspace{-0.5cm}
\caption{Phase diagram showing the gas temperature as a function of the gas density for a calculation that is identical to that portrayed in Fig.~\ref{GC2012a}, except that molecular depletion is turned on.   The black solid line is the equilibrium curve from Fig.~\ref{T_n}. }
\label{GC2012a_depletion}
\end{figure}

\begin{figure}
\centering \vspace{-0.3cm} \hspace{-0cm}
    \includegraphics[width=9cm]{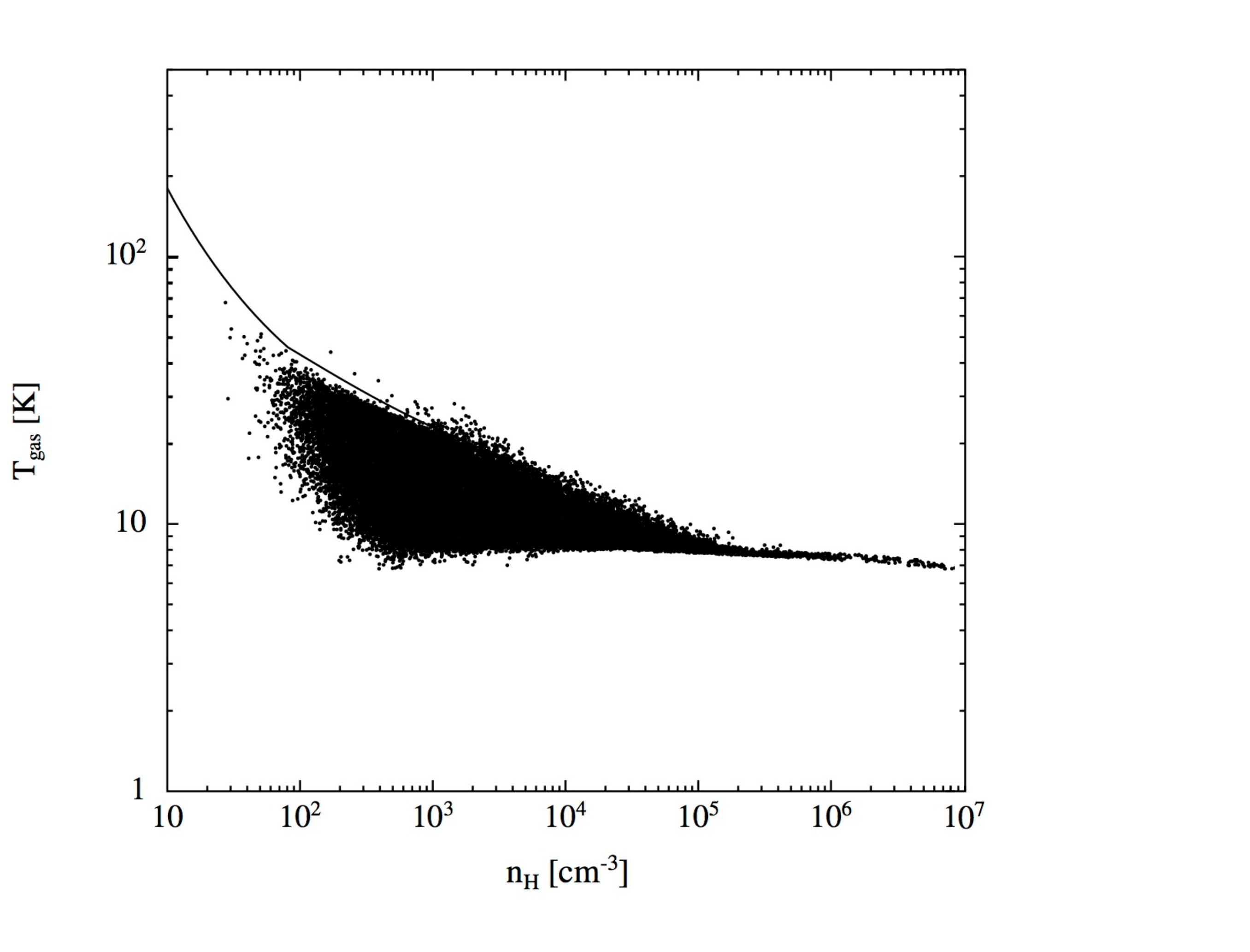} \vspace{-0.5cm}
\caption{Phase diagram showing the gas temperature as a function of the gas density for a calculation that is identical to that portrayed in Fig.~\ref{GC2012a_depletion}, except that all the carbon is assumed to be in the form of C$^+$.   The black solid line is the equilibrium curve from Fig.~\ref{T_n}. }
\label{GC2012a_cplus}
\end{figure}

\begin{figure}
\centering \vspace{-0.3cm} \hspace{-0cm}
    \includegraphics[width=9cm]{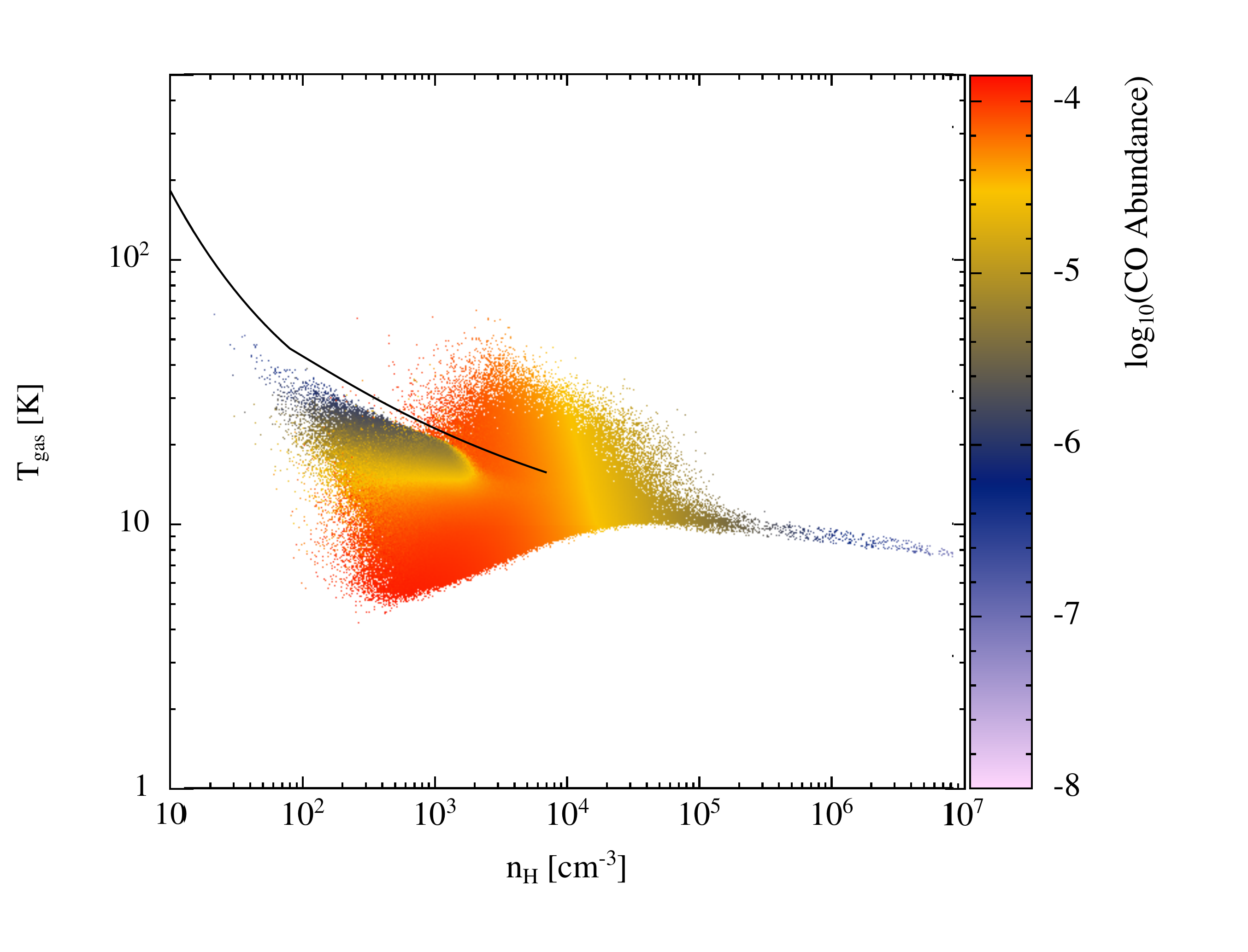} \vspace{-0.5cm}
\caption{Phase diagram showing the gas temperature as a function of the gas density for a calculation that is identical to that portrayed in Fig.~\ref{GC2012a_depletion}, except that the molecular line cooling rate has been increased by a factor of 3.  The colours in the gas temperature plot indicate the CO abundance of each point in temperature-density space.  The black solid line is the equilibrium curve from Fig.~\ref{T_n}. }
\label{GC2012a_line3}
\end{figure}

\begin{figure}
\centering \vspace{-0.3cm} \hspace{-0cm}
    \includegraphics[width=9cm]{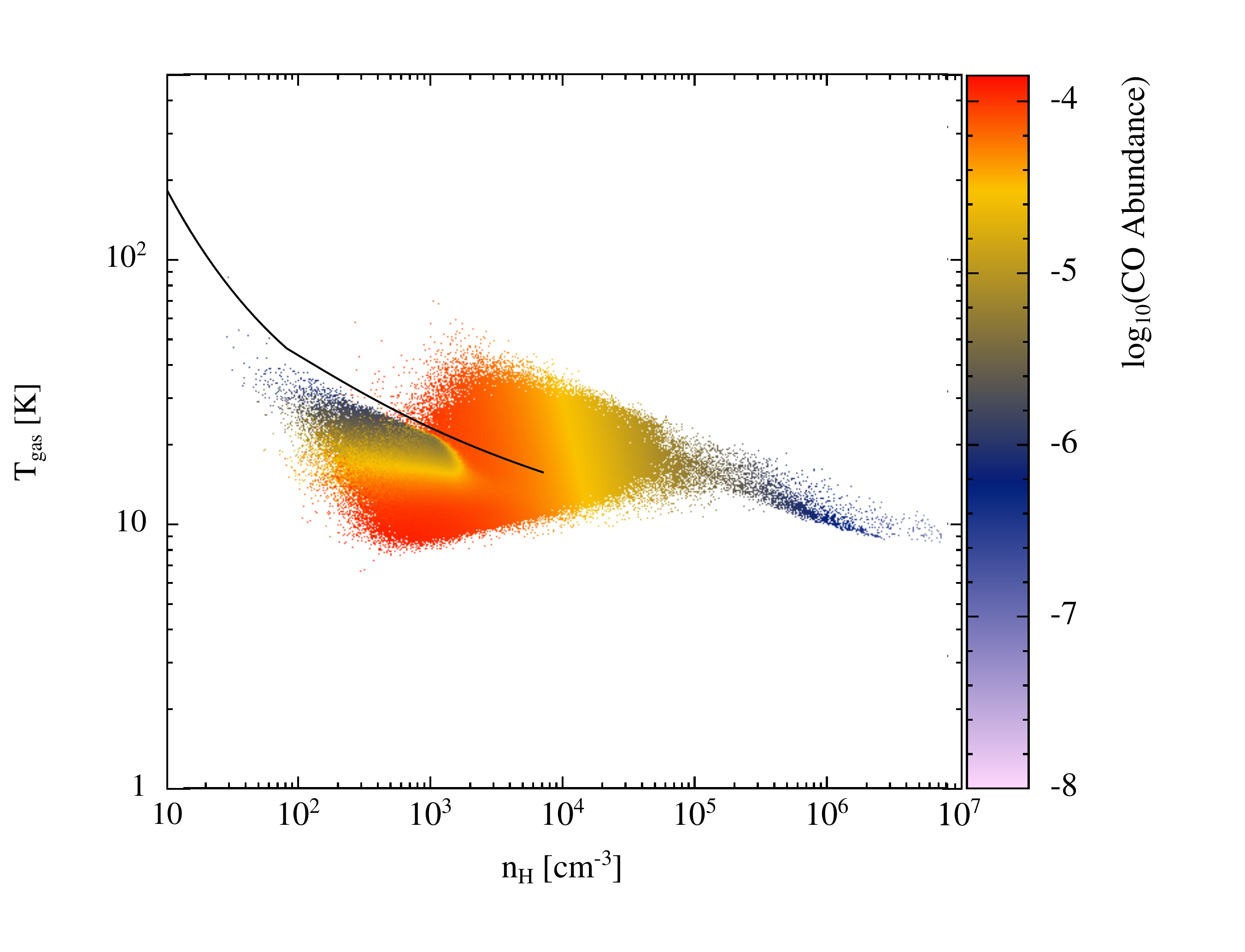} \vspace{-0.5cm}
\caption{Phase diagram showing the gas temperature as a function of the gas density for a calculation that is identical to that portrayed in Fig.~\ref{GC2012a_depletion}, except that the collisional thermal coupling between the dust and the gas is 15 times smaller (equation \ref{eq:gasdust} has been used rather than \ref{eq:gasdust2}).   Note that there is a larger dispersion in the gas temperatures at densities $n_{\rm H} \gsim 10^5$~cm$^{-3}$.  The black solid line is the equilibrium curve from Fig.~\ref{T_n}. }
\label{GC2012a_gasdust10}
\end{figure}

As mentioned above, \cite{GloCla2012a} do not treat depletion of CO onto dust grains.  Turning this on has little effect on the phase diagram (Fig.~\ref{GC2012a_depletion}).  As may be expected, the temperatures at densities of $n_{\rm H}\approx 10^3-10^5$~cm$^{-3}$ are slightly higher, but the effect is hardly noticeable.  At lower densities the CO is not depleted and the dominant coolant is C$^+$ rather than CO anyway.  At higher densities where the CO does become depleted, the dominant coolant is dust

We now explore the sensitivity of the results to several of the assumptions in our model.  We examine the dependence on changes of the chemistry, in particular the abundance of C$^+$ and the depletion of molecules like CO onto grains, as well as the effects of variations in the molecular line cooling rates and the strength of the coupling between the gas and the dust. 

The main effect of the carbon chemistry model described in Section \ref{sec:carbon} is to provide the abundances of the coolants C$^+$ at low-densities and CO at high densities.  Typically the transition between the dominance of the two coolants occurs at $n_{\rm H}\sim 10^3$~cm$^{-3}$ \citep[e.g.][]{GloCla2012a}.  The temperature of the gas is quite sensitive to this transition from C$^+$ to CO.  For example, in Fig.~\ref{GC2012a_cplus} we give the results obtained by assuming that the carbon remains as C$^+$ at all densities and we turn off all molecular cooling.  In this case, essentially all of the gas lies at temperatures below the equilibrium curve of Fig.~\ref{T_n} and there is little gas with temperatures greater than 20~K with densities $n_{\rm H}\approx 10^3-10^5$~cm$^{-3}$.  This extreme case shows that it is important to have at least a simple model that switches the carbon from C$^+$ to CO.

In Fig.~\ref{GC2012a_line3} we return to our standard model including molecular depletion, but we increase the molecular line cooling rates by an arbitrary factor of 3.  As expected, this decreases the temperatures of the gas with densities $n_{\rm H}\approx 10^3-10^5$~cm$^{-3}$.  The lowest temperature (at densities $n_{\rm H}\approx 10^3$~cm$^{-3}$) decrease from $T_{\rm g}\approx 8$ to $\approx 5$~K, but the highest temperatures (at densities $n_{\rm H}\approx 10^3$~cm$^{-4}$) are not significantly affected. 

As mentioned in Section \ref{sec:dust_equations}, equations \ref{eq:gasdust} and \ref{eq:gasdust2} for the gas-dust thermal coupling coefficient differ by a factor of 15 (the latter, used by \citeauthor{GloCla2012b}, gives stronger coupling).  Therefore, in Fig.~\ref{GC2012a_gasdust10} we examine the effect of using the weaker coupling provided by equation \ref{eq:gasdust} instead.  As expected, this only has an effect on the temperatures at high densities, $n_{\rm H}\gsim 10^5$~cm$^{-3}$.  The gas shows a somewhat increased spread of temperatures because the coupling with the dust is not strong enough to force the gas to adopt the dust temperature.

In summary, the gas temperatures in solar-metallicity turbulent clouds are relatively independent of many of the parameters of our diffuse ISM model.  The most important factors are the low-density equilibrium model (portrayed in Fig.~\ref{T_n}) which essentially sets an upper limit to the temperatures at densities $n_{\rm H}\lsim 10^3$~cm$^{-3}$, and the way in which the carbon transitions from C$^+$ to CO.  The exact molecular cooling rates, molecular depletion, and dust-gas thermal coupling parameters are less important.

\subsubsection{The effects of metallicity and hydrogen chemistry}

Following \cite{GloCla2012c}, we examine the dependence on metallicity and on hydrogen chemistry.  We perform full calculations (i.e. including the simple carbon chemical model and molecular depletion) for four different metallicities: 3, 1, 1/10, and 1/100 times the solar value.  For each metallicity, we perform calculations that exclude hydrogen chemistry and its associated heating terms (results from the solar-metallicity case have already been displayed in Fig.~\ref{GC2012a_depletion}).  We also perform calculations that include the evolution of hydrogen (i.e. its molecular fraction $x_{\rm H2}$) and heating due to H$_2$ formation on dust grains.  Two calculations are performed for each metallicity: one beginning with fully atomic hydrogen, and one beginning with fully molecular hydrogen.

\begin{figure}
\centering \vspace{-0.3cm} \hspace{-0cm}
    \includegraphics[width=8cm]{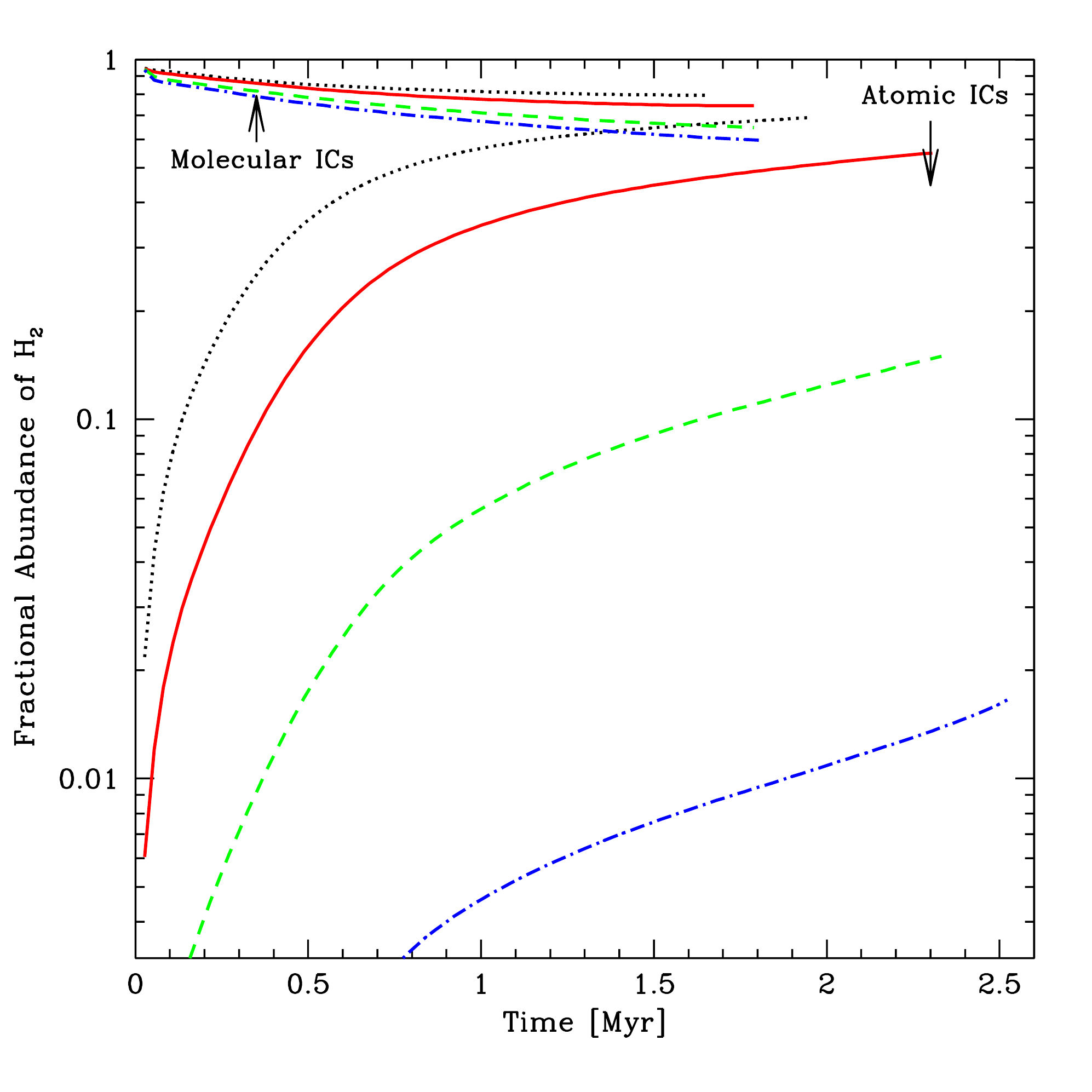} \vspace{-0.5cm}
\caption{We plot the evolution of the total fractional abundance of molecular hydrogen in calculations of turbulent clouds up until the time that stars begin to form.  For each metallicity, we perform two calculations: one with fully molecular initial conditions (upper lines), and one with fully atomic initial conditions (lower lines).   The metallicities are 3 times solar (black dotted lines), solar (red solid lines), 1/10 solar (green dashed lines), and 1/100 solar (blue dot-dashed lines).  By the time stars begin to form, clouds with solar or super-solar metallicity are mostly molecular, regardless of their initial H$_2$ abundance, but little molecular gas has formed in the low-metallicity clouds that consist of fully atomic hydrogen initially. }
\label{GC2012a_abundance}
\end{figure}

In Fig.~\ref{GC2012a_abundance} we plot the evolution of the fractional abundance of molecular hydrogen in the eight calculations that include hydrogen chemistry up until the formation of the first sink particle in each calculation.  This is the equivalent of Fig.~3 in \cite{GloCla2012c}.  Beginning with fully molecular hydrogen, the decrease in H$_2$ is primarily due to photodissociation in the outer parts of the clouds.  The destruction is greater at lower metallicities because the UV radiation is less attenuated by dust.  Beginning with fully atomic hydrogen, H$_2$ is formed on dust grains and the total abundance monotonically increases in each calculation.  However, the rate of increase strongly depends on the metallicity. The main reason for this is that there are more dust grains on which to produce H$_2$ at higher metallicity (equation \ref{eq:h2form1}) so the rate of formation is higher.  In addition, the extra dust attenuates the UV radiation that destroys H$_2$ so the destruction rate is also lower.  By the time stars begin to form, clouds with solar metallicity or higher are mostly molecular regardless of whether atomic or molecular initial conditions were adopted (particularly in the central regions where the stars actually form).  However, at lower metallicities the initial conditions matter much more.  When stars begin to form in the $1/10~{\rm Z}_\odot$ calculations, the H$_2$ abundance is 65\% when beginning with molecular initial conditions, but only 15\% beginning with atomic initial conditions.   For  $1/100~{\rm Z}_\odot$ the disparity is even stronger, with the two abundances being 60\% and 1.7\%, respectively.  

The behaviour of the molecular hydrogen evolution is qualitatively similar to that reported by  \cite{GloCla2012c}.  In particular, the growth rate of molecular hydrogen in the atomic solar-metallicity calculation matches theirs to within 10\%, and the decay rates of the abundances in all of the molecular calculations are very similar.  The clouds in  \cite{GloCla2012c} take longer to begin forming stars (2.7--4.0~Myr rather than 1.7--2.6~Myr).  This difference is most likely due to the different initial turbulent velocity field.  With fully molecular initial conditions, because the stars take longer to form in the calculations of  \cite{GloCla2012c}, the H$_2$ abundances when the star formation takes place are lower.  This is most important in the lowest metallicity case, where the total abundance is only $\approx 20$\% when the star formation begins in the  \cite{GloCla2012c} calculation, whereas it is 60\% in our calculation.  There are also differences in the abundances of molecular hydrogen in the lower metallicity calculations with atomic initial conditions.  The total H$_2$ abundances when beginning with fully atomic hydrogen remain low throughout both our calculations and those of \citeauthor{GloCla2012c}.  However, at any particular time our abundance is twice that obtained by \cite{GloCla2012c} with $1/10~{\rm Z}_\odot$ and it is about an order of magnitude higher for the $1/100~{\rm Z}_\odot$ case until shortly before stars begin to form (when the abundances in both calculations are $\approx 1$\%).  No doubt some of this difference is due to the different velocity field and dust opacities used in the calculations, though it is not clear whether or not this explains all of the discrepancy.

\begin{figure*}
\centering \vspace{0.0cm} \hspace{-0cm}
    \includegraphics[width=17cm]{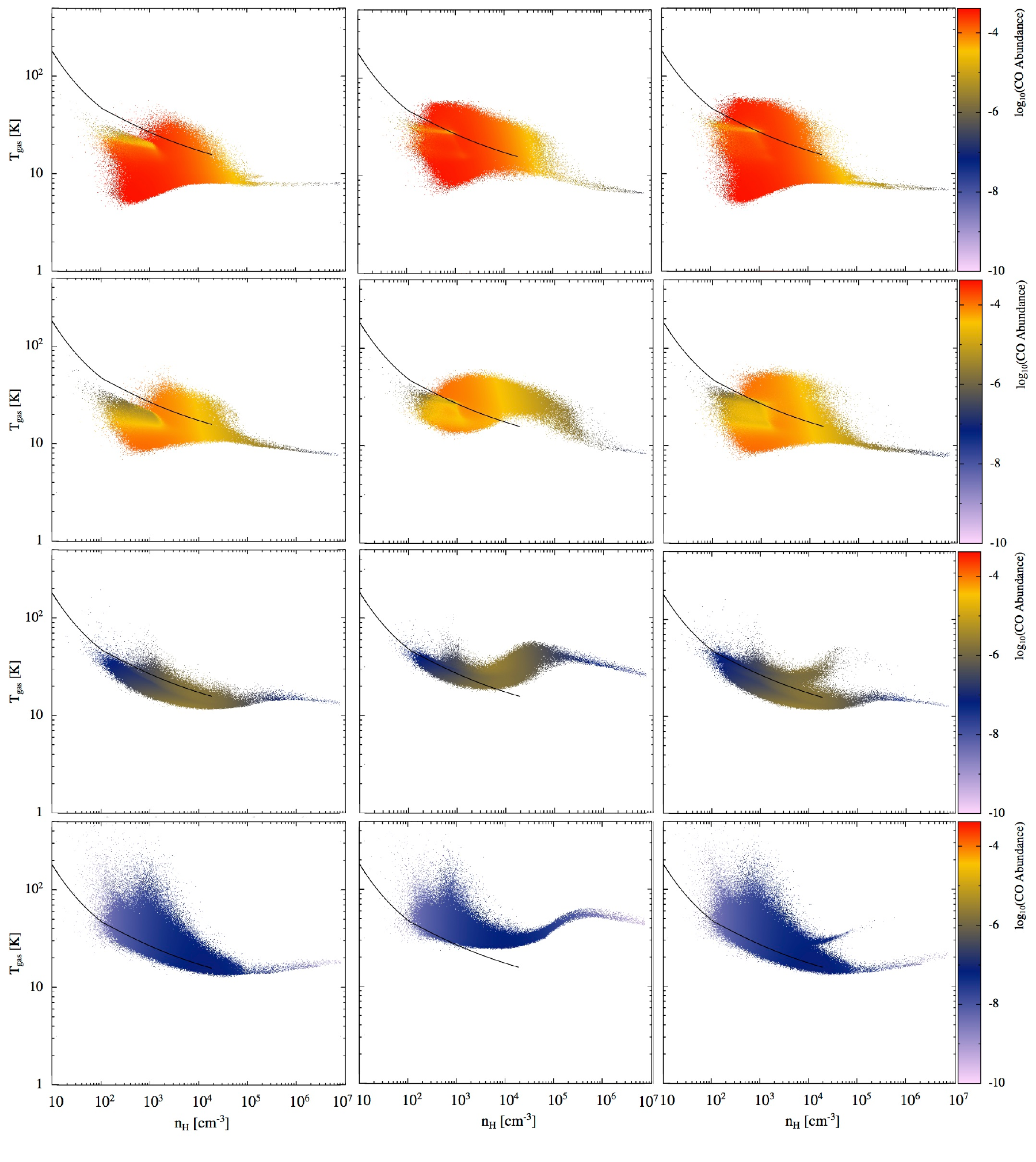}
\caption{Phase diagrams of the gas temperature versus the gas density in turbulent clouds with metallicities of 3, 1, 1/10, and 1/100 times solar (top to bottom, respectively).  The calculations in the left column were performed without any H$_2$ chemistry (assuming purely molecular hydrogen) and without heating from H$_2$ formation.  The other two columns included H$_2$ chemistry and heating from H$_2$ formation, but the calculations in the centre column began with fully atomic hydrogen, and the calculations in the right column began with fully molecular hydrogen.  The colour bars indicate the average CO abundance of each point in temperature-density space (note that the scales differ from those in Figs. \ref{GC2012a}, \ref{GC2012a_depletion}, \ref{GC2012a_line3}, and \ref{GC2012a_gasdust10}).  The black solid lines give the equilibrium curve from Fig.~\ref{T_n} for solar metallicity.  Beginning with atomic gas results in significantly higher gas temperatures for densities $n_{\rm H} \sim 10^4-10^6$~cm$^{-3}$ (or even higher densities in the low-metallicity cases) due to heating from H$_2$ formation on dust grains.  }
\label{GC2012_metal}
\end{figure*}

Turning to the phase diagrams of gas temperature in Fig.~\ref{GC2012_metal}, the general trends are again in agreement with those found by \cite{GloCla2012c}.  First we consider the calculations that exclude hydrogen chemistry (they assume all hydrogen is in molecular form) and its associated heating (left column of Fig~\ref{GC2012_metal}).  Increasing the metallicity above solar results in a reduction of the lower envelope of the gas temperatures at densities $n_{\rm H} = 10^2-10^4$~cm$^{-3}$ because the additional molecular cooling and increased extinction produces lower gas temperatures.  The upper temperature envelope is less affected.  Lowering the metallicity to 1/10 of solar results in a smaller range of temperatures because extinction has less of an effect at low and intermediate densities so the temperatures tend to lie close to the equilibrium curve of Fig.~\ref{T_n} (which does not depend significantly on metallicity because the photoelectric heating and the cooling due to electron recombination and fine structure emission are all assumed to be proportional to the metallicity).  At higher densities, the main heating sources (compressional heating and cosmic ray heating) are independent of the metallicity and although the gas is less well coupled to the dust it is still coupled well enough that the gas temperatures are kept around 10~K.  As the metallicity is reduced still further to 1/100 solar, the temperatures move above the solar-metallicity equilibrium curve of Fig.~\ref{T_n}.  This is because the cosmic ray heating (which is taken to be independent of metallicity) becomes as important as the photoelectric heating at low metallicity.  Furthermore, the magnitudes of the electron recombination and fine structure cooling rates are proportional to metallicity, so regions that are heated by adiabatic and shock heating take longer to cool.

At high densities $n_{\rm H} \gsim 10^5$~cm$^{-3}$ where the gas is thermally coupled to the dust in the solar metallicity case there is not much change in the super-solar metallicity case.  The temperatures are slightly cooler (about 1-2~K) because the coupling is even stronger and the dust has slightly cooler temperatures due to the increased extinction.  A larger effect is seen with reduced metallicities because the thermal coupling to the dust is substantially weaker.  At 1/10 solar metallicity, the gas has a maximum in the temperature distribution of about 15~K at $n_{\rm H} \approx 3 \times 10^5$~cm$^{-3}$ before the dust cooling takes over and reduces the temperatures at higher densities.  At 1/100 solar metallicity, the gas temperature rises with increasing density until $n_{\rm H} \approx 2 \times 10^7$~cm$^{-3}$ when it reaches a maximum of $\approx 20$~K before it begins to drop again at higher densities.

We now consider the effects of hydrogen chemistry on the gas temperature due to heating of the gas when molecular hydrogen is formed on dust grains (equation \ref{eq:h2dust}).  This has little effect in the calculations that began with fully molecular hydrogen (right column of Fig.~\ref{GC2012_metal}).  There are two small differences.  First, in the solar and super-solar calculations the upper envelope of gas temperatures is higher at densities $n_{\rm H} \approx 10^2-10^4$~cm$^{-3}$.  This is because some molecular hydrogen is destroyed by photodissociation and when it reforms on dust grains this adds a source of heating.  Second, in the low-metallicity calculations some regions of mostly atomic gas become molecular at densities $n_{\rm H} \approx 10^5-10^6$~cm$^{-3}$ and the extra heating makes this gas hotter than the molecular gas at the same densities.  Thus, a bifurcation of the gas temperature is apparent at these densities.

Beginning with fully atomic gas has a much greater effect on the gas temperature (centre column of Fig.~\ref{GC2012_metal}).  In these cases, heating due to H$_2$ formation on dust becomes significant at densities $n_{\rm H} \gsim 10^3$~cm$^{-3}$, and the gas temperature is generally much greater than in the calculations without hydrogen chemistry or with fully molecular hydrogen initially.  At solar and super-solar metallicities, the main effect is to raise the gas temperatures in the density range $n_{\rm H} \approx 10^3-10^6$~cm$^{-3}$.  Above these densities the gas is mostly molecular and the gas is thermally well-coupled to the dust.  At low metallicities, the gas is significantly hotter from densities of $n_{\rm H} \approx 10^3$~cm$^{-3}$ even up to $n_{\rm H} \approx 10^8$~cm$^{-3}$.  This is primarily because of the reduced dust abundance which has two main effects: the atomic hydrogen persists to higher densities, and the gas does not become thermally coupled to the gas until much higher densities.

As mentioned above, the general temperature trends seen here are in agreement with those found by \cite{GloCla2012c}.  However, there are also some differences.  At low-densites our temperatures tend to be lower because of the different equilibrium models that we have already mentioned (Fig.~\ref{T_n}).  But there are also differences at higher densities.  In the solar metallicity cases, \cite{GloCla2012c} obtain slightly lower minimum temperatures at  $n_{\rm H} \sim 10^3$~cm$^{-3}$.  This may be because their molecular cooling is somewhat stronger (c.f. the difference that increasing the molecular cooling makes in Figs.~\ref{GC2012a_depletion} and \ref{GC2012a_line3}).  In the calculations at 1/10~Z$_\odot$, the mean temperatures at each density are similar to those obtained by  \cite{GloCla2012c} for both atomic and molecular initial conditions, but the scatter in the temperatures obtained by \cite{GloCla2012c} is greater at $n_{\rm H} \gsim 10^5$~cm$^{-3}$.  This may be largely due to the different density structure of the clouds (caused by the different initial velocity fields).  In the lowest metallicity calculations (1/100~Z$_\odot$), the temperatures obtained by \cite{GloCla2012c} are generally higher than ours at $n_{\rm H} \gsim 10^4$~cm$^{-3}$ for the molecular initial conditions, and above $n_{\rm H} \gsim 10^6$~cm$^{-3}$ for the atomic initial conditions.  This may be related to the higher molecular hydrogen abundances that were commented on when we discussed Fig.~\ref{GC2012a_abundance}.  In our initially atomic calculation, the gas is mostly molecular for densities $n_{\rm H} \gsim 10^6$~cm$^{-3}$ meaning that there will be little heating from H$_2$ formation above these densities.  If \cite{GloCla2012c} have more atomic gas remaining at high densities this will provide additional heating.  In our initially molecular calculation, because the stars begin to form at 2.0 Myr instead of 3.6 Myr, more of the initial molecular hydrogen remains.  The total abundance is only 20\% in the calculation of \cite{GloCla2012c} when the temperature diagrams were plotted, whereas the total abundance in our calculation is 60\% and all of the high-density gas ($n_{\rm H} \gsim 10^5$~cm$^{-3}$) is essentially fully molecular (i.e. there is little heating from H$_2$ formation).  On the other hand, the temperature-density behaviour we obtain for the low-metallicity fully atomic initial conditions is in very good agreement with the results given in Fig.~1 of \cite{Omukaietal2005}.  For 1/10~Z$_\odot$ calculations, the temperature at high densities has a peak at $n_{\rm H} \sim 10^5$~cm$^{-3}$ at similar temperatures in both our calculations and those of \citeauthor{Omukaietal2005}  At 1/100~Z$_\odot$, both our results and those of \citeauthor{Omukaietal2005} display high-density temperature maxima of $T_{\rm g}\approx 60$~K at $n_{\rm H} \approx 10^6$~cm$^{-3}$.

\section{Conclusions}
\label{conclusions}

We have presented a new method for modelling the thermal evolution of star-forming molecular clouds.  The method combines a model for the thermodynamics of the diffuse interstellar medium with radiative transfer in the flux-limited diffusion approximation.  The former is required to correctly model the thermal behaviour of molecular clouds at low densities and metallicities, while the latter allows us to model protostar formation.  Unlike most previously published star formation calculations, our new model evolves the temperatures of the gas, dust, and radiation separately.  The code can also follow the evolution of atomic and molecular hydrogen.  We have compared our method with existing literature on the thermal behaviour of the ISM and molecular cloud cores and generally obtain good agreement.

We have also explored the sensitivity of the thermal structure of molecular cloud cores and turbulent clouds to many of the parameters that enter our diffuse ISM model.  For the gas at solar metallicities, the most important thermal processes at low densities ($n_{\rm H}\lsim 1-1000$~cm$^{-3}$) are photoelectric heating of electrons from dust grains and cooling due to electron recombination with small grains and PAHs and fine structure emission from C$^+$ and atomic oxygen.  At high densities  ($n_{\rm H}\gsim 10^5$~cm$^{-3}$), cosmic ray heating and the work done by hydrodynamical flows dominate the heating, while the cooling is dominated by continuum dust emission so that the gas and dust adopt similar temperatures.  At intermediate densities ($n_{\rm H}\lsim 10^3 - 10^5$~cm$^{-3}$), most processes have a significant effect and there tends to be a large dispersion of temperatures in turbulent clouds.    The abundance of C$^+$ changes rapidly at these densities and because C$^+$ is such an effective coolant it is important to have a model for the C$^+$ abundance.  However, the exact values of the molecular cooling rates, molecular depletion, and the strength of the thermal coupling between the dust and the gas are less important.  When beginning with low-density molecular clouds (which may not be purely molecular) the thermal evolution also depends significantly on relative abundances of atomic and molecular hydrogen due to heating from the formation of molecular hydrogen on dust grains.  This becomes more important at lower metallicities.

Our new method should allow more realistic radiation hydrodynamical calculations of star formation to be performed, particularly in molecular clouds that have low mean densities ($n_{\rm H}\lsim 10^4$~cm$^{-3}$) and/or sub-solar metallicities.  However, we also emphasise that this model is far from complete.  In particular, it relies on various parameterisations of the thermal effects of many physical processes, it does not calculate these processes explicitly.  Furthermore, it contains only extremely simple models of hydrogen and carbon chemistries and does not explicitly model the chemistry of other elements at all.  However, the basic model developed in this paper could easily be extended to increase the complexity of the chemical modelling (at the cost of increased computational expense).

\section*{Acknowledgments}

We thank Paul Clark for constructive criticism that helped improve the method, and the anonymous referee whose comments helped us improve the manuscript.
This work was supported by the European Research Council under the European Community's Seventh Framework Programme (FP7/2007-2013 Grant Agreement No. 339248).  The calculations for this paper were performed on the University of Exeter Supercomputer, 
a DiRAC Facility jointly funded by STFC, the Large Facilities Capital Fund of BIS, and the University of Exeter, and on the Complexity DiRAC Facility jointly funded by STFC and the Large Facilities Capital Fund of BIS.

\bibliography{mbate}

\end{document}